\newcounter{defthm}

\def\A             {\ensuremath{A}}
\def\AA            {{\!A}}
\def\alg           {algebra}

\def\Aut           {{\rm Aut}}

\def\bc            {boundary condition}

\def\be            {\begin{equation}}
\def\bea           {\begin{equation}\begin{array}l}
\def\beaa          {\begin{equation}\begin{array}{ll}}
\def\bearl         {\begin{array}{l}}
\def\bearll        {\begin{array}{ll}}

\def\C             {\ensuremath{\mathcal C}}
\def\CAA           {\ensuremath{\mathcal C_{\!A|A}}}

\def\calc          {\ensuremath{\mathcal C}}
\def\calca         {\ensuremath{\mathcal C_{\!A}}}

\def\calh          {\ensuremath{\mathcal H}}

\def\cat           {category}
\def\cats          {categories}
\def\Cf            {\mbox{\sl Cor}}

\def\cft           {conformal field the\-o\-ry}
\def\CFTA          {\mbox{\sc cft($A$)}}
\def\cfta          {{\footnotesize\Fbox{$A$}}}
\def\CFTB          {\mbox{\sc cft($B$)}}
\def\cftb          {{\footnotesize\Fbox{$B$}}}
\def\cftbtiny      {{\tiny\Fbox{$B$}}}
\def\CFTC          {\mbox{\sc cft($C$)}}
\def\cftc          {{\footnotesize\Fbox{$C$}}}

\def\cfts          {conformal field theories}

\def\chii          {\raisebox{.15em}{$\chi$}}

\def\cir           {\,{\circ}\,}
\newcommand\clc[2] {{\mathcal C}_{{#1}|{#2}}}

\def\complex       {\ensuremath{\mathbbm C}}
\def\con           {conformal }
\def\Con           {Conformal }
\def\Cong          {\,{\cong}\,}
\def\corfu         {correlation function}

\def\dim           {{\rm dim}}

\def\dimc          {{\rm dim}_{\scriptscriptstyle\complex}}

\def\dsty          {\displaystyle }

\def\eE            {{\rm e}}
\def\ee            {\end{equation}}
\def\eear          {\end{array}}

\def\End           {{\rm End}}
\newcommand\epicture[2] {\end{picture}\\{}\\[#1.#2em]\end{array}}
\def\eps           {\varepsilon}

\def\eq            {\,{=}\,}

\def\Equiv         {\,{\equiv}\,}
\newcommand\erf[1] {(\ref{#1})}

\newcommand\Fbox[1]{\shadowbox{#1}}

\def\FF            {{\sf F}}

\newcommand\Frac[2]{\mbox{\large$\frac{#1}{#2}$}}
\def\frob          {Fro\-be\-ni\-us algebra}
\newcommand\Fs[6]  {{\sf F}_{\,{#5}\,{#6}}^{\,({#1}\,{#2}\,{#3})\,{#4}}}

\def\ft            {field theory}

\def\furu          {fusion rule}

\def\GammA         {\Gamma}
\def\GammB         {\Lambda}

\def\Hom           {{\rm Hom}}
\def\HomA          {{\rm Hom}_{\!A}}
\def\HomAA         {{\rm Hom}_{\!A|A}}
\def\HomAB         {{\rm Hom}_{\!A|B}}
\def\HomAC         {{\rm Hom}_{\!A|C}}
\def\HomB          {{\rm Hom}_{\!B}}
\def\HomBA         {{\rm Hom}_{\!B|A}}
\def\HomBB         {{\rm Hom}_{\!B|B}}
\def\HomBC         {{\rm Hom}_{\!B|C}}
\newcommand\hsp[1] {\mbox{\hspace{#1 em}}}
\def\hy            {$\mbox{-\hspace{-.66 mm}-}$}
\def\I             {\mbox{$\II$}}

\def\ib            {{\bar\imath}}
\def\id            {\mbox{\sl id}}

\def\idsmall       {\mbox{\scriptsize\sl id}}

\def\II            {\mathcal I}

\def\iN            {\,{\in}\,}

\newcommand\includeourbeautifulpicture[2] {{\begin{picture}(0,0)(0,0)
            \scalebox{.38}{\includegraphics{pic_ffrs5_#1#2.eps}} \end{picture}}}
\newcommand\Includeourbeautifulpicture[3] {{\begin{picture}(0,0)(0,0)
            \scalebox{.38}{\includegraphics{pic_ffrs5_#1#2#3.eps}} \end{picture}}}
\newcommand\Includeournicelargepicture[3] {{\begin{picture}(0,0)(0,0)
            \scalebox{.28}{\includegraphics{pic_ffrs5_#1#2#3.eps}} \end{picture}}}

\newcommand\Includeoursmallnicepicture[3] {{\begin{picture}(0,0)(0,0)
            \scalebox{.65}{\includegraphics{pic_ffrs5_#1#2#3.eps}} \end{picture}}}
\newcommand\Includeourtinynicepicture[3] {{\begin{picture}(0,0)(0,0)
            \scalebox{.90}{\includegraphics{pic_ffrs5_#1#2#3.eps}} \end{picture}}}
\def\inda          {{\rm Ind}_{A\!}}
\def\indb          {{\rm Ind}_{B\!}}

\def\Itemize       {\def\leftmargini{1.8em}\begin{itemize}\addtolength\itemsep{-3pt}} 
\def\ITemize       {\def\leftmargini{1.0em}\begin{itemize}\addtolength\itemsep{-3pt}} 
\def\J             {\mbox{$\JJ$}}
\def\jb            {{\bar\jmath}}
\def\JJ            {\mathcal J}
\def\K             {{\rm K}}

\def\KK            {\mathcal{K}}

\newcommand\labl[1]{\label{#1}\ee}

\def\M             {{\rm M}}

\def\modinv        {modular invarian}

\def\MX            {{\rm M}_{\rm X}}

\def\nE            {\,{\ne}\,}
\def\nxt           {\raisebox{.08em}{\rule{.44em}{.44em}}\hsp{.4}}

\def\oa            {operator algebra}

\def\objc          {{\mathcal O}bj(\calc)}

\def\one           {{\bf1}}

\def\ota           {\,{\otimes}_{\!A}^{}\,}
\def\otA           {\,{\otimes}_{\!A}^{}\,}
\def\OtA           {{\otimes}_{\!A}^{}}
\def\otB           {\,{\otimes}_{\!B}^{}\,}
\def\OtB           {{\otimes}_{\!B}^{}}
\def\otC           {\,{\otimes}_{C}^{}\,}

\def\oti           {\,{\otimes}\,}
\def\Oti           {{\otimes}}
\def\otim          {\,{\otimes^-}\,}
\def\otip          {\,{\otimes^+}\,}
\def\otiP          {\,{\otimes^+}}
\def\parfu         {partition function}
\def\PicC          {\ensuremath{\mathrm{Pic}(\C)}}
\def\PicCAA        {\ensuremath{\mathrm{Pic}(\CAA)}}
\newcommand\ppmatrix[4]{\mbox{{\Large(}$\!\!\begin{array}{cc}{}\\[-1.8em]
                   \scs #1\!&\!\!\scs #2\\[-.39em]\scs #3\!&\!\!\scs #4
                   \\[-.19em]\eear\!\!${\Large)}}}
\newcommand\ppvector[2]{\mbox{{\Large(}$\!\!\begin{array}{cc}{}\\[-1.8em]
                   \!\scs #1\!\\[-.39em]\!\scs #2\!\\[-.19em]\eear\!\!${\Large)}}}
\def\q             {quantum }
\def\Q             {Quantum }
\def\qed           {\hfill\checkmark\medskip}

\def\qg            {quantum group}

\def\rep           {representation}
\def\Rep           {{\mathcal R}ep}

\def\rhs           {right hand side}

\def\reals         {\mbox{$\mathbb R$}}

\newcommand\Rs[4]  {{\sf R}^{#1\,(#2\,#3)\,#4}}

\def\SCA           {{\mathrm{S}}} 
\def\scs           {\scriptstyle}
\newcommand\sect[1]{\section{#1}\setcounter{equation}0\setcounter{defthm}0}

\def\SymGen        {\ensuremath{\mathrm{Sy}_{\!\mathrm{gen}}}}
\def\tcs           {tensor categories}

\def\tft           {topological field theory}

\def\Times         {\,{\times}\,}
\def\To            {\,{\to}\,}

\def\tr            {{\rm tr}\,}

\def\twodim        {two-di\-men\-si\-o\-nal}

\def\X             {\ensuremath{\mathrm X}}
\newcommand\xlabp[2]{}
\newcommand\Xlabp[3]{}

\def\zet           {\ensuremath{\mathbb Z}}

\documentclass[12pt]{article}

\usepackage{latexsym,amsmath,amssymb,amsfonts,epsf,color,colordvi,fancybox}
\usepackage{bbm}
\usepackage{graphics} 
\usepackage[mathscr]{eucal}
\usepackage{rotating}

\newtheorem{defthm}{\whattheorem}[section]
\newcommand\dt[1]  {\noindent\def\whattheorem{#1}\pagebreak[0]\begin{defthm}{}%
                   \samepage{$\!\!${\rm:}\nopagebreak\\[-1.91em]{}}\end{defthm}}
\newcommand\dtl[2] {\noindent\def\whattheorem{#1}\pagebreak[0]\begin{defthm}{}%
                   \samepage{$\!\!${\rm:}\label{#2}\nopagebreak\\[-1.91em]{}}%
                   \end{defthm}}

\def\frsi   {\cite{tft1}} 
 \def\frsiv  {\cite{tft4}}
\def\citeOtoV  {[0--V]}

\begin{document}
\thispagestyle{empty}

\begin{flushright}  {~} \\[-12mm]
{\sf hep-th/0607247}\\[1mm]{\sf KCL-MTH-06-08}\\[1mm]
{\sf Hamburger$\;$Beitr\"age$\;$zur$\;$Mathematik$\;$Nr.$\;$255}\\[1mm]
{\sf ZMP-HH/06-11}\\[1mm]
{\sf July 2006} \end{flushright}
\begin{center} \vskip 13mm
{\Large\bf DUALITY AND DEFECTS}\\[4mm]
{\Large\bf IN RATIONAL CONFORMAL FIELD THEORY}\\[18mm]
{\large J\"urg Fr\"ohlich$\;^1$ \ \ \ J\"urgen Fuchs$\;^2$ \ \ \ 
Ingo Runkel$\;^3$ \ \ \ Christoph Schweigert$\;^{4}$}
\\[8mm]
$^1\;$ Institut f\"ur Theoretische Physik, ETH Z\"urich\\
       CH\,--\,8093\, Z\"urich\\[5mm]
$^2\;$ Avdelning fysik, \ Karlstads Universitet\\
       Universitetsgatan 5, \ S\,--\,651\,88\, Karlstad\\[5mm]
$^3\;$ Department of Mathematics\\ King's College London, Strand\\
       London WC2R 2LS\\[5mm]
$^4\;$ Organisationseinheit Mathematik, \ Universit\"at Hamburg\\
       Schwerpunkt Algebra und Zahlentheorie\\
       and Zentrum f\"ur mathematische Physik, \\
       Bundesstra\ss e 55, \ D\,--\,20\,146\, Hamburg
\end{center}
\vskip 12mm

\begin{quote}{\bf Abstract}\\[1mm]
We study topological defect lines in two-dimensional 
rational conformal field theory. Continuous variation of the location 
of such a defect does not change the value of a correlator.
Defects separating different phases of local CFTs with the same
chiral symmetry are included in our discussion. We show how the
resulting one-dimensional phase boundaries can be used to extract
symmetries and order-disorder dualities of the CFT.
\\
The case of central charge $c=4/5$, for which there are two inequivalent 
world sheet phases corresponding to the tetra-critical Ising model and 
the critical three-states Potts model, is treated as an illustrative example.
\end{quote} \vfill \newpage

\tableofcontents


\sect{Introduction and summary}

Symmetries in general, and symmetry groups in particular, provide significant 
insight in the structure of physical theories. It is thus of much 
interest to gain a firm conceptual and computational handle on symmetries of 
conformal field theories. A recent formulation of rational CFT, the so-called 
TFT approach (see \citeOtoV\ and sections \ref{sec:chiral}, \ref{sec:cft-alg-rel} 
and \ref{sec:calc-tft} below) allows one to address these issues 
from a novel point of view. The TFT approach accounts for chiral symmetries 
which lead to conserved chiral quantities by working with the representation
category of the chiral symmetry algebra; in rational CFT, this is a modular 
tensor category \calc. One can typically construct several full CFTs which 
have a given chiral algebra as part of their chiral symmetry; the free boson 
compactified at different radii (squaring to a rational number) and the A-D-E 
modular invariants of the $\mathfrak{sl}(2)$ WZW model are examples of 
this. Therefore, beyond \calc\ one also needs an additional datum to 
characterise a full CFT model. This turns out to be a (Morita class of) 
Frobenius algebra(s) $A$ in \calc. Accordingly, also other symmetries than the 
chiral symmetries can be present; e.g.\ for the critical Ising model the 
$\zet_2$ transformation which exchanges spin up and spin down, or the 
$S_3$-symmetry for the critical three-states Potts model. We refer to such 
additional symmetries as {\em internal\/} symmetries of the full CFT.

On the basis of the TFT approach we show that the internal symmetries of a CFT 
$(\calc,A)$ are closely related to a certain class of conformal defect lines 
that preserve all the chiral symmetries. These defect lines can be used to 
construct nontrivial mappings of CFT data like boundary conditions and
bulk or boundary fields, and to establish invariance of correlation functions 
under those transformations. The construction uses a generalisation 
of contour deformation arguments familiar from complex analysis.

In more technical terms, the group of internal symmetries that are described 
by defects turns out to be the Picard group 
\PicCAA\ of the tensor category of $A$-bimodules. Naively one might instead 
have expected the symmetry group to be the group of automorphisms of the
algebra $A$. This would, however, be incompatible with the result that the 
CFT depends only on the Morita class of $A$ (%
    \!\!%
\cite{tft0}, for a proof see 
section \ref{sec:morita} of the present paper): Morita equivalent algebras 
have, in general, non-isomorphic automorphism groups. It turns out 
(see proposition 7 of \cite{fuRs11}, which extends a result in \cite{vazh})
that in fact the group of outer automorphisms of $A$ is a subgroup of \PicCAA.

This is not the end of the story:
Besides internal symmetries, it is known that certain CFTs
exhibit order-disorder dualities of Kramers\hy Wannier type.    
We show that their presence can be deduced with similar arguments from the 
existence of defect lines of a more general type. A relation between 
Kramers\hy Wannier dualities and defects might actually have been suspected, 
since such dualities relate bulk fields to disorder fields which, after all, 
are located at the end points of defect lines. However, the defect lines that 
we show to be relevant for obtaining the dualities are of a different 
type than the ones created by the dual disorder fields.

\medskip

Let us give a general description of defects in CFT in somewhat more detail.
Consider two CFTs defined on the upper and lower half of the complex plane, 
respectively, which are joined together along an interface -- or defect -- on 
the real line. To characterise the defect, one must consider the behaviour
of the holomorphic and antiholomorphic components 
$T^1(z)$, $\overline T{}^1(z)$ and $T^2(z)$, $\overline T{}^2(z)$ of the stress 
tensors of the two CFTs. If the stress tensors obey the conservation law
  \be
  T^1(x) - \overline T{}^1(x) =  T^2(x) - \overline T{}^2(x)
  \qquad \text{for all} ~~ x \iN \reals
  \labl{eq:conf-defect-bc}
then the defect is called {\em conformal\/}. This condition can be implemented 
by the so-called `folding trick', in which one
identifies  the lower and upper complex half planes, thereby giving rise to a tensor 
product of two CFTs on the upper half plane \cite{woaf}. 
The condition \erf{eq:conf-defect-bc} then states that the resulting boundary 
condition on the real line is conformal in the sense of \cite{card}. 
Conformal defects have been investigated with the help of the folding trick in 
e.g.\ \cite{osaf,osaf2,bddo,qusc}.

There are two obvious solutions to the condition \erf{eq:conf-defect-bc},
namely $T^1(x)\eq\overline T{}^1(x)$, $T^2(x) \eq \overline T{}^2(x)$ and
$T^1(x)\eq T^2(x)$, $\overline T{}^1(x) \eq \overline T{}^2(x)$. The former is
the {\em totally reflective\/} case, for which the defect can be regarded as
describing a conformal boundary for each of the two CFTs individually, so that
the two theories are completely decoupled. In the second case the two theories 
are maximally coupled in the sense that the defect is {\em totally transmissive\/}
for momentum. If $T^1(x)\eq T^2(x)$ on the real line, then the operator products
$T^1(x)T^1(y)$ and $T^2(x)T^2(y)$ must agree, so that in particular totally 
transmissive defects can only exist if the two CFTs have equal central charge.

In this article we shall investigate the totally transmissive case.
Such defects have been studied in the framework of rational CFT in e.g.\
\cite{watt8,pezu5,pezu7,coSc,tft1,tft2,grwa,tft3,ffrs3,tft4}.
Integrable lattice realisations have been found in \cite{cmop,otpe2}.
Totally transmissive defects commute by definition with the action of the 
generators $L_m$, $\overline L_m$ of conformal transformations; they
can be deformed without affecting the value of correlators,
as long as they are not taken across a field insertion.
Accordingly we will follow \cite{baGa} in referring to
totally transmissive defects also as {\em topological defects\/}.

Topological defects appear in fact very naturally in the context
of lattice models when one studies order-disorder dualities
\cite{krwa,savi,drwa}. In that context, a correlator of local fields is
equated to a correlator of lines of inhomogeneity in the lattice
couplings which end at the positions of the field insertions.
The path by which these lines connect the various insertion points is
a gauge choice and can be varied without changing the value of
the correlator. The continuum limit of such lines of inhomogeneity
provides one example of topological defects.

Correlators of disorder fields related to the internal $\zet_2$
symmetry of certain minimal models have been considered
in \cite{petk,fugp}, while correlators of disorder fields for
a $\zet_2$ symmetry which in addition obey a level-2 null vector
relation have recently been given a statistical interpretation
in terms of winding numbers in a loop gas model \cite{gaCa}.

Topological defects also appear when compactifying topologically twisted 
four-dimensional $N\eq4$ super Yang-Mills theory to two dimensions
\cite{kawi}. They arise as the images of certain topological Wilson line
operators in four dimensions.

\medskip

\begin{figure}[tb]
\begin{center}
  \begin{picture}(300,80)(0,0)
  \put(0,-10){
    \put(0,0){\scalebox{.90}{\includegraphics{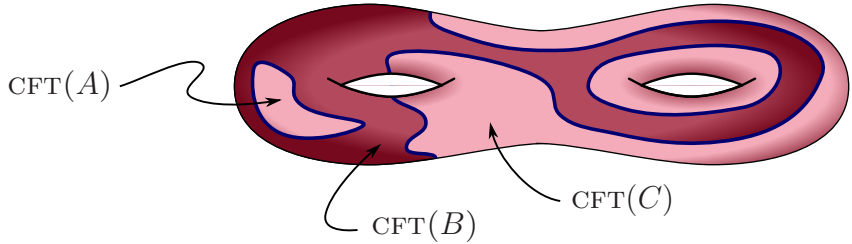}}}
    \setlength{\unitlength}{.90pt}
    \put(-45,58){$\CFTA$}
    \put(108,-1){$\CFTB$}
    \put(192,10){{\sc cft}($C$)}
    \setlength{\unitlength}{1pt}
    }  
  \end{picture}
\end{center}
\caption{A genus two world sheet with different phases
of full CFT. The phases meet along one-dimensional phase
boundaries which are topological defects.}
\label{fig:ws-phases}
\end{figure}

We are interested in situations like the one in figure \ref{fig:ws-phases},
where different parts of a world sheet can be in different `phases' which are 
joined by one-dimensional phase boundaries. By this we mean the following. 
The chiral symmetry of a rational CFT can be encoded in a rational vertex 
algebra $\mathcal{V}$. As mentioned above, one can fix a full CFT with 
symmetry $\mathcal{V}$ by giving a symmetric special Frobenius algebra $A$ in 
$\calc\eq\Rep(\mathcal{V})$. 
If the field content, OPEs, possible boundary conditions, etc.\ in some region 
of the world sheet are that of a full CFT obtained from an algebra $A$, we say 
that this part of the world sheet is in {\em phase\/} $\CFTA$. Different 
regions, with (generically different, but possibly also equal) phases, meet 
along phase boundaries. These phase boundaries can carry their own fields, 
called defect fields, which can in particular change the type of phase boundary.
Here we will only consider phase boundaries that are transparent to all fields 
in $\mathcal{V}$, so that they are described by topological defects.
An important aspect of the TFT approach
is that CFT quantities have a natural counterpart in the category \C.
For example, boundary conditions preserving $\mathcal{V}$
correspond to modules of the algebra $A$, and topological defects
that separate two phases $\CFTA$ and $\CFTB$ on a world sheet
are given by $A$-$B$-bimodules. If a defect cannot be written
as a superposition of other defects, it corresponds to a simple bimodule; 
accordingly we refer to such a defect also as a `simple' defect.

Consider two simple defects running parallel to each other. In the limit of 
vanishing distance they fuse to a single defect which is, in general,
not a simple defect. This gives rise to a fusion algebra of
defects \cite{pezu5,pezu6,cmop,coSc}.
In terms of the category \C, this corresponds to the tensor product of
bimodules \cite{tft1,ffrs3}. Specifically, if the two defects are
labelled by an $A$-$B$-bimodule $X$ and a $B$-$C$-bimodule $Y$,
then the fused defect is labelled by the $A$-$C$-bimodule $X \otB Y$. 

While there is a version of the TFT approach also for unoriented world sheets, 
apart from section \ref{sec:morita-unor} we will in this paper only consider 
oriented world sheets. Some further information on defects in the unoriented 
case can be found in section 3.8 of \cite{tft2}.

\medskip\noindent
The main results of this paper are the following.
\\[.2em]
\nxt 
To distinguish two topological defects labelled by non-isomorphic bimodules it 
suffices to compare their action on bulk fields (as opposed to considering e.g.\
also the action on disorder fields or on boundary conditions); 
the action on a bulk field consists in surrounding the field  
with a little defect loop and contracting that loop to zero 
size. This applies in particular to defects which separate different
phases $\CFTA$ and $\CFTB$, so that their action turns a bulk field
of $\CFTA$ into a bulk field of $\CFTB$ (which may be zero).
This is shown in proposition \ref{prop:g-distinct}.
\\[.2em]
\nxt 
While every symmetric special Frobenius algebra $A$ in 
$\calc\eq\Rep(\mathcal{V})$ gives rise to a full CFT, not all of
these full CFTs are distinct. As already announced in \cite{tft0}, Morita 
equivalent algebras give equivalent full CFTs on oriented world sheets. The 
proof of this claim, together with an explanation of the relevant concept of
equivalence of full CFTs, is provided in section \ref{sec:morita}; it requires 
appropriate manipulations of phase-changing defects. In section 
\ref{sec:morita-unor} analogous considerations
are carried out for unoriented world sheets, leading to the
notion of `Jandl-Morita equivalence' of algebras.
\\[.2em]
\nxt 
In addition to being labelled by a bimodule, defect lines also carry an 
orientation. A defect $X$ which has the same full CFT on either side
is called {\em group-like\/} iff, upon fusing two defects which are both 
labelled by $X$ but which have opposite orientation, one is left only with the 
invisible defect, i.e.\ the two defects disappear from the world sheet. Such 
defects form a group. It turns out that this is a group of internal symmetries 
for CFT correlators on world sheets of arbitrary genus; this is established in 
section \ref{sec:group-sym} (see also \cite{ffrs3}). Owing to the 
result mentioned in the first point, two different defect-induced symmetries 
can already be distinguished by their action on bulk fields.
\\[.2em]
\nxt 
In theorem \ref{thm:dual} we extend these results to order-disorder dualities.
We establish a simple characterisation of the defects that lead to such 
dualities: A defect $Y$ is a duality defect iff there is at least one other 
defect, labelled by $Y'$, say, such that fusing $Y$ and $Y'$ results in a 
superposition of only group-like defects. Moreover, in this case one can 
simply take $Y'\eq Y$, but with orientation opposite to that of the defect $Y$.
This allows one to read off duality symmetries of the CFT from the fusion 
algebra of defects. Such dualities can relate correlators of different
full CFTs, too. This is shown in the example of the tetracritical Ising model 
and the three-states Potts model in section \ref{sec:M56-phasechange}.
\\[.2em]
\nxt 
In proposition \ref{prop:orbifold} we show that if two phases $\CFTA$ and 
$\CFTB$ can be separated by a duality defect $Y$, then the torus partition 
function of $\CFTA$ can be written in terms of partition functions with defect 
lines of $\CFTB$ in a manner that resembles the way the partition function of an
orbifold is expressed as a sum over twisted partition functions. The relevant 
group elements are the group-like defects appearing in the fusion of the defect 
$Y$ with an orientation-reversed copy of itself. In the case $A\eq B$ this is 
precisely the `auto-orbifold property' observed in \cite{ruel'5}.
\\[.2em]
\nxt 
When the full CFT is of simple current type, one can use tools developed in 
\cite{tft3} to study topological defects; this is done in section
\ref{sec:sicu}. For instance, one can make general statements about the
symmetry group, and in particular determine a subgroup that is present
generically in all models of a given simple current type, see section 
\ref{sec:SymGen}.
\\[.2em]
\nxt 
In proposition \ref{prop:UiAUj-submod} we establish
a result for bimodules of simple symmetric Frobenius algebras in modular tensor
categories, namely that every $A$-$A$-bimodule is a submodule of the tensor 
product of two $\alpha$-induced bimodules with opposite braiding in the two 
$\alpha$-inductions. An analogous statement has been proved in \cite{boek} in 
the context of subfactors, and has been conjectured for modular tensor 
categories in general in \cite{ostr}. This result has a simple interpretation 
in terms of the CFT associated to the algebra $A$: it is equivalent to the 
statement that every defect line that has the phase $\CFTA$ on both sides 
can end in the bulk at an appropriate disorder field.


\sect{Algebras and CFT}

As mentioned in the introduction, in the TFT approach to RCFT a full CFT is
given by two pieces of data: a rational vertex algebra $\mathcal{V}$, which
encodes the chiral symmetries of the CFT, and a symmetric special Frobenius
algebra $A$ in the representation category $\calc \eq \Rep(\mathcal{V})$, which
determines the full CFT whose chiral symmetries include $\mathcal{V}$.

In section \ref{sec:chiral} we review some aspects of algebra in braided tensor
categories, in section \ref{sec:cft-alg-rel} the relation of these quantities
to CFT is described, and in section \ref{sec:calc-def} we describe some 
identities
that are useful in calculations with correlators involving topological defect
lines. These rules will be derived later in section \ref{sec:calc-tft} with the
help of the TFT approach.  Finally, in section \ref{sec:distinct} it is 
established that all defects can be distinguished by their action on bulk fields.


\subsection{Algebraic preliminaries}\label{sec:chiral}

In order to investigate the properties of topological defects we will need the
notion of algebras, modules and bimodules in tensor categories and, more
specifically, in modular tensor categories. We adopt the following definition,
which is slightly more restrictive than the original one in \cite{TUra}.
\\[-2.2em]

\dtl{Definition}{def:modular}
A tensor category is called {\em modular\/} iff
\\[.2em]
(i)\phantom{ii}
  The tensor unit is simple.\,\footnote{~%
In a semisimple \complex-linear tensor category every object $U$ that is
simple (i.e., does not have any proper subobject), is also absolutely
simple, i.e.\ satisfies $\End(U) \Cong \complex\,\id_U$.}
\\[.2em]
(ii)\phantom{i}
  $\calc$ is abelian, $\complex$-linear and semisimple.
\\[.2em]
(iii)~\,%
  $\calc$ is ribbon:
  There are families $\{c_{U,V}\}$ of braiding, $\{\theta_U\}$ of twist, and
  $\{d_U,b_U\}$ of evaluation and coevaluation morphisms satisfying the
  relevant properties (see e.g.\ definition 2.1 in \cite{ffrs}).
\\[.2em]
(iv)
  $\calc$ is Artinian (or `finite'), i.e.\ the number of isomorphism
  classes of simple objects is finite.
\\[.2em]
(v)~\,%
  The braiding is maximally non-degenerate: the numerical matrix $s$ with
  entries $s_{i,j}$ ${:=}$ $(d_{U_j}^{}\oti\tilde d_{U_i}^{}) \cir
  [\id_{U_j^\vee}\oti(c_{U_i,U_j}^{}{\circ}\,c_{U_j,U_i}^{})\oti\id_{U_i^\vee}]
  \cir (\tilde b_{U_j}^{}\oti b_{U_i}^{})\,$
  for $i,j\iN\I\,$ is invertible.
\\[.2em]
Here we denote by $\{U_i\,|\,i\iN\I\}$ a (finite) set of representatives
of isomorphism classes of simple objects; we also take $U_0\,{:=}\,\one$ as the
representative for the class of the tensor unit.
Further, $\tilde d_U$ and $\tilde b_U$ denote the left duality morphisms
constructed from right duality, braiding and twist.

\dtl{Remark}{rem:modular}
(i)~\,The relevance of modular tensor categories for the present discussion
derives from the fact that for certain conformal vertex algebras $\mathcal{V}$,
the category $\Rep(\mathcal{V})$ of representations is modular.
It is this class of vertex algebras to which we refer as being `rational';
sufficient conditions on $\mathcal{V}$ for $\Rep(\mathcal{V})$ to be modular
have been derived in \cite{huan24}.
\\
However, it is important to stress that for the algebraic computations
involving modular categories we do not assume that the category in question
can be realised as $\Rep(\mathcal{V})$ for some $\mathcal{V}$.
\\[2pt]
(ii) We do not require the modular tensor categories we consider to be unitary
(in the sense of \cite[Sect.\,II:5.5]{TUra}). In particular, we allow for the
possibility that some simple objects have negative quantum dimension, and
that some non-simple objects have zero quantum dimension.

\medskip

We will make ample use of the graphical notation of \cite{joSt5} to represent
morphisms in braided tensor categories. Our conventions, as well as more
references, are given in \cite[sect.\,2]{tft1} and \cite[sect.\,2.1]{ffrs}.

Let $\calc$ be a tensor category.
A {\em Frobenius algebra\/} $A\eq(A,m,\eta,\Delta,\eps)$ in $\calc$ is, by
definition, an object of $\calc$ carrying the structures of a unital associative
algebra $(A,m,\eta)$ and of a counital coassociative coalgebra $(A,\Delta,\eps)$
in $\calc$, with the algebra and coalgebra structures satisfying the
compatibility requirement that the coproduct $\Delta{:}\ A \,{\to}\, A\oti A$
is a morphism of $A$-bimodules (or, equivalently, that the product $m{:}\ A\oti
A\,{\to}\,A$ is a morphism of $A$-bi-comodules). A Frobenius algebra is called
{\em special\/} iff a scalar multiple of the coproduct is a right-in\-ver\-se to
the product -- this means in particular that the algebra is separable -- and a
multiple of the unit $\eta{:}\ \one\,{\to}\,A$ is a right-inverse to the counit
$\eps{:}\ A\,{\to}\,\one$. Suppose now that $\calc$ is also sovereign (i.e.\ it 
has compatible left/right dualities). There are two isomorphisms $A\,{\to}\, 
A^\vee$ that are naturally induced by product, counit and duality; $A$ is called
{\em symmetric\/} iff these two isomorphisms coincide. For more details and 
references, as well as graphical representations of the above conditions and 
morphisms we refer the reader to \cite[sect.\,3]{tft1} and 
\cite[sect.\,2.3]{ffrs}.

Actually a special Frobenius algebra has a one-parameter family of coproducts, 
obtained by rescaling coproduct and counit by an invertible scalar;
out of this family, we always take the element for which
  \be
  m \circ \Delta = \id_A
  \qquad {\rm and} \qquad \eps \circ \eta = \dim(A)\, \id_\one
  \ee
holds.

\medskip

Modules and bimodules of an algebra are defined analogously to the case of
vector spaces, too. Since bimodules will be used frequently
let us state explicitly
\\[-2.3em]

\dtl{Definition}{def:bimod}
Let $A \eq (A,m_A,\eta_A)$ and $B \eq (B,m_B,\eta_B)$ be (unital, associative)
algebras in a strict\,\footnote{~%
  Recall that by Mac Lane's coherence theorem, every tensor category
  is equivalent to a strict one.}
tensor category $\calc$. An $A$-$B$-{\em bimodule\/} is a triple
$X\eq (\dot X,\rho_l,\rho_r)$, where $\dot X$ is an object of $\calc$,
$\rho_l \iN \Hom(A \oti \dot X,\dot X)$, and $\rho_r \iN \Hom(\dot X \oti B,\dot X)$
such that the following equalities hold:
\\
(i)~~Unit property:\,
$\rho_l \cir (\eta_A \oti \id_{\dot X}) \eq \id_{\dot X}$ and
$\rho_r \cir (\id_{\dot X} \oti \eta_B) \eq \id_{\dot X}$.
\\
(ii)\,\,\,Representation property:\,
$\rho_l \cir (m_A \oti \id_{\dot X}) \eq \rho_l \cir (\id_A \oti \rho_l)$ and
$\rho_r \cir (\id_{\dot X} \oti m_B) \eq \rho_r \cir (\rho_r \oti \id_B)$.
\\
(iii)~Left and right action commute:\,
$\rho_l \cir (\id_A \oti \rho_r) \eq \rho_r \cir (\rho_l \oti \id_B)$.

\medskip

The definition for non-strict tensor categories involves
associators and unit constraints in the appropriate places.
Note also that an $A$-left module is the same as an $A$-$\one$-bimodule.

Given two algebras $A$, $B$, and two bimodules $X$, $Y$, the
space of bimodule intertwiners from $X$ to $Y$ is given by
  \be
  \HomAB(X,Y) = \big\{ f \iN \Hom(\dot X,\dot Y) \,\big|\,
  f \cir \rho_l \eq \rho_l \cir (\id_A \oti f) \,,~
  f \cir \rho_r \eq \rho_r \cir (f \oti \id_B) \big\} \,.
  \ee
In this way we obtain the category $\clc{A}{B}$ whose objects
are $A$-$B$-bimodules and whose morphism spaces are given by
$\HomAB(\,\cdot\,,\,\cdot\,)$. If the algebras are also
special Frobenius, many properties pass from $\calc$ to $\clc{A}{B}$.

\dtl{Proposition}{prop:CAB-properties}
Let $A$ and $B$ be special Frobenius algebras in a strict tensor category \C.
\\[2pt]
(i)~~If \C\ is idempotent complete, then so is $\clc{A}{B}$.
\\[2pt]
(ii)~\,If \C\ is abelian, then so is $\clc{A}{B}$.
\\[2pt]
(iii)~If \C\ is semisimple, then so is $\clc{A}{B}$.
\\
If in addition, \C\ only has a finite number of
non-isomorphic simple objects, then so has $\clc{A}{B}$.

\medskip

The proof is analogous to the proof of the corresponding
statements for $\C_A$ in section 5 of \cite{fuSc16}.

\medskip

If the category $\calc$ is ribbon,\,\footnote{~%
  In fact it is enough to require that $\calc$ is spherical. But later on we
  will only be interested in  categories $\calc$ that are ribbon,
  and even modular.\label{fot:spherical}}
then we get a contravariant functor $(\,\cdot\,)^\vee {:}\ \clc{A}{B} \To
\clc{B}{A}$. For an object $X$ of $\clc AB$ we set
  \be
  X^\vee := ({\dot X}^\vee, \tilde \rho_l, \tilde \rho_r)
  \labl{eq:Xvee-action}
where ${\dot X}^\vee$ is the dual in $\calc$ and
  \be
  \tilde\rho_l := (\tilde d_B{\otimes}\id_{\dot X^\vee}) \circ
    (\id_B{\otimes} \rho_r^\vee) \,,\quad
  \tilde\rho_r := (\id_{\dot X^\vee}{\otimes}d_A) \circ
    (\rho_l^\vee{\otimes}\id_A) \,.
  \ee
On morphisms, $(\,\cdot\,)^\vee$ just takes $f$ to $f^\vee$, where we understand
the morphism spaces of $\clc AB$ and $\clc BA$ as subspaces
of morphism spaces in $\calc$ and use the action of the duality of $\calc$
on morphisms. One verifies that $X^\vee$ is indeed a $B$-$A$-bimodule.
In graphical notation, the left/right action \erf{eq:Xvee-action}
looks as follows (compare equation (2.40) of \cite{tft2})
  \bea \begin{picture}(85,52)(0,42)
  \put(20,0)     {\Includeourtinynicepicture 82{}}
  \put(20,0){
     \setlength{\unitlength}{.90pt}\put(-34,-30){
     \put(30,21)   {\scriptsize$ B $}
     \put(53,21)   {\scriptsize$ X^\vee $}
     \put(54,130)  {\scriptsize$ X^\vee $}
     \put(78.7,21) {\scriptsize$ A $}
     }\setlength{\unitlength}{1pt}}
  \epicture25\labl{pic-ffrs5-82}

Another important concept is the tensor product of two bimodules over the
intermediate algebra. We will define it in terms of idempotents. Suppose that
$\calc$ is an idempotent complete ribbon category\,${}^{\ref{fot:spherical}}$.
Let $A$, $B$ and $C$ be special Frobenius algebras,
let $X$ be an $A$-$B$-bimodule and let $Y$ be a $B$-$C$ bimodule.
Consider the morphism $P_{X,Y} \iN \Hom(X \oti Y, X\oti Y)$ given by
  \bea \begin{picture}(44,35)(0,40)
  \put(-30,30){$P_{X,Y} = $}
  \put(20,0)     {\Includeourtinynicepicture 83{}}
  \put(20,0){
     \setlength{\unitlength}{.90pt}\put(-15,-16){
     \put(36,67)  {\scriptsize$ B $}
     \put(11.2,7) {\scriptsize$ X $}
     \put(48,7)   {\scriptsize$ Y $}
     \put(12,101) {\scriptsize$ X $}
     \put(48,101) {\scriptsize$ Y $}
     }\setlength{\unitlength}{1pt}}
  \epicture20\labl{pic-ffrs5-83}
Noting that $X \oti Y$ is an $A$-$C$-bimodule, one checks that in fact
$P_{X,Y} \iN \HomAC(X \oti Y, X\oti Y)$. Furthermore, using that $B$ is special
Frobenius, one can verify that $P_{X,Y}$ is an idempotent (this is analogous
to the calculation in e.g.\ equation (5.127) of \cite{tft1}). By proposition
\ref{prop:CAB-properties}, the category $\clc AC$ is idempotent complete, so
there exists an object $\text{Im}(P_{X,Y})$ in $\clc AC$, a monomorphism
$e_{X,Y} \iN \HomAC(\text{Im}(P_{X,Y}) , X \oti Y)$ and an epimorphism
$r_{X,Y} \iN \HomAC(X \oti Y , \text{Im}(P_{X,Y}))$ such that
$P_{X,Y} \eq e_{X,Y} \cir r_{X,Y}$. We define
  \be
  X \otB Y := \text{Im}(P_{X,Y}) \,.
  \labl{eq:ImPXY}
In fact we get a bifunctor $\otB{:}\ \clc AB\Times\clc BC \To \clc AC$,
by also defining the action on morphisms
$f \iN \HomAB(X,X')$ and $g \iN \HomBC(Y,Y')$ as
  \be
  f \otB g := r_{X',Y'}^{} \circ (f \oti g) \circ e_{X,Y}^{} \,.
  \ee
When verifying functoriality one needs that $(f \oti g) \cir P_{X,Y} \eq 
P_{X',Y'} \cir (f \oti g)$, which follows from the fact that $f$ and $g$ commute
with the action $B$. This bifunctor also admits natural associativity 
constraints and unit constraints, see e.g.\ \cite{stre8,laud} for more details.

Given objects $U$ and $V$ of a category and idempotents $p \iN \End(U)$ and 
$q\iN\End(V)$, the following subspaces of $\Hom(U,V)$ will be of interest to us:
  \bea
  \Hom_{(p)}(U,V) := \big\{ f \iN \Hom(U,V) \,|\, f \cir p \eq f \big\}
  \quad \text{and} \\{}\\[-.7em]
  \Hom^{(q)}(U,V) := \big\{ f \iN \Hom(U,V) \,|\, q \cir f \eq f \big\} \,.
  \eear\labl{eq:Homp-Homq}
Note that the embedding and restriction morphisms defined before \erf{eq:ImPXY} 
give isomorphisms 
  \bea
  {\HomAC}_{(P_{X,Y})}(X \oti Y,Z) \overset{\cong}{\longrightarrow}
  \HomAC(X \otB Y,Z) \qquad \text{and} \\{}\\[-.7em]
  {\HomAC}^{(P_{X,Y})}(Z,X \oti Y) \overset{\cong}{\longrightarrow}
  \HomAC(Z,X \otB Y) \,,
  \eear\labl{eq:HomP-HomTens}
for $X, Y$ bimodules as above and $Z$ an $A$-$C$-bimodule.
The first isomorphism is given by $f \,{\mapsto}\, f \cir e_{X,Y}$,
and the second by $g \,{\mapsto}\, r_{X,Y} \cir g$.

\dtl{Remark}{rem:bimod_dual_et_al}
(i)~~We have seen in \erf{eq:Xvee-action} that, given an $A$-$B$-bimodule $X$,
the duality of $\calc$ can be used to define a dual bimodule $X^\vee$ that is
a $B$-$A$-bimodule. For these bimodule dualities analogous rules are valid
as for ordinary dualities (where the dual object lies in the same category).
In particular, for $A$ and $B$ simple symmetric special Frobenius algebras and 
$X$ a simple $A$-$B$-bimodule, the tensor product $X\otB X^\vee$ contains $A$ 
with multiplicity one, while $X^\vee\otA X$ contains $B$ 
with multiplicity one, see lemma \ref{lem:duals} below.
\\[2pt]
(ii)~\,Note that in this way $\clc AA$ becomes a tensor category with dualities.
If $A$ is simple, the quantum dimension $\dim_A(\,\cdot\,)$ in $\clc AA$ is
given in terms of the quantum dimension $\dim(\,\cdot\,)$ of $\calc$ by
$\dim_A(X) \eq \dim(X){/}{\dim}(A)$ (this follows for example from writing out
the definition of $\dim_A(\,\cdot\,)$ and using lemma \ref{lem:X-insert} below).
\\[2pt]
(iii) For a modular tensor category $\calc$ one can consider the 2-category
${\mathcal F}\!\mbox{\sl rob}_\calc$ whose objects are symmetric special
Frobenius algebras in $\calc$ (compare also \cite{muge8,yama12,laud} and section
3 of \cite{scfr2}). The categories ${\mathcal F}\!\mbox{\sl rob}_\calc(A,B)$
(whose objects are the 1-morphisms $A \To B$) are the categories $\clc AB$
of $A$-$B$-bimodules and the 2-morphisms are morphisms of bimodules. As we will
see in the next section, the objects of ${\mathcal F}\!\mbox{\sl rob}_\calc$
can be interpreted as possible phases of a CFT with given chiral symmetry, and
the 1-mor\-phisms as the allowed types of topological boundaries between different
phases of the theory. In fact, from this point of view it is not surprising that
CFT world sheets with topological phase boundaries (but without field
insertions) look very similar to 2-categorial string diagrams, see
e.g.\ \cite{stre3} and section 2.2 of \cite{laud2}.

\medskip

Given an $A$-$B$-bimodule $X \eq (\dot X,\rho_l,\rho_r)$ in a ribbon category
\C, and two objects $U$, $V$ of $\calc$, we can use the braiding to define
several bimodules structures on the object $U \oti \dot X \oti V$ of \C. The
one which we will use is
  \be\begin{array}{l}
  U \otip X \otim V := \big( U \oti \dot X \oti V,
  \\[4pt]\hspace*{5em}
  (\id_{U}\oti\rho_l \oti\id_{V})\cir (c^{-1}_{U,A}\oti\id_{\dot X}\oti\id_{V}),
  (\id_{U}\oti\rho_r \oti\id_{V})\cir (\id_{U}\oti\id_{\dot X}\oti c^{-1}_{B,V})
  \big) \,.  \end{array}
  \labl{eq:UxXxV}
Pictorially, the left and right actions of $A$ and $B$, respectively, on
$U \otip X \otim V$ are given by
  \bea \begin{picture}(70,57)(0,40)
  \put(0,0)     {\Includeourtinynicepicture 84{}}
  \put(0,0){
     \setlength{\unitlength}{.90pt}\put(-27,-30){
     \put(24,22)   {\scriptsize$ A $}
     \put(48,22)   {\scriptsize$ U $}
     \put(48,136)  {\scriptsize$ U $}
     \put(63,22)   {\scriptsize$ X $}
     \put(64,136)  {\scriptsize$ X $}
     \put(80,22)   {\scriptsize$ V $}
     \put(81,136)  {\scriptsize$ V $}
     \put(104,22)  {\scriptsize$ B $}
     }\setlength{\unitlength}{1pt}}
  \epicture25\labl{pic-ffrs5-84}


\subsection{Relation between algebras and CFT quantities}\label{sec:cft-alg-rel}

The construction of correlation functions of a full rational CFT on oriented
world sheets with (possibly empty) boundary can be described in terms of
a modular tensor category \C\ and a symmetric special Frobenius algebra
$A$ in \C. These data can be realised using a rational conformal
vertex algebra $\mathcal{V}$ with $\C \,{\simeq}\, \Rep(\mathcal{V})$
and for $A$ an `open-string vertex algebra' in the sense of \cite{huko}. We
denote this full CFT by $\CFTA$. Recall that we denote by $\{U_i\,|\,i\iN\I\}$
a set of representatives for the isomorphism classes of simple objects in \C.

Quantities of interest for the CFT do have natural algebraic counterparts. For
example, conformal boundary conditions (which preserve also $\mathcal{V}$) are
labelled by $A$-left modules\,,\footnote{~%
   The use of left modules rather than right modules is dictated by the
   conventions for the relative orientation of bulk and boundary.}
see \cite[sect.\,4.4]{tft1} for details. By proposition
\ref{prop:CAB-properties}, the category $\calca$ of $A$-left modules (which are
the same as $A$-$\one$-bimodules) is again semisimple and has a finite number 
of simple objects. We denote by
  \be
  \{ M_\mu \,|\, \mu \iN \JJ_A \}
  \labl{eq:Amod-rep}
a set of representatives of isomorphism classes of simple $A$-left modules.
Simple $A$-modules correspond to boundary conditions that
cannot be written as a direct sum of other boundary conditions.

The multiplicity space of boundary fields and boundary changing fields
which join a segment of boundary labelled $M$ to a segment labelled $N$
and that transform in the representation $U_k$ of $\mathcal{V}$ is given by
$\Hom_A(M\oti U_k,N)$, more details can be found in \cite[sect.\,3.2]{tft4}.
In fact, the algebra $A$ is a left module over itself and thus a preferred 
boundary condition. The multiplicity spaces for the boundary fields are then
$\Hom_A(A\oti U_k,A) \Cong \Hom(U_k,A)$ and the OPE
of boundary fields is just given by the multiplication on $A$
(see \cite[sect.\,3.2]{tft1} and \cite[remark\,6.2]{tft4}).

Thus, the pair $(\mathcal{V},A)$ does in fact correspond to a full CFT defined
on oriented surfaces with boundary, {\em together with a distinguished boundary
condition\/}, namely the one labelled by the $A$-module $A$. The algebras of
boundary fields on other boundary conditions lead to
symmetric special Frobenius algebras, too. These are all Morita equivalent
to $A$ (see section 4.1 of \cite{tft1} and theorem 2.14 in \cite{tft2}).
In section \ref{sec:morita} we will show, by manipulating topological defects,
that, conversely, two Morita equivalent algebras lead to equivalent full CFTs.

As mentioned in the introduction, we are interested in situations in which
different parts of a world sheet can be in different {\em phases\/}, i.e.\
different full CFTs are assigned to different parts of a world sheet. We refer
to the lines along which such phases meet as phase boundaries or defect lines.
We demand all these full CFTs to have the same underlying rational vertex
algebra $\mathcal{V}$, and we demand that correlators are continuous when a bulk
field for which both the left moving and right moving degrees of freedom
transform in the defining representation $\mathcal{V}$ of the vertex algebra
is taken across a phase boundary. In particular, the phase boundaries we
consider are transparent to the holomorphic and anti-holomorphic components of
the stress tensor and are therefore realised by topological defect lines. We
can then label the various
parts of a world sheet by symmetric special Frobenius algebras $A_1$, $A_2$,
$A_3$, $\dots$ in $\Rep(\mathcal{V})$.  A phase boundary between phases
$\CFTA$ and $\CFTB$ is labelled by an $A$-$B$-bimodule $X$. We call such
a phase boundary an $A$-$B$-{\em defect line of type\/} $X$, or an
$A$-$B$-{\em defect\/} $X$, for short. The {\em invisible defect\/} in phase
$\CFTA$ is labelled by the algebra $A$ itself. The category $\clc AB$ of
$A$-$B$-bimodules is again semisimple with a finite number of simple objects
\cite{fuSc16}. We denote by
  \be
  \{ X_\mu \,|\, \mu \in \KK_{AB} \}
  \labl{eq:ABbimod-rep}
a set of representatives of isomorphism classes of simple $A$-$B$-bimodules.

\medskip

The multiplicity space of {\em bulk fields\/} that transform in the
representation $U_i \Times U_j$ of the left- and right-moving copies of 
$\mathcal{V}$ (or in short, in the left/right representation $(i,j)$)
is given by $\HomAA(U_i \otiP A \otim U_j,A)$. Analogously, the space of
{\em defect fields\/}, which join a defect labelled by an $A$-$B$-bimodule $X$
to a defect labelled by another $A$-$B$-bimodule $Y$ is given by
$\HomAB(U_i \otip X \otim U_j,Y)$, see \cite[sect.\,3.3\,\&\,3.4]{tft4} for
details in the case $A\eq B$. By the term {\em disorder field\/} we refer to
a defect field that changes a defect labelled by an $A$-$A$-bimodule $X$ to
the invisible defect, labelled by $A$. In other words, a defect line starts
(or ends) at a disorder field.

Topological defects can be deformed continuously on the world sheet with the
field insertion points removed without changing the value of a correlator.
For brevity, we will refer in the sequel to the situation when two defect lines
can be taken close to each other as dealing with two {\em parallel\/} defect
lines. The dimensions of the multiplicity spaces are encoded in the partition
function of a torus on which two parallel defect lines of opposite orientation,
labelled $X$ and $Y$, are inserted along a non-contractible cycle. We will only
need the case $A\eq B$, in which case the coefficients of the expansion of the
partition function in terms of simple characters are given by
  \be
  Z(A)_{ij}^{X|Y} = \dim_{\mathbbm C} \HomAA( U_i{\otimes^+}X{\otimes^-}U_j,Y) \,.
  \labl{eq:ZijXY}
Such partition functions were first investigated in \cite{pezu5,pezu7}.
Further details can be found in section 5.10 of \frsi\ and section 2.2 of \frsiv.
Note also that the matrix $Z(A) \,{\equiv}\, Z(A)^{A|A}$,
which gives the multiplicities of bulk fields, describes
the modular invariant torus partition of the CFT in phase \CFTA.

\medskip

For some of the calculations later on we need to choose bases in various
morphism spaces. First, given an $A$-$B$-bimodule $X$, we choose a basis for
the morphisms which describe its decomposition
into a finite direct sum of simple bimodules. Let
  \be
  e_{X,\mu}^{\alpha} \in \HomAB(X_\mu,X) \qquad{\rm and}\qquad
  r_{\!X,\mu}^{\alpha} \in \HomAB(X,X_\mu) \,,
  \labl{eq:Xmu-X-basis}
with $\mu \iN \KK_{AB}$ and $\alpha \eq 1,...\,,\dimc\HomAB(X_\mu,X)$, be
elements of dual bases
in the sense that $r_{X,\mu}^{\alpha} \cir e_{X,\mu}^{\beta} \eq
\delta_{\alpha,\beta} \,\id_{X_\mu}$. In particular,
$(X_\mu,e_{X\mu}^{\alpha},r_{X\mu}^{\alpha})$ is a retract of $X$.
Similarly we choose
  \be
  {\Lambda}_{(\rho\sigma)\mu}^{\alpha}
  \in \HomAC(X_\rho \otB X_\sigma,X_\mu)  \qquad{\rm and}\qquad
  {\overline\Lambda}_{(\rho\sigma)\mu}^{\alpha}
  \in \HomAC(X_\mu,X_\rho \otB X_\sigma) \,,
  \labl{eq:Xmu-XY-basis}
where $\mu \iN \KK_{AC}$, $\rho \iN \KK_{AB}$, $\sigma \iN \KK_{BC}$
and $\alpha \eq 1,...\,,\dimc\HomAC(X_\mu,X_\rho \otB X_\sigma)$. Again
we demand that ${\Lambda}_{(\rho\sigma)\mu}^{\alpha} \cir {\overline\Lambda}
_{(\rho\sigma)\mu}^{\beta} \eq \delta_{\alpha,\beta} \, \id_{X_\mu}$. 
Via the isomorphisms \erf{eq:HomP-HomTens}, the bases \erf{eq:Xmu-XY-basis} 
also yield bases
  \be
  {\Lambda}_{(\rho\sigma)\mu}^{\alpha}
  \in {\HomAC}_{(P_{X_\rho,X_\sigma})}(X_\rho \oti X_\sigma,X_\mu)
  \quad{\rm and}\quad
  {\overline\Lambda}_{(\rho\sigma)\mu}^{\alpha}
  \in {\HomAC}^{(P_{X_\rho,X_\sigma})}(X_\mu,X_\rho \oti X_\sigma) \,.
  \labl{eq:Xmu-XY-basis-HomP}
We will use the same symbols as in \erf{eq:Xmu-XY-basis} for these bases; it 
will be clear from the context to which morphism space we refer. Finally 
we need a basis for the multiplicity space of defect fields; we denote them by
  \be
  \xi_{(i\mu j)\nu}^{\alpha} \in
  \HomAB(U_i{\otimes^+}X_\mu{\otimes^-}U_j,X_\nu)  \quad{\rm and}\quad
  {\overline\xi}_{(i\mu j)\nu}^{\alpha} \in
  \HomAB(X_\nu,U_i{\otimes^+}X_\mu{\otimes^-}U_j) \,,
  \labl{eq:field-basis}
respectively, where $i,j \iN \II$, $\mu,\nu \iN \KK_{AB}$ and
$\alpha$ runs from $1$ to the dimension of the space. And again we require the
two sets of vectors to be dual to each other, $\xi_{(i\mu j)\nu}^{\alpha}
\cir {\overline\xi}_{(i\mu j)\nu}^\beta \eq \delta_{\alpha,\beta}\,\id_{X_\nu}$.

Since we allow for categories \C\ in which non-simple objects can have
zero quantum dimension, the following result is useful.
\\[-2.2em]

\dtl{Lemma}{lem:dim(simple)}
Let $A$ and $B$ be simple symmetric special Frobenius algebras in \C,
and let $X$ be a simple $A$-$B$-bimodule. Then
$\dim(X)\nE 0$ (as an object in \C).

\medskip\noindent
Proof:\\
The argument is analogous to the one that shows that the quantum dimension
$\dim(U)$ of a simple object in a semisimple tensor category with dualities
cannot be zero, see e.g.\ section 2 of \cite{caet}. Consider the two morphisms
  \bea \begin{picture}(360,70)(0,49)
  \put(58,0)     {\Includeourbeautifulpicture 37a }
  \put(58,0){
     \setlength{\unitlength}{.38pt}\put(0,0){
     \put(18,-22)  {\scriptsize$ A $}
     \put(14,313)  {\scriptsize$ X $}
     \put(142,313) {\scriptsize$ X^\vee $}
     \put(90,190)  {\scriptsize$ B $}
     }\setlength{\unitlength}{1pt}}
  \put(253,0)    {\Includeourbeautifulpicture 37b }
  \put(251,0){
     \setlength{\unitlength}{.38pt}\put(0,0){
     \put(0,313)   {\scriptsize$ A $}
     \put(57,-20)  {\scriptsize$ X $}
     \put(185,-20) {\scriptsize$ X^\vee $}
     \put(136,121) {\scriptsize$ B $}
     }\setlength{\unitlength}{1pt}}
  \put(2,56)     {$\beta_X ~:=$}
  \put(164,56)   {and}
  \put(203,56)   {$\tilde \delta_X ~:=$}
  \epicture29 \labl{pic-ffrs5-37}
The morphism in the dashed boxes is the idempotent $P_{X,X^\vee}$ introduced in
\erf{pic-ffrs5-83} whose image is the tensor product over $B$. 
Note that the morphisms 
\be
  b_X = r_{X,X^\vee} \cir \beta_X \iN \HomAA(A,X \otB X^\vee)
  ~~~ \text{and} ~~~
  \tilde d_X = \tilde \delta_X \cir e_{X,X^\vee} \iN \HomAA(X \otB X^\vee,A)
\labl{eq:bimod-dual-morphs}
are the duality morphisms in the bimodule categories, 
cf.\ remark \ref{rem:bimod_dual_et_al}.
Since $X$ is simple, it follows from lemma \ref{lem:duals} below that both 
morphism spaces in \erf{eq:bimod-dual-morphs} are one-dimensional. 
Thus they are spanned by $b_X$ and by $\tilde d_X$, respectively, provided 
that these morphisms are nonzero.
Writing $P_{X,X^\vee} \eq e_{X,X^\vee} \cir r_{X,X^\vee}$ we see that
$\beta_X$ and $\tilde \delta_X$ can be written as
$\beta_X \eq e_{X,X^\vee} \cir b_X$ and
$\tilde \delta_X \eq \tilde d_X \cir r_{X,X^\vee}$.
To show that e.g.\ $b_X$ is nonzero, it is therefore enough
to show that $\beta_X$ is nonzero. This in turn follows from
  \bea \begin{picture}(240,142)(0,48)
  \put(20,0)     {\Includeourbeautifulpicture 38a }
  \put(20,0){
     \setlength{\unitlength}{.38pt}\put(0,0){
     \put(115,104) {\scriptsize$ A $}
     \put(249,-21) {\scriptsize$ X $}
     \put( 22,493) {\scriptsize$ X $}
     \put(100,301) {\scriptsize$ B $}
     }\setlength{\unitlength}{1pt}}
  \put(200,0)    {\Includeourbeautifulpicture 38b }
  \put(198.5,0){
     \setlength{\unitlength}{.38pt}\put(0,0){
     \put(-4.2,-21){\scriptsize$ X $}
     \put(-2 ,493) {\scriptsize$ X $}
     }\setlength{\unitlength}{1pt}}
  \put(154,89)   {$=$}
  \epicture30 \labl{pic-ffrs5-38}
where the morphism in the dashed box is $\beta_X$ and we used properties of $A$ and 
$B$ as well as of the duality in \C. It follows that $\tilde d_X \cir b_X \eq 
\lambda\, \id_A$ for some $\lambda \iN \complex^\times$. By taking the trace 
one determines $\lambda \eq \dim(X)/{\dim}(A)$, so that in particular
$\dim(X)\nE 0$. (The dimension of a special algebra $A$ is required to be 
nonzero, see definition 3.4\,(i) of \cite{tft1}.)
\qed


\subsection{Calculating with defects}\label{sec:calc-def}

In this section we present a number of rules for how to compute
with topological defects. Later on, these rules will be deduced
with the help of the TFT construction, see section \ref{sec:tft-form} below.

First of all, two defects can fuse to produce another defect,
and a defect can act on a boundary, thereby changing the boundary condition.
These notions were first introduced (for $A$-$A$-defects, in our terminology)
from the CFT point of view in \cite{pezu5,pezu6,grwa}
and from a lattice perspective in \cite{cmop,chmp}\,\footnote{~%
Topological defects joining different phases can also be realised 
in integrable lattice models \cite{ppc}.}.

Consider an $A$-$B$-defect labelled $X$ and a $B$-$C$-defect labelled $Y$ running
parallel to each other. We can move these defects close to each other and then
replace them by a single new defect of type $A$-$C$. This defect is then labelled
by the $A$-$C$-bimodule $X \OtB Y$. Similarly, if an $A$-$B$-defect $X$ is
running close to a $B$-boundary $M$, it can be replaced by an $A$-boundary
labelled by the $A$-module $X \OtB M$. Graphically,
  \begin{eqnarray}\begin{picture}(420,86)(13,0)
        \put(0,0){\begin{picture}(0,0)
  \put(20,0)     {\Includeourbeautifulpicture 01a }
  \put(171,0)    {\Includeourbeautifulpicture 01b }
  \put(-9,64)    {\cfta}
  \put(31,64)    {\cftb}
  \put(69,64)    {\cftc}
  \put(25,18)    {\scriptsize$ X $}
  \put(63,18)    {\scriptsize$ Y $}
  \put(109,37)   {$ \stackrel{\rm fuse}\leadsto $}
  \put(145,64)   {\cfta}
  \put(183,64)   {\cftc}
  \put(177,18)   {\scriptsize$ X \OtB Y $}
        \end{picture}}
        \put(252,37){ and }
        \put(308,0){\begin{picture}(0,0)
  \put(27,0)     {\Includeourbeautifulpicture 01c }
  \put(144,0)    {\Includeourbeautifulpicture 01d }
  \put(1,64)     {\cfta}
  \put(41,64)    {\cftb}
  \put(20,18)    {\scriptsize$ X $}
  \put(56,18)    {\scriptsize$ M $}
  \put(114,18)   {\scriptsize$ X \OtB M $}
  \put(111,37)   {$ \stackrel{\rm fuse}\leadsto $}
        \end{picture}}
  \end{picture}
  \nonumber\\[1pt]~\label{pic-ffrs5-01}
  \end{eqnarray}

The algebra $A$ is a bimodule over itself, and it obeys $A \otA X \Cong X$ for 
any $A$-$B$-bimodule $X$. In fact, $A$ labels the {\em invisible defect\/} in 
\CFTA; defect lines labelled by $A$ can be omitted without changing the value 
of a correlator.

Defects can also be joined. The junction is labelled by an element of the
relevant morphism space of bimodules. For example, when joining two
$A$-$B$-defects $X$ and $X'$, or an $A$-$B$-defect $X$ and a $B$-$C$-defect
$Y$ to an $A$-$C$-defect $Z$, according to
  \bea \begin{picture}(190,45)(0,38)
  \put(20,0)     {\Includeourbeautifulpicture 02a }
  \put(135,0)    {\Includeourbeautifulpicture 02b }
  \put(-4,65)    {\cfta}
  \put(37,65)    {\cftb}
  \put(28,8)     {\scriptsize$ X $}
  \put(28,56)    {\scriptsize$ X' $}
  \put(32,43)    {\scriptsize$ \alpha $}
  \put(78,40)    {and}
  \put(122,65)   {\cfta}
  \put(147,-2)   {\cftb}
  \put(169,65)   {\cftc}
  \put(132,13)   {\scriptsize$ X $}
  \put(157.5,61) {\scriptsize$ Z $}
  \put(162,42)   {\scriptsize$ \beta $}
  \put(171,13)   {\scriptsize$ Y $}
  \epicture15 \labl{pic-ffrs5-02}
the junctions get labelled by morphisms $\alpha\iN\HomAB(X',X)$ and
$\beta\iN\HomAC(Z,X \OtB Y)$, respectively.\,\footnote{~%
  Note that e.g.\ in the first picture in \erf{pic-ffrs5-02}, the arrows on
  the defect lines point from $X$ to $X'$, while the bimodule morphism labelling
  the point where they join goes from $X'$ to $X$. This is just a choice of
  convention entering the TFT construction of an RCFT correlator (see sections
  3.1 and 3.4 of \cite{tft4} as well as section \ref{sec:calc-tft} below for
  more details), and does not have any deeper significance.}
Note also that a junction linking an $A$-$A$-defect $X$ to the invisible
defect $A$ is labelled by an element $\alpha$ of $\HomAA(A,X)$. In particular,
  \bea \begin{picture}(370,7)(0,9)
  \put(0,0)     {\includeourbeautifulpicture 20 }
  \put(0,0){
     \setlength{\unitlength}{.38pt}\put(0,0){
     \put( 63, 26) {\scriptsize$ X $}
     \put(0.3, 34) {\scriptsize$ \alpha $}
     }\setlength{\unitlength}{1pt}}
  \put(86,4){$=0~~~$ if $A$ and $X$ are simple bimodules and $A \not\cong X\,.$}
  \epicture-5 
  \labl{eq:X-not-end}
As a consequence, a nontrivial simple defect $X$ cannot just end in the interior
of a world sheet.

An arbitrary defect can be decomposed into a sum of simple defects by using the
direct sum decomposition of the corresponding bimodules (recall from section
\ref{sec:cft-alg-rel} that $\CAA$ is semisimple). Via the fusion procedure,
this applies likewise to the situation that defects are running
parallel to one another. In particular, we have
  \bea \begin{picture}(420,49)(9,40)
  \put(20,0)     {\Includeourbeautifulpicture 03a }
  \put(100,0)    {\Includeourbeautifulpicture 03b }
  \put(240,0)    {\Includeourbeautifulpicture 03c }
  \put(360,0)    {\Includeourbeautifulpicture 03d }
  \put(12,40)  {\scriptsize$ X $}
  \put(-2,66)  {\scriptsize$ \cfta $}
  \put(29,66)  {\scriptsize$ \cftb $}
  \put(48,41)  {$\dsty = \sum_{\mu,\alpha}$}
  \put(77,66)  {\scriptsize$ \cfta $}
  \put(117,49) {\scriptsize$ \cftb $}
  \put(109,74) {\scriptsize$ X $}
  \put(109,39) {\scriptsize$ X_{\mu} $}
  \put(109,8)  {\scriptsize$ X $}
  \put(93,56)  {\scriptsize$ \bar{\alpha} $}
  \put(93,29)  {\scriptsize$ {\alpha} $}
  \put(167,42) {and}
 \put(10,0){
  \put(205,60) {\scriptsize$ \cfta $}
  \put(242,60) {\scriptsize$ \cftb $}
  \put(275,60) {\scriptsize$ \cftc $}
  \put(219,36) {\scriptsize$ X_{\rho} $}
  \put(273,36) {\scriptsize$ X_{\sigma} $}
  \put(299,42) {$\dsty = \sum_{\nu,\beta}$}
  \put(340,35) {\scriptsize$ \cfta $}
  \put(390,35) {\scriptsize$ \cftc $}
  \put(361,80) {\scriptsize$ \cftb $}
  \put(363,-7) {\scriptsize$ \cftb $}
  \put(345,14) {\scriptsize$ X_{\rho} $}
  \put(383,14) {\scriptsize$ X_{\sigma} $}
  \put(358,24) {\scriptsize$ \bar{\beta} $}
  \put(358,58) {\scriptsize$ {\beta} $}
  \put(345,72) {\scriptsize$ X_{\rho} $}
  \put(385,72) {\scriptsize$ X_{\sigma} $}
  \put(372,40) {\scriptsize$ X_{\nu} $}
 }
  \epicture16 \labl{pic-ffrs5-03}
where the $\mu$-summation is over the label set $\KK_{AB}$ of (isomorphism
classes of) simple $A$-$B$-bi\-modules, $\alpha$ runs over the basis
\erf{eq:Xmu-X-basis} of $\HomAB(X_\mu,X)$, while $\nu$ runs over $\KK_{AC}$,
and $\beta$ over the basis \erf{eq:Xmu-XY-basis} of
$\HomAB(X_\rho\OtB X_\sigma,X_\nu)$. Another useful identity is
  \bea \begin{picture}(290,57)(0,56)
  \put(20,0)     {\Includeourbeautifulpicture 12a }
  \put(210,0)    {\Includeourbeautifulpicture 12b }
  \put(-5, 80) {\scriptsize$ \cfta $}
  \put(68, 80) {\scriptsize$ \cftc $}
  \put(33, 80) {\scriptsize$ \cftb $}
  \put(9,10)   {\scriptsize$ X_{\rho} $}
  \put(63,10)  {\scriptsize$ X_{\sigma} $}
 \put(20,0){
  \put( 85,53) {$\dsty =\sum_{\mu,\gamma}
                \frac{\dim X_{\mu}}{\dim X_{\rho}} $}
  \put(169,40) {\scriptsize$ \cfta $}
  \put(206,80) {\scriptsize$ \cftb $}
  \put(230,52) {\scriptsize$ \cftc $}
  \put(206,20) {\scriptsize$ \cftb $}
  \put(181,102){\scriptsize$ X_{\rho} $}
  \put(181,7)  {\scriptsize$ X_{\rho} $}
  \put(184,89) {\scriptsize$ \bar{\gamma} $}
  \put(184,27) {\scriptsize$ {\gamma} $}
  \put(179,70) {\scriptsize$ X_{\mu} $}
  \put(234,102){\scriptsize$ X_{\sigma} $}
  \put(234,7)  {\scriptsize$ X_{\sigma} $}
 }
  \epicture28 
  \labl{eq:2-opp-sum}

An $A$-$B$-defect $X_\nu$ can also wrap around a $B$-bulk field,
changing it into a disorder field of \CFTA\ by shrinking the defect loop:
  \bea \begin{picture}(280,23)(0,18)
  \put(20,0)     {\Includeourbeautifulpicture 07a }
  \put(20,0){
     \setlength{\unitlength}{.38pt}
     \put(40,70)   {\cftb\put(0,0){\line(3,-1){50}}\put(0,0.2){\line(3,-1){50}}}
     \put(220,-5)  {\cfta}
     \put(7,46)    {\scriptsize$ X_\mu $}
     \put(200,78)  {\scriptsize$ X_\nu $}
     \put(174,39)  {\scriptsize$ \phi $}
     \put(133,-10) {\scriptsize$ \alpha $}
     \setlength{\unitlength}{1pt}}
  \put(160,20)   {\Includeourbeautifulpicture 07b }
  \put(160,20){
     \setlength{\unitlength}{.38pt}
     \put(40,-60)  {\cfta}
     \put(15,20)   {\scriptsize$ X_\mu $}
     \put(130,-19) {\scriptsize$ D_{\mu\nu\alpha}(\phi) $}
     \setlength{\unitlength}{1pt}}
  \put(135,19)   {$=$}
  \epicture03 
  \labl{eq:def-bulk-act}
Here $\alpha$ is an element of $\HomAB(X_\mu\OtA X_\nu,X_\nu)$, and the bulk
field is labelled by the morphism $\phi \iN \HomBB(U \otiP B \otim V, B)$. The
resulting disorder field starts the $A$-$A$-defect $X_\mu$ and is labelled
by $D_{\mu\nu\alpha}(\phi)$, where $D_{\mu\nu\alpha}$ is a linear map
  \be
  D_{\mu\nu\alpha} :\quad \Hom_{B|B}(U \otip B \otim V, B)
  \longrightarrow \HomAA(U \otip X_\mu \otim V, A) \,.
  \labl{eq:D-def}
This map can be obtained explicitly in the TFT construction, see equation 
\erf{pic-ffrs5-80} below.  In the special case that $X_\mu \eq A$ and that 
$\alpha \eq \rho_{X_\nu}$ is given by the representation morphism, we abbreviate
$D_\nu \Equiv D_{\mu\nu\alpha}$, i.e.\ write
  \bea \begin{picture}(100,19)(0,23)
  \put(20,0)     {\Includeourbeautifulpicture 30a }
  \put(120,15)   {\Includeourbeautifulpicture 30b }
  \put(91,15)    {$=$}
  \put(-20,28)   {\cftb \put(0,0){\line(3,-1){30}}\put(0,0.2){\line(3,-1){30}} }
  \put(60,30)  {\scriptsize$\cfta$}
  \put(8,10)   {\scriptsize$ X_\nu $}
  \put(42,19)  {\scriptsize$ \phi $}
  \put(115,30) {\scriptsize $ \cfta$}
  \put(126,12) {\scriptsize$ D_{\nu}(\phi)$}
  \epicture04 
  \labl{eq:D_nu-def}
If, still for $X_\mu \eq A$ and $\alpha$ the representation morphism, the
defect line that wraps around the bulk field is labelled by an arbitrary
(not necessarily simple) $A$-$B$-bimodule $X$ instead of $X_\nu$, we write
analogously $D_X$ for the resulting linear map. These maps obey, for $Y$ a
$B$-$C$-defect and $\phi$ a bulk field of \CFTC,
  \be
  D_X \cir D_Y(\phi) = D_{X\OtB\! Y}(\phi) \,.
  \labl{eq:defect-repn}
When $A\eq B\eq C$, this gives a representation of the fusion algebra of
$A$-$A$-defects on each of the spaces $\HomAA(U \otip A \otim V, A)$.

An $A$-$B$-defect $X$ can also start and end on the boundary of the world sheet.
The corresponding junctions are again labelled by appropriate morphisms,
  \bea \begin{picture}(320,48)(0,40)
  \put(20,0)     {\Includeourbeautifulpicture 08a }
  \put(220,0)    {\Includeourbeautifulpicture 08b }
  \put(25,40)  {\scriptsize$\cfta$}
  \put(25,-5)  {\scriptsize$\cftb$}
  \put(57,70)  {\scriptsize$M$}
  \put(57.8,10){\scriptsize$N$}
  \put(38,23)  {\scriptsize$ X$}
  \put(60,49.5){\scriptsize$\alpha$}
  \put(145,44) {and}
  \put(220,20) {\scriptsize$\cfta$}
  \put(220,70) {\scriptsize$\cftb$}
  \put(257,10) {\scriptsize$M$}
  \put(257.8,70){\scriptsize$N$}
  \put(222,46) {\scriptsize$ X$}
  \put(252,32) {\scriptsize$\beta$}
  \epicture22 \labl{pic-ffrs5-08}
respectively, where $M$ is an $A$-module, i.e.\ a boundary condition for the
phase \CFTA, $N$ a $B$-module, $\alpha$ a morphism in $\Hom_A(M,X \OtB N)$, and
$\beta \iN \HomA(X \OtB N,M)$.  In this way one also obtains an action of
defects on boundary fields, in the same spirit as in \erf{eq:def-bulk-act}:
  \bea \begin{picture}(220,69)(0,48)
  \put(20,0)     {\Includeourbeautifulpicture 09a }
  \put(180,0)    {\Includeourbeautifulpicture 09b }
  \put( 0, 80) {\scriptsize$\cfta$}
  \put(-5,32)  {\cftb \put(0,13.2){\line(4,1){27}} }
  \put(12,57)  {\scriptsize$ X$}
  \put(33,5)   {\scriptsize$M_1$}
  \put(50,41)  {\scriptsize$N_1$}
  \put(33,105) {\scriptsize$M_2$}
  \put(50,75)  {\scriptsize$N_2$}
  \put(37,59.1){\scriptsize$\psi$}
  \put(32,90)  {\scriptsize$\alpha$}
  \put(32,23)  {\scriptsize$\beta$}
  \put(110,55) {$=$}
  \put(169,20) {\scriptsize$M_1$}
  \put(169,100){\scriptsize$M_2$}
  \put(172.5,58.5) {\scriptsize$\widetilde{\psi}$}
  \put(145,73) {\scriptsize$\cfta$}
  \epicture25 
  \labl{eq:bnd-act}
We will not use this transformation explicitly in the present paper, though, and
accordingly we do not introduce a separate notation for the resulting boundary
field (labelled $\widetilde\psi$), unlike what we did in \erf{eq:D-def} for
the bulk case. The fusion of defect lines to boundaries, and in particular
their action on boundary fields, has also been studied in \cite{grwa}, where
topological defects were used to deduce relations between boundary
renormalisation group flows.

Finally, in a region of the world sheet in the phase \CFTA\ we can insert
a little bubble of phase \CFTB, via a small circular $B$-$A$-defect $Y$.
If the algebra $A$ is simple, then this merely changes the correlator by a
factor $\dim(A)/\dim(Y)$. Hence in this case we obtain the identity
  \bea \begin{picture}(40,20)(0,18)
  \put(80,0)     {\includeourbeautifulpicture 05 }
  \put(80,0){
    \put(8,8)    {\cftb}
    \put(34,25)  {\scriptsize$ Y $}
    \put(-26,15) {\cfta}
    \put(-153,10){\cfta}
    \put(-175,13){$\dsty \Bigg[ \hspace{4em} \Bigg]
         ~=~ \frac{\dim(A)}{\dim(Y)} \cdot \Bigg[ \hspace{7em} \Bigg]$}
  }
  \epicture10 
  \labl{eq:insertloop}
This leads to the notion of ``{\em inflating a $B$-$A$-defect in a world sheet in
phase\/} \CFTA'', by which we refer to the following procedure. Let now both 
algebras $A$ and $B$ be simple, let $Y$ be a $B$-$A$-bimodule, and let \X\ be a
connected world sheet in phase \CFTA. We start by inserting a little circular
defect labelled $Y$ as in \erf{eq:insertloop}. The $Y$-loop separates \X\ in
regions `$A$' and `$B$'. Now deform the loop until the `$A$' area has shrunk to
zero (this is only possible if \X\ is connected). For example, on a genus one
world sheet with connected boundary and one bulk insertion, we have
  \begin{eqnarray}\begin{picture}(420,81)(21,0)
  \put(-30,0)    {\Includeoursmallnicepicture 10a }
  \put(150,0)    {\Includeoursmallnicepicture 10b }
  \put(330,0)    {\Includeoursmallnicepicture 10c }
  \put(40,20)  {\scriptsize$\cfta$}
  \put(122,30) {$\dsty =~ \frac{\dim A}{\dim Y}$}
  \put(220,20) {\scriptsize$\cfta$}
  \put(201,20) {\scriptsize$ Y$}
  \put(170,55) {\cftb\put(0,0  ){\line(1,-3){7}}
                     \put(0,0.2){\line(1,-3){7}} }
  \put(350,55) {\cftb\put(0,0  ){\line(1,-3){6}}
                     \put(0,0.2){\line(1,-3){6}} }
  \put(302,30) {$\dsty =~ \frac{\dim A}{\dim Y}$}
  \put(480,10) {\cfta \put(-18,13  ){\line(-7,0){29}}
                      \put(-18,13.2){\line(-7,0){29}} }
  \end{picture}
  \nonumber\\[-.8em]~\label{eq:appy-defect} 
  \end{eqnarray}
In more detail, here we used the deformations
  \begin{eqnarray}\begin{picture}(420,92)(21,0)
  \put(0,0)      {\Includeoursmallnicepicture 11a }
  \put(167,0)    {\Includeoursmallnicepicture 11b }
  \put(334,0)    {\Includeoursmallnicepicture 11c }
  \put(60,75)  {\cfta \put(-0.3,-16.2){\line(0,4){20}}
                      \put(-0.1,-16.2){\line(0,4){20}} }
  \put(0,32)   {\scriptsize$\cftb$}
  \put(77,3)   {\scriptsize$Y$}
  \put(244,3)  {\scriptsize$Y$}
  \put(411,12) {\scriptsize$Y$}
  \put(152,40) {$=$}
  \put(320,40) {$=$}
  \put(195,-7) {\scriptsize$\cftb$}
  \put(224,75) {\cfta \put(-0.3,-16.2){\line(0,4){20}}
                      \put(-0.1,-16.2){\line(0,4){20}} }
  \put(346,35) {\scriptsize$\cftb$}
  \put(376,75) {\cfta \put(-0.3,-16.2){\line(0,4){20}}
                      \put(-0.1,-16.2){\line(0,4){20}} }
  \end{picture}
  \nonumber\\[-.4em]~\label{pic-ffrs5-11}
  \end{eqnarray}
for the handle. The concept of inflating a topological defect in a
world sheet will play a central role in the discussions below.

\dt{Remark}
Topological defects also play an important role in the recently established
connection between the geometric Langlands program and dimensionally reduced
topologically twisted $N\eq4$ four-dimensional super Yang\hy Mills theory
\cite{kawi}. These Yang\hy Mills theories contain topological Wil\-son loop 
operators or, at
different points in  moduli space, topological 't\,Hooft operators, both
supported on oriented lines  embedded in a four-manifold. One can consider such
a theory on a product $\Sigma \Times C$ of two two-manifolds, where one thinks
of $C$ as being `small', so  that at low energies one deals with an effective
two-dimensional sigma-model on $\Sigma$. The image of the Wilson and 't Hooft
operators under the projection $\Sigma \Times C \To \Sigma$ gives rise to
(possibly point-like) topological defect lines on $\Sigma$.
\\
Other phenomena discussed above, most prominently the action of defects on
boundary conditions and boundary fields, play a crucial role in the
identification of Hecke eigensheaves in the setting of \cite{kawi}, see section
6 there. In our context, a more general
notion of `eigenbrane' seems to be natural: an $A$-module $M$ -- corresponding
to a boundary condition of the conformal field theory -- is called an
eigenbrane for the $A$-$A$-bimodule $X$ -- describing a defect -- iff there
exists an object $U$ of $\calc$ such that
  \be
  X\otA M \cong M \oti U \,.
  \ee
Notice that here the role of the eigenvalues in \cite{kawi} is taken over by
objects of $\calc$; the case of Chan\hy Paton multiplicities considered in
\cite{kawi} amounts to requiring $U$ to be of the form $\one^{\oplus n}$.
\\
While the generalised eigenvalue equation above can be formulated for
all $A$-$A$-bimodules $X$, at the present stage it is not clear to us
whether for a general CFT there is a distinguished subset of bimodules
that play the role of Hecke operators.


\subsection{Non-isomorphic bimodules label distinct defects}\label{sec:distinct}

In this section we establish the following result. Let $Y$ and $Y'$ be 
$B$-$A$-bimodules. If labelling a given defect line by $Y$ or by $Y'$ gives 
the same result for all correlators, then already $Y \Cong Y'$ as bimodules. 
This is a consequence of the next proposition, which makes the stronger
statement that non-isomorphic $B$-$A$-defects differ in their action on at 
least one bulk field of $\CFTA$, where the action is given by the maps 
$D_X$ and $D_Y$ defined below \erf{eq:D_nu-def}.
\\[-2.2em]

\dtl{Proposition}{prop:g-distinct}
Let $X$ and $Y$ be $B$-$A$-bimodules.
If for all $i,j \iN \II$ and all $\phi \iN \HomAA(U_i \otiP A \otim U_j, A)$
we have $D_X(\phi) \eq D_Y(\phi)$, then already $X \Cong Y$ as bimodules.

\dt{Remark}
Topological defects labelled by non-isomorphic bimodules can also be
distinguished by their action on the collection of all boundary conditions and
boundary fields. To see this one uses the fact that a $B$-$A$-bimodule gives
rise to a module functor from $\C_A$ to $\C_B$. Note that a module functor also 
acts on morphisms, which corresponds to the action of the $B$-$A$-defect on 
boundary fields. The important point in proposition \ref{prop:g-distinct} is 
that knowing the action on boundary fields is not required, but rather
it suffices to consider the action on bulk fields.

\medskip

As a preparation for the proof of proposition \ref{prop:g-distinct}, we need to
consider a certain two-point correlator on the Riemann sphere (which we identify
with $\complex \,{\cup}\, \{\infty\}$). The correlator we are interested in is
  \bea \begin{picture}(130,34)(0,29)
  \put(20,0)     {\includeourbeautifulpicture 17 }
  \put(14,11)  {\scriptsize$X_{\nu}$}
  \put(43,37)  {\scriptsize$\cfta$}
  \put(37,24)  {\scriptsize$\phi_{ij,\alpha}$}
  \put(61,0)   {\scriptsize$\gamma$}
  \put(106,25) {\scriptsize$X_{\mu}$}
  \put(117,12) {\scriptsize$\theta_{\bar\imath\bar\jmath,\beta}$}
  \put(85,45)  {\scriptsize$\cftb$}
  \put(-22,25) {$C:=$}
  \epicture12 
  \labl{eq:A-corr}
where the bulk field in \CFTA, inserted at some point $z\iN\complex$, is
labelled by a morphism $\phi_{ij,\alpha} \iN \HomAA(U_i \otiP A \otim U_j,A)$,
and the disorder field in \CFTB, inserted at $w\iN\complex$, is labelled by
$\theta_{\bar\imath \bar\jmath,\beta}\iN \HomBB(U_{\bar\imath} \otiP X_\mu \otim
U_{\bar\jmath},B)$.  The label $\gamma$ stands for a basis element of the 
morphism space $\HomBA(X_\nu,X_\mu\otB X_\nu)$, and we take $\phi_{ij,\alpha}$ 
and $\theta_{\bar\imath\bar\jmath,\beta}$ to be given by the basis elements 
\erf{eq:field-basis}, i.e.\ $\phi_{ij,\alpha} \eq \xi^{\alpha}_{(i0j)0}$ and
$\theta_{\bar\imath\bar\jmath,\beta}\eq\xi^\beta_{(\bar\imath\mu\bar\jmath)0}$.
As a two-point correlator on the Riemann
sphere, $C$ can be written as a product of conformal two-point blocks
  \be
  C = \GammA(\mu)_{ij\alpha\beta}^{\nu\gamma}\,
  \beta[i,\bar\imath](z,w)\, \beta[j,\bar\jmath](z^*\!,w^*) \,,
  \labl{eq:A-def}
where $\GammA(\mu)_{ij\alpha\beta}^{\nu\gamma} \iN \complex$ is a constant and
$\beta[k,\bar k](\zeta_1,\zeta_2)$ denotes a conformal two-point block with
insertions at $\zeta_1,\zeta_2$, i.e.\ a bilinear map from the two
representation spaces $\SCA_k \times \SCA_{\bar k}$ of the chiral
algebra to the complex numbers; see sections 5 and 6
of \cite{tft4} for more details. Here we are only interested
in the prefactor $\GammA(\mu)_{ij\alpha\beta}^{\nu\gamma}$.

Let $R_1$ be the set of all tuples $(ij\alpha\beta)$ for
which $\alpha$ and $\beta$ have a non-empty range, i.e.\ for
which $Z(A)_{ij}\,{>}\,0$ and $Z(B)^{X_\mu|B}_{\bar\imath\bar\jmath}\,{>}\,0$
(these numbers were defined in \erf{eq:ZijXY} above).
Further, let $R_2$ be the set of all tuples $(\nu\gamma)$ for which
the index $\gamma$ has a nonzero range, i.e.\ for which
$\dim\HomAA(X_\nu,X_\mu \otB X_\nu) \,{>}\, 0$.
\\[-2.3em]

\dtl{Lemma}{lem:non-deg}
For each choice of $\mu \iN \KK_{BB}$, either
$R_1 \eq \emptyset \eq R_2$, or else the $|R_1| \Times |R_2|$\,-matrix
$\GammA(\mu)$ with entries $\GammA(\mu)_{ij\alpha\beta}^{\nu\gamma}$
is nondegenerate. In particular, we have $|R_1| \eq |R_2|$.

\medskip

This assertion will be proved in section \ref{sec:proof} with the help of the
TFT approach. It is actually a
corollary to the more general theorem \ref{thm:gen-non-deg}.

\medskip\noindent
{\bf Proof of proposition \ref{prop:g-distinct}:}\\[2pt]
We need the special case $\mu\eq 0$, for which $R_2 \eq \KK_{BA}$.
Lemma \ref{lem:non-deg} then implies that $|R_2| \eq |R_1|$, and that
the $|\KK_{BA}| \Times |\KK_{BA}|$-matrix
$\GammA(0)_{ij\alpha\beta}^{\nu}$ is non-degenerate (the index $\gamma$ in
\erf{eq:A-def} can only take a single value as $\dimc\Hom_{BA}(B \otB X_\nu,
X_\nu) \eq 1$, and has been omitted). Suppose now that
  \be
  X \cong \bigoplus_{\mu \in \KK_{BA}} X_\mu^{\oplus n(X)_\mu}
  \qquad \text{and} \qquad
  Y \cong \bigoplus_{\mu \in \KK_{BA}} X_\mu^{\oplus n(Y)_\mu}
  \ee
as $B$-$A$-bimodules for some $n(X)_\mu, n(Y)_\mu \iN \zet_{\ge 0}$.
We abbreviate $\phi_{ij,\alpha} \,{\equiv}\, \xi^{\alpha}_{(i0j)0}$ and
$\theta_{\bar\imath\bar\jmath,\beta} \,{\equiv}\,
\xi^{\beta}_{(\bar\imath\mu\bar\jmath)0}$.
If $D_X(\phi_{ij,\alpha}) \eq D_Y(\phi_{ij,\alpha})$, then also
$\sum_{\mu \in \KK_{BA}}\! n(X)_\mu\, D_\mu(\phi_{ij,\alpha})
\eq \sum_{\mu \in \KK_{BA}}\! n(Y)_\mu\, D_\mu(\phi_{ij,\alpha})$, i.e.
  \be
  \sum_{\mu \in \KK_{BA}}
  (n(X)_\mu-n(Y)_\mu)\, D_\mu(\phi_{ij,\alpha}) = 0 \,.
  \ee
Inserting this identity in
the two-point correlator \erf{eq:A-corr} (with $\mu\eq0$) leads to
  \be
  \sum_{\nu \in \KK_{BA}}
  (n(X)_\nu-n(Y)_\nu)\, \GammA(0)_{ij\alpha\beta}^{\nu} = 0 \,.
  \ee
Since the matrix $\GammA(0)_{ij\alpha\beta}^{\nu}$ is non-degenerate, this
can only hold if $n(X)_\nu \eq n(Y)_\nu$ for all $\nu \iN \KK_{BA}$,
i.e.\ if $X \Cong Y$ as bimodules.
\qed


\sect{Group-like and duality generating defects}

In this section we investigate two interesting subclasses of topological 
defects, the {\em group-like defects\/} and the {\em duality defects\/}.
The former can be seen to give symmetries of CFT correlators.
The latter, which include the former as a special case, result in
order-disorder duality relations between correlators of a 
single phase \CFTA, or of two different phases \CFTA\ and \CFTB.

Throughout this section we assume that the symmetric special Frobenius 
algebras under consideration are in addition simple. That an algebra $A$ 
is simple means, by definition, that it is a simple $A$-$A$-bimodule; in this 
case we reserve for it the label $0 \iN \KK_{AA}$, i.e.\ write $X_0 \eq A$.
In CFT terms, simplicity of $A$ means that the coefficient $Z_{0,0}$ of the
torus partition function is equal to 1, i.e.\ that there is a unique bulk
vacuum. In other words, non-simple algebras correspond to a superposition of 
several CFTs rather than a single CFT (see definition 2.26 and remark 2.28(i) 
of \cite{ffrs}, as well as section 3.2 of \cite{tft1} for details).


\subsection{Defects generating symmetries}\label{sec:group-sym}

Let us start by defining the notion of a group-like defect.
In the remainder of this section we then study some of their properties.
\\[-2.3em]

\dtl{Definition}{def:group-like}
(i)~\,An $A$-$A$-bimodule $X$ is called 
{\em group-like\/} iff $X \otA X^\vee \Cong A$ as bimodules.
\\[2pt]
(ii) A topological defect is called group-like iff it is labelled by a
group-like $A$-$A$-bimodule.

\medskip

In other words, group-like bimodules are the invertible objects of \CAA, 
see e.g.\ definition 2.1 in \cite{tft3}.
In the fusion of two group-like $A$-$A$-defects labelled by $X$
which have opposite orientation, only the invisible defect $A$ appears:
  \bea \begin{picture}(250,45)(0,44)
  \put(20,0)     {\Includeourbeautifulpicture 06a }
  \put(200,0)    {\Includeourbeautifulpicture 06b }
  \put(120,40)   {$ \stackrel{\rm fuse}\leadsto $}
  \put(12,24)    {\scriptsize$X$}
  \put(51,24)    {\scriptsize$X$}
  \put(-5,70)    {\scriptsize$\cfta$}
  \put(34,70)    {\scriptsize$\cfta$}
  \put(68,70)    {\scriptsize$\cfta$}
  \put(176,70)   {\scriptsize$\cfta$}
  \put(208,70)   {\scriptsize$\cfta$}
  \put(205,24)   {\scriptsize$A$}
  \epicture19 \labl{pic-ffrs5-06}
Note that we do not include the possibility of phase-changing defects in the 
definition of group-like defects. As will be explained in section 
\ref{sec:morita} below, for an $A$-$B$-defect the corresponding property 
implies that the algebras $A$ and $B$ are Morita equivalent. We will show, 
also in section 3.3, that \CFTA\ and \CFTB\ are then equivalent, too.

The following three lemmas will be useful in the discussion of group-like 
defects and of Morita equivalences. The proofs of lemma \ref{lem:simple} and 
lemma \ref{lem:XAB-simple} use the semisimplicity of the relevant categories.
\\[-2em]

\dtl{Lemma}{lem:simple}
Let $U$ and $V$ be two objects in a semisimple \complex-linear tensor category 
such that $U \oti V$ is simple. Then also $U$ and $V$ are simple.

\medskip\noindent
Proof:\\
Since $U \oti V$ is simple, we have $\End(U \oti V) \Cong \complex$. 
Let $U \Cong \bigoplus m_i\, U_i$ and $V \Cong \bigoplus n_i\, U_i$ be
decompositions of $U$ and $V$ into simple objects $U_i$. Then (canonically)
  \be
  \bearll\dsty \dim_\complex \End(U \oti V) \!\!&\dsty
   = \sum_{i,j,k,l} m_i\, m_k\, n_j\, n_l \,
     \dim_\complex \Hom(U_i \oti U_j, U_k \oti U_l)
  \\{}\\[-.9em]& \dsty 
   \ge \sum_{i,j} m_i^2\, n_j^2 \, \dim_\complex \Hom(U_i\oti U_j, U_i\oti U_j)
   \ge\, \sum_{i,j} m_i^2\, n_j^2 \,.
  \eear\ee
Thus in order that $\dim_\complex\End(U\oti V)$ is equal to 1, precisely one of
the coefficients $m_i$ and one of the $n_j$ must be equal to 1, while all
other coefficients must be zero. This means that $U$ and $V$ are simple.
\qed

\dtl{Lemma}{lem:duals}
Let $A$, $B$ and $C$ be (not necessarily simple) symmetric special Frobenius 
algebras, and let $X$ be an $A$-$B$-bimodule, $Y$ a $B$-$C$-bimodule, and
$Z$ an $A$-$C$-bimodule. Then
\\[.3em]
(i)~~$\HomAC(X \otB Y , Z) \cong \HomBC(Y,X^\vee \otA Z)$
\\[.2em]\mbox{$\quad~$}
 and~ $\HomAC(X \otB Y , Z) \cong \HomAB(X,Z \otC Y^\vee)$.
\\[.3em]
(ii)~$\HomAC(Z,X \otB Y) \cong \HomBC(X^\vee \otA Z,Y)$
\\[.2em]\mbox{$\quad~$}
 and~ $\HomAC(Z,X \otB Y) \cong \HomAB(Z \otC Y^\vee,X)$.

\medskip\noindent
Proof:\\
We denote by $e$ and $r$ the embedding and restriction morphisms
for tensor products, as introduced above \erf{eq:ImPXY}.
Then the first isomorphism in (i) is given by
  \be
  \HomAC(X \OtB Y , Z) \,{\ni}~ f \mapsto 
  r_{\!X^\vee\Oti Z}^{} \cir (\id_{X^\vee}\oti (f \cir r_{X \Oti Y}^{})) 
  \cir (\tilde b_X \oti \id_Y) \,.
  \labl{eq:lem-duals-aux1}
To check that \erf{eq:lem-duals-aux1} is indeed an isomorphism one verifies
that the morphism $\HomBC(Y,X^\vee \otA Z) 
   $\linebreak[0]$
{\ni}\, g \,{\mapsto}\,
(\tilde d_X \oti \id_Z) \cir (\id_X \oti ( e_{X^\vee\Oti Z} \cir  g )) \cir
e_{X \Oti Y}$ is its inverse. The projectors $e\cir r$ resulting in the 
composition can be
left out because all relevant morphisms are intertwiners of bimodules.
\\
The other three isomorphisms are similar.
\qed

\dtl{Lemma}{lem:XAB-simple}
Let $X$ be an $A$-$B$-bimodule and $Y$ be a $B$-$A$-bimodule
such that $X \otB Y \Cong A$. Then
\\[.3em]
(i)~~$X$ and $Y$ are simple bimodules.
\\[.3em]
(ii)~\,$Y \Cong X^\vee$ as $B$-$A$-bimodule, and
     $X \Cong Y^\vee$ as $A$-$B$-bimodule.
\\[.3em]
(iii)~$Y \otA X \Cong B$.

\medskip\noindent
Proof:\\
(i)~~can be seen by the same reasoning as in the proof of lemma 
\ref{lem:simple}, if we use that $A$ is simple and
hence $\End_{A|A}(A) \Cong \complex$. 
\\[.3em]
(ii) and (iii) are consequences of lemma \ref{lem:duals}.
\\
For obtaining (ii) note the isomorphisms $\complex \Cong \HomAA(X \OtB Y,A) 
\Cong \HomBA(Y,X^\vee)$, which follow by using that $X^\vee \OtA A 
\Cong X^\vee$. Since by (i) $X$ and $Y$ are simple, their duals
$X^\vee$ and $Y^\vee$ are simple as well. Thus
$\HomBA(Y,X^\vee) \Cong \complex$ implies that the simple bimodules
$Y$ and $X^\vee$ are isomorphic. $X \Cong Y^\vee$ is seen analogously.
\\[.3em]
For (iii), consider the isomorphisms
  \be
  X \otB (Y \otA X) \cong  (X \otB Y) \otA X \cong  A \otA X \cong X \,.
  \ee
Since $X$ is simple, $Y \OtA X$ must be simple (again by the same reasoning as 
in the proof of lemma \ref{lem:simple}, applied to the two bimodules $X$ and 
$Y \OtA X$). Further, by (ii) we have $Y \Cong X^\vee$. But then 
  \be
  \HomBB(Y \otA X,B) \cong 
  \HomBB(X^\vee \otA X,B) \cong \HomAB(X,X) \cong \complex
  \ee
by lemma \ref{lem:duals}.
As both $Y \otA X$ and $B$ are simple, this shows that they are isomorphic.
\qed

As an immediate consequence of lemma \ref{lem:XAB-simple}(i), 
all group-like bimodules are simple and hence
isomorphic to one of the $X_\mu$ for $\mu \iN \KK_{AA}$. We denote by
  \be
  \mathcal{G}_\AA \subseteq \KK_{AA}
  \ee
the subset labelling group-like bimodules.
Also, from lemma \ref{lem:XAB-simple}\,(iii) it follows
that $X \OtA X^\vee \Cong A$ iff $X^\vee \OtA X \Cong A$,
so that in definition \ref{def:group-like} equivalently one could use
the condition $X^\vee \OtA X \Cong A$  to characterise group-like defects.

\dtl{Lemma}{lem:group}
The tensor product $X_g \otA X_h$ of two group-like $A$-$A$-bimodules
$X_g, X_h$ is again group-like.

\medskip\noindent
Proof:\\
Let $Y \eq X_g \otA X_h$. Then $Y \otA Y^\vee \Cong  X_g \otA X_h \otA
X_h^\vee \otA X_g^\vee \Cong A$.
\qed

This observation can be used to define a group structure on $\mathcal{G}_\AA$ 
with the product $gh$ of two elements $g,h \iN \mathcal{G}_\AA$ defined by the 
condition $X_g \otA X_h \Cong X_{gh}$. The unit is given by $X_e\eq X_0\eq A$
and the inverse $g^{-1}$ is defined via $X_{g^{-1}} \Cong (X_g)^\vee_{}$. That 
this is a left and right inverse follows from definition \ref{def:group-like} 
and from lemma \ref{lem:XAB-simple}\,(iii). In fact, $\mathcal{G}_\AA$
is nothing but the Picard group of the tensor category of $A$-$A$-bimodules 
(see definition 2.5 and remark 2.6 of \cite{tft3}),
  \be
  \mathcal{G}_\AA = \PicCAA \,.
  \ee 
Because \CAA\ is not, in general, braided, the Picard group of \CAA\
can be nonabelian, in contrast to ${\rm Pic}(\calc)$.
An example for a nonabelian bimodule Picard group is found in the 
critical three-states Potts model, see section \ref{sec:potts-def}.

Taking the trace in the defining condition $X \otA X^\vee \Cong A$
of a group-like bimodule $X_g$, we see that 
  \be
  \big( \dim_A(X_g) \big)^2_{} = 1  \,,  
  \ee
where $\dim_A(X) \eq \dim(X)/{\dim}(A)$ is the
quantum dimension in \CAA. The assignment
  \be
  \eps :\quad \mathcal{G}_\AA \to \{ \pm 1 \} \cong \zet_2
  \,, \quad\ \eps(g) := \dim(X_g){/}{\dim(A)} 
  \ee
defines a group character on $\mathcal{G}_\AA$. The following result shows 
that if tensoring with $X_g$ leaves any simple bimodule fixed, then we are 
guaranteed that $\eps(g) \eq 1$.

\dtl{Proposition}{prop:eps(g)=1}
Let $A$ and $B$ be simple symmetric special Frobenius algebras in \C. 
If for $g \iN \mathcal{G}_A$ there exists a simple
$A$-$B$-bimodule $Y$ or a simple $B$-$A$-bimodule $Y'$ such that $X_g \otA Y 
\Cong Y$ or $Y' \otA X_g \Cong Y'$, respectively, then $\eps(g)\eq1$.

\medskip\noindent
Proof:\\
Suppose there exists a simple $A$-$B$-bimodule $Y$ such that $X_g\otA Y\Cong Y$. 
Taking the trace we obtain $\dim(X_g) \dim(Y) /{\dim}(A) \eq \dim(Y)$. By lemma 
\ref{lem:dim(simple)} $\dim(Y)\nE 0$, hence $\dim(X_g) \eq \dim(A)$, i.e.\ 
$\eps(g)\eq 1$. The reasoning in the case of $Y'$ is analogous.
\qed 

\dtl{Remark}{rem:M35}
(i)~\,As the proof of proposition \ref{prop:eps(g)=1} shows, in the first case
   only the left $A$-module structure of $Y$ is needed, and in the second case
   only the right $A$-module structure of $Y'$. The algebra $B$ is introduced
   in the statement merely because later on we want to use the result
   in the discussion of bimodule stabilisers.
\\[2pt]
(ii)~To find examples for $\eps(g)\eq{-}1$, consider the simplest case, i.e.\ 
$A\eq\one$. We are then looking for a chiral CFT which has simple currents of 
quantum dimension $-1$. It turns out that certain non-unitary Virasoro minimal 
models in the series M$(p,q)$, $p,q\,{\ge}\,2$ have this property.\,%
  \footnote{~For these models it has not yet been proven that the
  representation category of the vertex algebra is modular.
  However, the fusing and braiding matrices can be extracted
  from the monodromies of the conformal four-point blocks, which
  in turn are found by solving differential equations resulting
  form null vectors \cite{dofa2}.}
Recall that the quantum dimension $\dim(U)$ of an object $U$ is the trace of 
the identity morphism $\id_U$. Another notion of dimension is provided by 
the Perron-Frobenius dimension (see e.g.\ \cite{caet}), which is always
positive. Here we are interested in the factors that appear
when inserting or removing small circular defects, and these
are given in terms of traces of identity morphisms, i.e.\ by quantum dimensions. 
For a simple object $U_k$, the quantum dimension can be computed from the fusing
and braiding matrices and the twist as in equation (2.45) of \cite{tft1}:
  \be
  \dim(U_k) = \theta_k 
  \big( \Fs k{\bar k}kk00 \Rs{-}{\bar k}k0 \big)^{-1}.
  \ee
For Virasoro minimal models, this simplifies to $\dim(U_k) \eq \Fs kkkk00$.
Inserting the $\FF$-matrices\,\footnote{~%
  The relevant formulas have been gathered e.g.\ in formula (A.6) of
  \cite{grrw2}, which in the present notation gives the $\FF$-matrix element
  $\Fs JKLIPQ$.}
of \cite{dofa2}, one finds, for instance, that in the minimal model M(3,5) the 
field with Kac labels $(2,1)$ has quantum dimension $-1$.
\\[2pt]
(iii)\,Note that once negative quantum dimensions occur, there can also be
objects of dimension zero even if all simple objects have non-zero quantum 
dimension.  For M(3,5), for example, the object 
$U \eq (1,1) \,{\oplus}\, (2,1)$ has $\dim(U) \eq 0$.  
Recall from lemma \ref{lem:dim(simple)} that a simple $A$-$B$-bimodule 
necessarily has nonzero dimension. In particular, the object $U$ introduced
above cannot carry the structure of a simple $A$-$B$-bimodule, irrespective of
which simple symmetric special Frobenius algebras $A$ and $B$ are considered.  

\medskip

If we apply the identity \erf{eq:2-opp-sum} to the case of a group-like defect
$X_\sigma \eq X_\rho \eq X_g$, we get the relation
  \bea \begin{picture}(230,45)(0,44)
  \put(20,0)     {\Includeourbeautifulpicture 13a }
  \put(140,0)    {\Includeourbeautifulpicture 13b }
  \put(140,0){
    \put(35,65)  {\scriptsize$X_g$}
    \put(35,18)  {\scriptsize$X_g$}
  }
  \put(10,30)  {\scriptsize$X_g$}
  \put(57,30)  {\scriptsize$X_g$}
  \put(85,44)  {$ = ~ \epsilon (g)$ }
  \epicture19 
  \labl{eq:group-def-id}
since only $\mu\eq0$ can contribute to the summation coming from 
\erf{eq:2-opp-sum}. Using this identity it is easy to see what happens when a 
group-like defect $X_g$ is inflated in a world sheet \X\ in phase \CFTA.
For example, figure \erf{eq:appy-defect} then simplifies to
  \bea \begin{picture}(410,45)(0,38)
  \put(20,0)     {\Includeoursmallnicepicture 14a }
  \put(210,0)    {\Includeoursmallnicepicture 14b }
  \put(239,60)   {\cfta \put(   0,-16  ){\line(0,4){20}}
                        \put(-0.1,-16.2){\line(0,4){20}} }
  \put( 52,60)   {\cfta \put(   0,-16  ){\line(0,4){20}}
                        \put(-0.1,-16.2){\line(0,4){20}} }
  \put(200,40)   {$=$}
  \put(250.5,30) {\scriptsize$X_g$}
  \put(325,36)   {\scriptsize$X_g$}
  \epicture17 \labl{pic-ffrs5-14}
Note that here an even number of factors $\eps(g)$ occurs. 
In general, by the identity \erf{eq:group-def-id} we have the following 
result for taking a group-like defect $X_g$ past a bulk field.
  \bea \begin{picture}(340,44)(0,34)
  \put(20,0)     {\Includeourbeautifulpicture 86a }
  \put(20,0){
     \setlength{\unitlength}{.38pt}\put(0,0){
     \put(15,157)  {\scriptsize$ X_g $}
     \put(51, 82)  {\scriptsize$ \phi $}
     }\setlength{\unitlength}{1pt}}
  \put(100,0)    {\Includeourbeautifulpicture 86b }
  \put(100,0){
     \setlength{\unitlength}{.38pt}\put(0,0){
     \put(136,136)  {\scriptsize$ X_g $}
     \put(46, 82)  {\scriptsize$ \phi $}
     }\setlength{\unitlength}{1pt}}
  \put(240,0)    {\Includeourbeautifulpicture 86c }
  \put(240,0){
     \setlength{\unitlength}{.38pt}\put(0,0){
     \put( 14,155)  {\scriptsize$ X_g $}
     \put(100, 44)  {\scriptsize$ X_g $}
     \put( 53, 82)  {\scriptsize$ \phi $}
     }\setlength{\unitlength}{1pt}}
  \put(66,36)    {$\equiv$}
  \put(180,36)   {$=~~\eps(g)$}
  \epicture15 \labl{pic-ffrs5-86}
A similar identity holds for taking a group-like defect past a
`hole' in the world sheet $\X$, i.e.\ a connected boundary component. 
Applying also the moves \erf{pic-ffrs5-11} for each handle of $\X$, we find,
for an orientable world sheet in phase \CFTA\ of genus $h$ with $m$ bulk field 
insertions labelled by morphisms $\phi_1,\dots,\phi_m$ and and $b$ boundary 
components with boundary conditions $M_1,M_2,...\,,M_b$,
  \bea
  C\big(\,\X[\phi_1,...\,,\phi_m;M_1,...\,,M_b]\,\big)
  \\[5pt] \hspace*{6em}
  = \eps(g)^{m+b}\, C\big(\,\X[D_g(\phi_1),...\,,D_g(\phi_m);
  X_g \OtA M_1,...\,,X_g \OtA M_b]\,\big) \,,
  \eear\ee
where $D_g\,{\equiv}\,D_{X_g}$ is the map defined in \erf{eq:D_nu-def}. In the 
presence of boundary fields one obtains a similar identity, in which there is 
in addition an action of the
$X_g$-defect (as in \erf{eq:bnd-act}) on each boundary field. 

\medskip

We have thus arrived at our first notable conclusion:
\\[-1.5em]~
\begin{quote}
    {\em Group-like defects give rise to symmetries of CFTs on 
    oriented world sheets}.
\end{quote}

\noindent
Further, recall from proposition \ref{prop:g-distinct} that two non-isomorphic 
simple defects never act in the same way on all bulk fields. This 
implies that defects labelled by non-isomorphic group-like bimodules
in fact describe {\em distinct\/} symmetries.


\subsection{Defects generating order-disorder dualities}\label{sec:dual}

In this section we consider phase changing defects between \CFTA\ and 
\CFTB, which includes $A\eq B$ as a special case. 
Similarly as in \erf{pic-ffrs5-86} we may want to take an arbitrary
$B$-$A$-defect $Y$ past a bulk field of \CFTA. But when doing so
we now generically obtain a sum over disorder fields in \CFTB:
  \bea \begin{picture}(280,42)(0,30)
  \put(20,0)     {\Includeourbeautifulpicture 15a }
  \put(180,0)    {\Includeourbeautifulpicture 15b }
  \put(-9,20)    {\scriptsize$\cftb$}
  \put(195,10)   {\scriptsize$\cftb$}
  \put(250,40)   {\scriptsize$\cfta$}
  \put(77,40)    {\scriptsize$\cfta$}
  \put(40,40)    {\scriptsize$\phi$}
  \put(40,65)    {\scriptsize$Y$}
  \put(236,10)   {\scriptsize$Y$}
  \put(239,60)   {\scriptsize$Y$}
  \put(239,30)   {\scriptsize$\alpha$}
  \put(205,56)   {\scriptsize$X_{\mu}$}
  \put(175,43)   {\scriptsize$\theta_{\mu\alpha}$}
  \put(117,36)   {$\dsty =~~\sum_{\mu,\alpha}$}
  \epicture09 
  \labl{eq:X-past-bulk}
Here first \erf{eq:2-opp-sum} is applied, and then \erf{eq:def-bulk-act}. In 
the sum, $\alpha$ runs over a basis of $\HomAA(X_\mu\OtA Y,Y)$ and $\theta
_{\mu\alpha}$ denotes
the resulting disorder field at the end of the $B$-$B$-defect $X_\mu$. 

Repeating the above procedure with an $A$-$B$-defect $Y'$ will in general
again result in a sum over disorder fields. In case it is nevertheless possible 
to transform \erf{eq:X-past-bulk} back into an order correlator, 
i.e.\ into a correlator involving only bulk fields, but no disorder fields,
we have obtained an order/disorder symmetry. This motivates the 
\\[-2.3em]

\dtl{Definition}{def:duality}
A $B$-$A$-defect $Y$ is called a {\em duality defect\/} iff there exists
an $A$-$B$-defect $Y'$ such that, for every bulk field of \CFTA, taking first 
$Y$ past that bulk field and then $Y'$ past the resulting sum over
disorder fields, gives a sum over bulk fields of \CFTA.
\\
The bimodule underlying a duality defect is called a {\em duality bimodule\/}.

\medskip

Graphically, the requirement in the definition means that for every choice of 
bulk field label $\phi$ one has
  \bea \begin{picture}(420,75)(10,47)
  \put(20,0)     {\Includeourbeautifulpicture 16a }
  \put(180,0)    {\Includeourbeautifulpicture 16b }
  \put(340,0)    {\Includeourbeautifulpicture 16c }
  \put(100,70)   {\cftb \put( -19,10  ){\line(-3,1){20}}
                        \put( -19,10.2){\line(-3,1){20}} }
  \put(100,100)  {\scriptsize$\cfta$}
  \put(15,100)   {\scriptsize$\cfta$}
  \put(55,58)    {\scriptsize$\phi$}
  \put(87,95)    {\scriptsize$Y$}
  \put(87,15)    {\scriptsize$Y$}
  \put(63,15)    {\scriptsize$Y'$}
  \put(63,95)    {\scriptsize$Y'$}
  \put(125,60)   {$\dsty = ~\sum_{\mu,\alpha}$}
  \put(213,64)   {\scriptsize$X_{\mu}$}
  \put(190,51)   {\scriptsize$\theta_{\mu\alpha}$}
  \put(253,45)   {\scriptsize$\alpha$}
  \put(265,66)   {\cftb \put( -19,10  ){\line(-3,1){20}}
                        \put( -19,10.2){\line(-3,1){20}} }
  \put(190,100)  {\scriptsize$\cfta$}
  \put(265,100)  {\scriptsize$\cfta$}
  \put(251,95)   {\scriptsize$Y$}
  \put(248,15)   {\scriptsize$Y$}
  \put(221,15)   {\scriptsize$Y'$}
  \put(221,95)   {\scriptsize$Y'$}
  \put(285,60)   {$\dsty =~ \sum_{\mu,\alpha,\beta}$}
  \put(346,62)   {\scriptsize$\phi_{\beta}$}
  \put(340,100)  {\cfta}
  \put(420,100)  {\cfta}
  \put(382,87)   {\cftb }
  \put(380,12)   {\cftb }
  \put(411,90)   {\scriptsize$Y$}
  \put(409,15)   {\scriptsize$Y$}
  \put(359,15)   {\scriptsize$Y'$}
  \put(362,90)   {\scriptsize$Y'$}
  \put(385,68)   {\scriptsize$X_{\mu}$}
  \put(363,72)   {\scriptsize$\beta$}
  \put(412,44)   {\scriptsize$\alpha$}
  \epicture27 
  \labl{eq:duality-prop}
That is, on the right hand side we have a sum over bulk fields only,
rather than in addition over defect fields. The dashed box indicates that
after taking the two defects $Y$ and $Y'$ past the bulk field, they are 
no longer separated, but linked by a morphism in $\HomAA(Y' \OtB Y,Y' \OtB Y)$.

We write
  \be
  \mathcal{D}_{BA} := \big\{\, \mu \iN \KK_{BA} \,\big|\, X_\mu 
  {\rm ~is~a~duality~bimodule}\,\big\} \,.
  \labl{eq:Dset-def}
Note that in definition \ref{def:duality} we do not demand that
a duality defect is simple, and indeed non-simple duality defects
can appear. However, all duality defects can be obtained as
appropriate superpositions of simple duality defects. This will be shown
in lemma \ref{lem:sum-dual} below. Before, however, we give a more
convenient characterisation of duality defects, since in practise
it would be tedious to check property \erf{eq:duality-prop} directly. This 
characterisation only uses the fusion algebra of the defect lines:
\\[-2.2em]

\dtl{Theorem}{thm:dual}
A $B$-$A$-bimodule $Y$ is a duality bimodule 
if and only if $Y^\vee \otB Y$ is a direct sum of group-like bimodules,
  \be
  Y^\vee \otB Y \cong \bigoplus_{g \in \mathcal{G}_\AA} n_g X_g
  \labl{eq:XvX-decomp}
for suitable $n_g \iN \zet_{\ge 0}$.

\noindent
Proof:\\
``$\Leftarrow$'': Suppose that $Y^\vee \OtB Y \Cong \bigoplus_g\! n_g X_g$. 
The following transformations show that $Y$ is a duality defect, with the 
$A$-$B$-defect $Y'$ in definition \ref{def:duality} given by $Y^\vee$:
  \bea \begin{picture}(420,85)(16,59)
  \put(0,0)      {\Includeourbeautifulpicture 18a }
  \put(170,0)    {\Includeourbeautifulpicture 18b }
  \put(340,0)    {\Includeourbeautifulpicture 18c }
  \put(78,95)    {\cftb \put(-38,4){\line(4,0){20}}
                        \put(-38,4.2){\line(4,0){20}} }
  \put(25,120)   {\scriptsize$\cfta$}
  \put(70,120)   {\scriptsize$\cfta$}
  \put(33,68)    {\scriptsize$\phi$}
  \put( 8,40)    {\scriptsize$Y$}
  \put(25,85)    {\scriptsize$Y$}
  \put(120,70)   {$\dsty =~\sum_{g,\alpha}$}
  \put(220,5)    {\scriptsize$Y$}
  \put(250,5)    {\scriptsize$Y$}
  \put(250,130)  {\scriptsize$Y$}
  \put(231,122)  {\scriptsize$Y$}
  \put(203,47)   {\scriptsize$\alpha$}
  \put(203,95)   {\scriptsize$\widetilde{\alpha}$}
  \put(217,68)   {\scriptsize$\phi$}
  \put(183,70)   {\scriptsize$X_g$}
  \put(255,109)  {\cftb \put(-38,4){\line(4,0){20}}
                        \put(-38,4.2){\line(4,0){20}} }
  \put(205,130)  {\scriptsize$\cfta$}
  \put(230, 70)  {\scriptsize$\cfta$}
  \put(270,70)   {$\dsty =~\sum_{g,\alpha}\epsilon(g)$}
  \put(15,0){
    \put(340,100)  {\cfta}
    \put(428,100)  {\cfta}
    \put(380,89)   {\cftb }
    \put(380,33)   {\cftb } }
  \put(360,72)   {\scriptsize$\phi$}
  \put(343,92)   {\scriptsize$X_g$}
  \put(378,125)  {\scriptsize$Y$}
  \put(426,129)  {\scriptsize$Y$}
  \put(378,4)    {\scriptsize$Y$}
  \put(425,4)    {\scriptsize$Y$}
  \put(426,40)   {\scriptsize$\alpha$}
  \put(425,113)  {\scriptsize$\widetilde{\alpha}$}
  \put(411,70)   {\scriptsize$X_g$}
  \epicture33 
  \labl{eq:dual-proof-1}
In the first step the bimodule $Y^\vee \OtB Y$ is decomposed into a direct sum
of simple $A$-$A$-bimodules, similarly to \erf{eq:2-opp-sum} (all 
prefactors have been absorbed into the definition of the morphisms $\alpha$ 
and $\tilde\alpha$). By assumption, only group-like bimodules appear in the 
decomposition. The resulting group-like defect can then be moved past the bulk 
field by using \erf{pic-ffrs5-86};
this is done in step two. Comparing the \rhs\ of \erf{eq:dual-proof-1}
with \erf{eq:duality-prop}, we conclude that $Y$ is a duality defect.
\\[.2em]
``$\Rightarrow$'': Suppose now that the $B$-$A$-defect $Y$ is a duality defect,
i.e.\ that there exists an $A$-$B$-defect $Y'$ such that for any choice
of bulk field label $\phi$ in \CFTA\ we have
  \bea \begin{picture}(320,64)(0,55)
  \put(20,0)     {\Includeourbeautifulpicture 19a }
  \put(200,0)    {\Includeourbeautifulpicture 19b }
  \put(97,83)    {\cftb \put(-38,4.5){\line(4,0){20}}
                        \put(-38,4.3){\line(4,0){20}} }
  \put(-5,70)    {\scriptsize$\cfta$}
  \put(90,50)    {\scriptsize$\cfta$}
  \put(53,60)    {\scriptsize$\phi$}
  \put(61,102)   {\scriptsize$Y'$}
  \put(89,102)   {\scriptsize$Y$}
  \put(146,60)   {$\dsty =~\sum_\sigma$}
  \put(208,60)   {\scriptsize$\phi_{\sigma}$}
  \put(263,90)   {\cftb \put(-38,4.5){\line(4,0){20}} 
                        \put(-38,4.3){\line(4,0){20}} }
  \put(190,90)   {\cfta}
  \put(265,40)   {\cfta}
  \put(210.8,19) {\scriptsize$Y'$}
  \put(210.8,111){\scriptsize$Y'$}
  \put(253,19)   {\scriptsize$Y$}
  \put(253,111)  {\scriptsize$Y$}
  \put(243.3,57.5){\scriptsize$\sigma$}
  \epicture29 \labl{pic-ffrs5-19}
where the sum is over elements $\sigma \iN \HomAA(Y' \otB Y,Y' \otB Y)$
and the $\phi_\sigma$ label again bulk fields of \CFTA.
Then in particular, the following identity must hold for the two-point 
correlator of one bulk field and one defect field on the Riemann sphere:
  \begin{eqnarray} \begin{picture}(430,160)(0,0)
  \put(0,0)      {\Includeourbeautifulpicture 21a }
  \put(270,0)    {\Includeourbeautifulpicture 21b }
  \put(115,110)  {\scriptsize$\cfta$}
  \put(14,140)   {\scriptsize$\cfta$}
  \put(10,40)    {\scriptsize$\cftb$}
  \put(198,70)   {\scriptsize$\theta_{\bar\imath\bar\jmath,\beta}$}
  \put(30,94)    {\scriptsize$Y$}
  \put(23,127)   {\scriptsize$Y'$}
  \put(34,79)    {\scriptsize$\phi_{ij,\alpha}$}
  \put(170,84)   {\scriptsize$X_{\mu}$}
  \put(136,55)   {\scriptsize$\gamma$}
  \put(104,9)    {\scriptsize$X_{\nu}$}
  \put(130,142)  {\scriptsize$X_{\nu}$}
  \put(63,14)    {\scriptsize$\tau$}
  \put(38,20)    {\scriptsize$X_{\rho_1}$}
  \put(68,32)    {\scriptsize$X_{\rho_2}$}
  \put(38,35)    {\scriptsize$\varepsilon_1$}
  \put(72,45)    {\scriptsize$\varepsilon_2$}
  \put(63,141)   {\scriptsize$\overline{\tau}$}
  \put(36,130)   {\scriptsize$X_{\rho_1}$}
  \put(68,123)   {\scriptsize$X_{\rho_2}$}
  \put(38,120)   {\scriptsize$\overline{\varepsilon}_1$}
  \put(72,110)   {\scriptsize$\overline{\varepsilon}_2$}
  \put(355,110)  {\scriptsize$\cfta$}
  \put(400,140)  {\scriptsize$\cfta$}
  \put(265,45)   {\cftb \put(0,4.5){\line(4,0){20}} }
  \put(265,94)   {\cftb \put(0,4.5){\line(4,0){20}} }
  \put(225,77)   {$\dsty=\sum_{\sigma}$}
  \put(277,79)   {\scriptsize$\phi_{\sigma}$}
  \put(421,84)   {\scriptsize$X_{\mu}$}
  \put(448,70)   {\scriptsize$\theta_{\bar\imath\bar\jmath,\beta}$}
  \put(387,55)   {\scriptsize$\gamma$}
  \put(352,9)    {\scriptsize$X_{\nu}$}
  \put(377,143)  {\scriptsize$X_{\nu}$}
  \put(311,14)   {\scriptsize$\tau$}
  \put(288,20)   {\scriptsize$X_{\rho_1}$}
  \put(317,27)   {\scriptsize$X_{\rho_2}$}
  \put(288,34)   {\scriptsize$\varepsilon_1$}
  \put(320,39)   {\scriptsize$\varepsilon_2$}
  \put(311,141)  {\scriptsize$\overline{\tau}$}
  \put(285,133)  {\scriptsize$X_{\rho_1}$}
  \put(317,126)  {\scriptsize$X_{\rho_2}$}
  \put(288,122)  {\scriptsize$\overline{\varepsilon}_1$}
  \put(320,114)  {\scriptsize$\overline{\varepsilon}_2$}
  \put(315,77)   {\scriptsize$\sigma$}
  \end{picture}
  \nonumber\\[-1.8em]
  ~\label{eq:2bulk-id} 
  \end{eqnarray}
for all choices of simple objects $i,j$ and simple bimodules $\rho_1, \rho_2, 
\nu, \mu$, and for all choices of basis labels $\alpha,\beta, \eps_1,\eps_2,
\tau,\gamma$. Here $\eps_{1,2}$ and $\bar\eps_{1,2}$ are dual basis elements
as in \erf{eq:Xmu-X-basis} and $\tau,\bar\tau$ are dual basis elements as in 
\erf{eq:Xmu-XY-basis}. By \erf{eq:X-not-end}, the right hand side of 
\erf{eq:2bulk-id} is zero if $\mu \nE 0$. For the left hand side we find
  \bea \begin{picture}(420,93)(16,28)
  \put(111,0)    {\includeourbeautifulpicture 22 }
  \put(0,59)     {l.h.s~\,of\,~\erf{eq:2bulk-id}~~=}
  \put(285,57)   {$ =~\,\GammA(\mu)_{ij\alpha\beta}^{\nu\gamma}
                    \cdot (\mbox{2-point-blocks}) $}
  \put(220,110)  {\scriptsize$\cfta$}
  \put(145,90)   {\scriptsize$\cfta$}
  \put(112,50)   {\cftb }
  \put(155,58)   {\scriptsize$\phi_{ij,\mu}$}
  \put(202,95)   {\scriptsize$X_{\nu}$}
  \put(265,50)   {\scriptsize$\theta_{\bar\imath\bar\jmath,\beta}$}
  \put(203,37)   {\scriptsize$\gamma$}
  \put(250,67)   {\scriptsize$X_{\mu}$}
  \put(173,14)   {\scriptsize$X_{\nu}$}
  \put(127,14)   {\scriptsize$\tau$}
  \put(98,80)    {\scriptsize$X_{\rho_1}$}
  \put(131,80)   {\scriptsize$X_{\rho_2}$}
  \put(127,102)  {\scriptsize$\overline{\tau}$}
  \epicture07 \labl{pic-ffrs5-22}
Here in the first step one uses the property that 
$\eps_{1,2}$ and $\bar\eps_{1,2}$ are dual to each other, and in the second
step that $\tau$ and $\bar\tau$ are dual, together with the
definition \erf{eq:A-def}. Thus the identity \erf{eq:2bulk-id}
implies that for every simple $A$-$A$-bimodule $X_\nu$ contained in 
$Y' \otB Y$, and for every $\mu \iN \KK_{AA}{\setminus}\{0\}$, we have
  \be
  \GammA(\mu)_{ij\alpha\beta}^{\nu\gamma} = 0
  \qquad {\rm for~all~}\, \gamma {\rm~such~that~} (\nu\gamma) \iN R_2
  {\rm~~and~all~} (ij\alpha\beta) \iN R_1 \,.
  \ee
Since by lemma \ref{lem:non-deg} the matrix $\GammA(\mu)$ is non-degenerate,
this is only possible if, for every $\mu \nE 0$ and every
$X_\nu$ contained in $Y' \otB Y$, the index set $R_2$ does not
contain any element of the form $(\nu\gamma)$, i.e.\ only if $\dim_\complex
\HomAA(X_\nu,X_\mu \otA X_\nu) \eq 0$. In other words, we learn that
  \be
  \HomAA(X_\nu , Y' \otB Y) \neq \{0\} \ \  \Longrightarrow \ \ 
  \HomAA(X_\nu \otA X_\nu^\vee,X_\mu) = \{0\} 
  ~~{\rm for~all~} \mu \nE 0 \,. 
  \ee
This is nothing but the statement that all simple sub-bimodules
of $Y' \OtB Y$ are group-like.
\\[2pt]
It remains to show that not only $Y'$ has this property, but also $Y^\vee$.
Let $Y \eq \bigoplus_r\! Y_r$ and $Y' \eq \bigoplus_s\! Y_s'$ be decompositions
of $Y$ and $Y'$ into simple bimodules. The fact that $Y' \OtB Y$ is
a direct sum of group-like bimodules implies in particular that
$Y_1' \OtB Y_r$ is a direct sum of group-like bimodules. Thus there is a 
$g_r \iN \mathcal{G}_A$ such that $X_{g_r} \OtA Y_1' \OtB Y_r$ contains $A$, 
and hence $X_{g_r}\OtA Y_1'\,{\cong}\,Y_r^\vee$. It follows that for all $r,s$,
$Y_r^\vee \OtB Y_s \eq X_{g_r} \OtA Y_1' \OtB Y_s$ is a direct sum of group-like
bimodules. Thus also $Y^\vee \OtB Y$ is a direct sum of group-like bimodules.
\qed

Theorem \ref{thm:dual} can be reformulated as the statement that $Y$ is a 
duality defect iff $Y^\vee \otB Y$ is in the Picard category of \CAA\
(see definition 2.1 of \cite{tft3}). 

\medskip

Up to now we did not demand the $B$-$A$-defect $Y$ to be simple. Suppose that 
$Y$ decomposes into simple $B$-$A$-bimodules 
$X_\mu$ as $Y \Cong \bigoplus_{\mu \in \KK_{BA}}\! m_\mu X_\mu$.
The statement that $Y$ is a duality defect then amounts to
  \be
  \bigoplus_{\mu,\nu \in \KK_{BA}} \!\!  m_\mu m_\nu\, X_\mu^\vee \otB X_\nu
  \cong \bigoplus_{g \in \mathcal{G}_A} n_g\, X_g \,.
  \labl{eq:dual-decomp}
This implies in particular that $X_\mu^\vee \otB X_\nu$ is a direct sum of 
group-like bimodules for any pair of $X_\mu$ and $X_\nu$ that are sub-bimodules
of $Y$. But then there exists a $g \iN \mathcal{G}_\AA$ such that
$X_g \otA X_\mu^\vee \otB X_\nu$ contains the tensor
unit $A$, and hence $X_\mu \otA X_{g^{-1}} \Cong X_\nu$. This implies
\\[-2.3em]

\dtl{Lemma}{lem:sum-dual}
Let $\mu_1,\mu_2,...\,,\mu_m \iN \KK_{BA}$. Then 
$Y \eq X_{\mu_1} \,{\oplus} \cdots {\oplus}\, \X_{\mu_m}$ is a duality
bimodule iff $\mu_1 \iN \mathcal{D}_{BA}$ and 
$\mu_1,\mu_2,...\,,\mu_m$ lie on one and the same orbit of the right action of
$\mathcal{G}_\AA$, i.e.\ there exist $g_2,...\,,g_m \iN \mathcal{G}_\AA$ such 
that $X_{\mu_1} \otA X_{g_k} \Cong X_{\mu_k}$ for all $k\eq2,3,...\,,m$.

\bigskip

We have learned that all $B$-$A$-duality defects can be obtained
as direct sums of the simple duality defects $\mathcal{D}_{BA}$.
Moreover, we are only allowed to superimpose such simple duality defects
that lie on one and the same orbit of the right $\mathcal{G}_\AA$-action
on $\mathcal{D}_{BA}$. To describe all defect-induced order-disorder
dualities it is therefore sufficient to consider simple duality defects only.

In order to better understand which group-like defects 
appear on the right hand side of \erf{eq:XvX-decomp}, we introduce
the notion of a left and a right stabiliser. This is a straightforward
generalisation of the corresponding notion for the braided category $\calc$
(in which the left and a right stabilisers coincide, see definition 4.1\,(i) 
of \cite{tft3}) to the sovereign category \CAA.
\\[-2.2em]

\dtl{Definition}{def:lr-stab}
The {\em left and right stabilisers\/} $\mathcal{S}^{l,r}(Y)$ of 
an $A$-$B$-bimodule $Y$ are given by
  \be
  \mathcal{S}^l(Y) := \big\{\, g \iN \mathcal{G}_\AA \,\big|\,
  X_g \OtA Y \Cong Y \, \big\} \quad~{\rm and}~\quad
  \mathcal{S}^r(Y) := \big\{\, h \iN \mathcal{G}_B \,\big|\,
  Y \OtB X_h \Cong Y \, \big\} \,.
  \ee

\medskip

It is straightforward to check that 
$\mathcal{S}^l(Y) \,{\subseteq}\, \mathcal{G}_\AA$ and
$\mathcal{S}^r(Y) \,{\subseteq}\, \mathcal{G}_B$ are in fact subgroups. With the
notion of a stabiliser, for a simple $B$-$A$-defect $X$ we can slightly 
strengthen the condition of theorem \ref{thm:dual}
for $X$ to be a duality defect:
\\[-2.2em]

\dtl{Proposition}{prop:simple-dual}
Let $Y$ be a simple $B$-$A$-defect. Then $Y$ is a duality defect iff
  \be
  Y^\vee \otB Y \cong \bigoplus_{h \,\in\, \mathcal{S}^r(Y)}\! X_h \,.
  \labl{eq:simple-dual-def}

\noindent
Proof:\\
First note that $Y \OtA X_h$ is again simple, since by
lemma \ref{lem:duals} we have $\End_{B|A}(Y \OtA X_h) \Cong \HomBA
   $\linebreak[0]$
(Y , Y \OtA X_h \OtA X_h^\vee) \Cong \HomBA(Y,Y)$.
Further, the multiplicity $n_h$ of $X_h$ in $Y^\vee \OtB Y$ is
given by (again using lemma \ref{lem:duals})
  \be
  n_h = \dim_\complex\HomAA(Y^\vee \OtB Y,X_h)
  = \dim_\complex\HomBA(Y,Y \OtA X_h) \,.
  \ee
Since $Y$ and $Y \OtA X_h$ are simple, it follows that
$n_h \eq 1$ iff $h \iN \mathcal{S}^r(Y)$ and $n_h \eq 0$ otherwise.
\qed

\medskip

Let $Y$ be a simple $B$-$A$-duality bimodule and let
$g,h \iN \mathcal{S}^r(Y)$ such that $gh\eq hg$. Consider the morphism 
$y_{gh} \iN \HomAA(X_h \OtA X_g, X_g \OtA X_h)$ that is given by
  \bea \begin{picture}(380,70)(0,43)
  \put(40,0)     {\Includeourtinynicepicture 6{7-5}a }
  \put(150,0)    {\Includeourtinynicepicture 6{7-5}b }
  \put(280,0)    {\Includeourtinynicepicture 6{7-5}c }
  \put(340,70)   {\cfta}
  \put(290,47)   {\cftb}
  \put(322,12)   {\scriptsize$h$}
  \put(322,53)   {\scriptsize$Y$}
  \put(322,78)   {\scriptsize$g$}
  \put(306,92)   {\scriptsize$h$}
  \put(305,40)   {\scriptsize$Y$}
  \put(305,69)   {\scriptsize$Y$}
  \put(274,65)   {\scriptsize$Y$}
  \put(306,4)    {\scriptsize$g$}
  \put(207,58)   {$\dsty := ~\frac{\dim A}{\dim Y}$}
  \put(182,12)   {\scriptsize$h$}
  \put(182,93)   {\scriptsize$g$}
  \put(145,12)   {\scriptsize$g$}
  \put(145,93)   {\scriptsize$h$}
  \put(160,56)   {\scriptsize$y_{gh}$}
  \put(108,58)    {$\equiv$}
  \put(72,12)    {\scriptsize$h$}
  \put(72,98)    {\scriptsize$g$}
  \put(37,12)    {\scriptsize$g$}
  \put(36,98)    {\scriptsize$h$}
  \epicture20 
  \labl{eq:group-crossing}
The morphisms in $\HomBA(Y \otA X_x, Y)$ and $\HomBA(Y, Y \otA X_x)$, for 
$x \eq g,h$, that label the junctions in \erf{eq:group-crossing} are chosen to 
be dual to each other. Since the morphism spaces are one-dimensional, the 
resulting morphism $y_{gh}$ is independent of these choices.
The normalisation factor in \erf{eq:group-crossing} has been introduced 
in such a way that \erf{eq:group-crossing} obeys
  \bea \begin{picture}(220,45)(0,40)
  \put(20,0)     {\Includeourbeautifulpicture 64a }
  \put(170,0)    {\Includeourbeautifulpicture 64b }
  \put(100,50)  {$=$}
  \put(-10,58)  {\scriptsize$\cfta$}
  \put(15,4)    {\scriptsize$g$}
  \put(13,40)   {\scriptsize$h$}
  \put(15,85)   {\scriptsize$g$}
  \put(52,4)    {\scriptsize$h$}
  \put(53,40)   {\scriptsize$g$}
  \put(52,85)   {\scriptsize$h$}
  \put(42,64)   {\scriptsize$y_{hg}$}
  \put(42,23)   {\scriptsize$y_{gh}$}
  \put(145,58)  {\scriptsize$\cfta$}
  \put(174,43)  {\scriptsize$g$}
  \put(202,43)  {\scriptsize$h$}
  \epicture16 
  \labl{eq:group-cross}
This can be seen by the following chain of equalities
(we abbreviate $\zeta \eq \dim(A)/{\dim}(Y)$).
  \begin{eqnarray} \begin{picture}(430,243)(12,0)
  \put(0,120)    {l.h.s. $\dsty=~ \zeta^2 \!\!\!
                  \sum_{g' \in \mathcal{S}^r(Y)}$}
  \put(201,120)  {$=~ \zeta$}
  \put(338,120)  {$=~ \zeta$}
 \put(-27,0){
  \put(128,0)    {\Includeourbeautifulpicture 70a }
  \put(100,200)  {\scriptsize$\cfta$}
  \put(145,170)  {\scriptsize$\cftb$}
  \put(145,50){\scriptsize$\cftb$}
  \put(198,16)  {\scriptsize$h$}
  \put(198,240)  {\scriptsize$h$}
  \put(195,57)  {\scriptsize$Y$}
  \put(195,176)  {\scriptsize$Y$}
  \put(198,130)  {\scriptsize$g'$}
  \put(171,16)  {\scriptsize$g$}
  \put(172,120)  {\scriptsize$h$}
  \put(135,70)  {\scriptsize$Y$}
  \put(135,190)  {\scriptsize$Y$}
  \put(171,242)  {\scriptsize$g$}
  }
\put(-20,0){
  \put(266,0)    {\Includeourbeautifulpicture 70b }
  \put(240,200)  {\scriptsize$\cfta$}
  \put(280,170)  {\scriptsize$\cftb$}
  \put(335,16)  {\scriptsize$h$}
  \put(335,240)  {\scriptsize$h$}
  \put(332,143)  {\scriptsize$Y$}
  \put(313,30)  {\scriptsize$Y$}
  \put(310,210)  {\scriptsize$Y$}
  \put(310,16)  {\scriptsize$g$}
  \put(310,120)  {\scriptsize$h$}
  \put(273,70)  {\scriptsize$Y$}
  \put(273,190)  {\scriptsize$Y$}
  \put(310,240)  {\scriptsize$g$}
  }
\put(-20,0){
  \put(400,0)    {\Includeourbeautifulpicture 70c }
  \put(405,215)  {\scriptsize$\cfta$}
  \put(425,120){\scriptsize$\cftb$}
  \put(470,12)  {\scriptsize$h$}
  \put(470,238)  {\scriptsize$h$}
  \put(467,118)  {\scriptsize$Y$}
  \put(405,130)  {\scriptsize$Y$}
  \put(443,238)  {\scriptsize$g$}
  \put(443,12)  {\scriptsize$g$}
  }
  \end{picture}
  \nonumber\\[-2pt]
  ~\label{eq:group-cross-aux1} 
  \end{eqnarray}
In the first step, the intermediate group-like defect $X_g$ is replaced by 
$X_{g'}$, with the label $g'$ summed over $\mathcal{S}^r(Y)$. This is allowed 
because for $gh\eq hg$, $\HomAA(X_g \OtA X_h, X_{h} \OtA X_{g'})$ is zero
unless $g'\eq g$. In the second step the $Y$-defect is deformed and the 
summation over $g'$ performed according to (a reflected version of) the 
identity \erf{eq:2-opp-sum}. The $X_h$-defect can then be omitted, as 
is easily verified with the help of the TFT 
formulation in section \ref{sec:tft-form}, using also that $Y$ is simple (see 
equation \erf{eq:fusion2-aux1}). That the \rhs\ of \erf{eq:group-cross-aux1}
equals the \rhs\ of \erf{eq:group-cross} can be seen by noting that 
these morphism spaces are one-dimensional and comparing traces.

A crossing of two group-like defects of the form just considered is implicit
in the formulation of the following 
\\[-2.3em]

\dtl{Proposition}{prop:orbifold}
Assume that $B$-$A$-duality defects exist, and let $Y$ be a simple 
$B$-$A$-duality defect. Then the torus partition function of 
\CFTB\ can be expressed in terms of torus amplitudes with defect lines of 
\CFTA\ as follows. (This is just the way in which the partition function of 
an orbifold theory is expressed as a sum over twisted sectors.) 
  \bea \begin{picture}(330,65)(0,25)
  \put(0,0)     {\Includeourtinynicepicture 66a }
  \put(210,0)    {\Includeourtinynicepicture 66b }
  \put(105,45)   {$=~\displaystyle \frac{1}{|\mathcal{S}^r(Y)|} ~
                  \underset{gh=hg}{\sum_{g,h \in \mathcal{S}^r(Y)}}$}
 \put(235,30)  {\scriptsize$\cfta$}
\put(25,55)  {\scriptsize$\cftb$}
 \put(238,60)  {\scriptsize$h$}
\put(281,76)  {\scriptsize$g$}
 \put(290,60)  {\scriptsize$h$}
\put(281,20)  {\scriptsize$g$}
  \epicture15 \labl{pic-ffrs5-66}
Proof:\\
The statement follows from the equalities
  \begin{eqnarray}\begin{picture}(430,94)(0,0)
  \put(0,0)      {\Includeourtinynicepicture 71a }
  \put(175,0)    {\Includeourtinynicepicture 71b }
  \put(110,45)  {$=~\displaystyle\frac{\dim(B)}{\dim(Y)}$}
  \put(179,67)  {\scriptsize$\cfta$}
  \put(228,70)  {\scriptsize$Y$}
  \put(25,45)   {\scriptsize$\cftb$}
  \put(212,34)  {\scriptsize$\cftb$}
  \end{picture}
  \nonumber\\ \begin{picture}(430,105)(0,0)
  \put(150,0)    {\Includeourtinynicepicture 71c }
  \put(370,0)    {\Includeourtinynicepicture 71d }
  \put(169,67)  {\scriptsize$\cfta$}
  \put(202,53)  {\cftbtiny}
  \put(400,30)  {\scriptsize$\cfta$}
  \put(220,30)  {\scriptsize$g$}
  \put(216,80)  {\scriptsize$Y$}
  \put(178,40)  {\scriptsize$h$}
  \put(160,70)  {\scriptsize$Y$}
  \put(400,60)  {\scriptsize$h$}
  \put(441,30)  {\scriptsize$g$}
  \put(445,60)  {\scriptsize$h$}
  \put(441,76)  {\scriptsize$g$}
  \put(0,45)    {$=~\displaystyle \frac{\dim(B)}{\dim(Y)}
                   \cdot\Big(\frac{\dim(A)}{\dim(Y)}\Big)^2 \sum_{g,h}$}
  \put(250,45)  {$=~\displaystyle \frac{\dim(A)\dim(B)}{\dim(Y)^2} \sum_{g,h}$}
  \end{picture}
  \nonumber\\~ \label{pic-ffrs5-71}
  \end{eqnarray}
together with the identity 
  \be
  \dim(Y)^2 / \dim(B) = |S^r(Y)| \dim(A) \,,
  \labl{YY=ABS}
which follows from taking the trace of \erf{eq:simple-dual-def}.
That only commuting pairs $(g,h)$ can give a nonzero contribution to the sum 
is due to the fact that otherwise the relevant coupling space is zero.
\qed

\dt{Remark}
In \cite{ruel'5} it is pointed out that
in the class of models investigated there (minimal models and $\mathfrak{sl}(2)$
WZW models), the models admitting a duality symmetry are precisely those
which can be described as their own orbifold, i.e.\ those which possess an 
``auto-orbifold'' property. Applying proposition \ref{prop:orbifold} for the 
special case $A\eq B$ shows that indeed an RCFT possessing a duality
defect automatically also has the auto-orbifold property.
For certain lattice models in the universality classes 
of the $c\,{<}\,1$ A-D-E minimal models, the $c\eq1$ compactified free 
boson, or its $\zet_2$-orbifolds, a related lattice construction, which also 
works off criticality and for $A\,{\ne}\,B$, is described in \cite{fegi}.


\subsection{Duality defects and Morita equivalence}\label{sec:morita}

Suppose that the simple symmetric special Frobenius algebras $A$ and $B$ are 
related by $X \otB Y \Cong A$ via an $A$-$B$-bimodule $X$ and a $B$-$A$-bimodule
$Y$. Then by lemma \ref{lem:XAB-simple}\,(iii) also $Y \OtA\, X \Cong B$, so 
that the algebras $A$ and $B$ are Morita equivalent \cite{pare14,pare16}. In 
this section we show in which sense Morita equivalent algebras lead to 
equivalent CFTs.

\medskip

Note that by lemma \ref{lem:XAB-simple}(ii), $X \otB Y \Cong A$ implies that
$Y \otA Y^\vee \Cong B$ and $Y^\vee \otB Y \Cong A$. Just as in 
\erf{eq:group-def-id}, let us apply the identity \erf{eq:2-opp-sum} to the case
$X_\sigma \eq X_\rho \eq Y$. This results in
  \bea \begin{picture}(250,44)(0,39)
  \put(20,0)     {\Includeourbeautifulpicture 29a }
  \put(180,0)    {\Includeourbeautifulpicture 29b }
  \put(30,60)    {\scriptsize$\cfta$}
  \put(-3,60)    {\scriptsize$\cftb$}
  \put(188,63)   {\scriptsize$\cfta$}
  \put(188,7)    {\scriptsize$\cfta$}
  \put(65,60)    {\scriptsize$\cftb$}
  \put(210,37)   {\scriptsize$\cftb$}
  \put(100,42)   {$\dsty =~~\frac{\dim (B)}{\dim (Y)}$}
  \put(25,30)    {\scriptsize$Y$}
  \put(63,30)    {\scriptsize$Y$}
  \put(172,18)   {\scriptsize$Y$}
  \put(214,70)   {\scriptsize$Y$}
  \epicture15 
  \labl{eq:morita-move}
Next inflate the defect $Y$ in an oriented connected world sheet \X\ as in 
\erf{eq:appy-defect}. Whenever two $Y$-defects run parallel to each other we 
can make use of the identity \erf{eq:morita-move}. In this way one obtains a 
world sheet with only small circular defects labelled by $Y$, each of
which contributes a factor $\dim(Y){/}{\dim}(A)$. Since 
$Y \otA Y^\vee \Cong B$ we have $\dim(Y){/}{\dim}(A) \eq \dim(B){/}{\dim}(Y)$,
so that the factors from circular $Y$-defects can cancel against the
factors in \erf{eq:morita-move}. Abbreviating
  \be
  \dim(B)\,/\,{\dim}(Y) =: \gamma \,,
  \ee
the net effect is\\[2pt]
{\bf--} a factor of $\gamma^2$ for each handle of $\X$;\\[2pt]
{\bf--} a factor of $\gamma$ for each connected boundary component of $\X$;\\[2pt]
{\bf--} a factor of $\gamma$ -- to be absorbed into a redefinition of
        bulk fields -- for each bulk insertion of \X.
\\[2pt]
Altogether we obtain the following relation between correlators of \CFTA\ and 
\CFTB\ on a connected, oriented world
sheet $\X$ with genus $h$, $m$ bulk insertions and $b$ boundary components:
  \be
  \Cf_A(\X) = \gamma^{-\chi(\X)}\, \Cf_B(\X') \qquad{\rm with}\qquad
  \gamma = \Frac{\dim(B)}{\dim(Y)} = \Frac{\dim(Y)}{\dim(A)}\,,
  \labl{eq:or-morita-eq}
where $\chi(\X) \eq 2-2h-b$ is the Euler character of \X.
\\
The world sheet $\X'$ is obtained from $\X$ by changing the labelling of bulk 
fields and boundary conditions. If $\phi\iN\HomAA(U \otip A \otim V, A)$ 
labels a bulk field of \CFTA\ on $\X$, then on $\X'$ this label
is replaced by $\,\gamma\, D_Y(\phi) \iN \HomBB(U \otip B \otim V, B)$,
where $D_Y \Equiv D_{0\mu\,{\bf \cdot}}$ is the map used in 
\erf{eq:defect-repn}. Using $\gamma D_Y$ instead of $D_Y$ ensures that the 
identity field of $\CFTA$ gets mapped to the identity field of $\CFTB$.    
For a boundary component of $\X$ labelled by an $A$-module
$M$, this label is replaced in $\X'$ by the $B$-module $Y \OtA M$. In the 
presence of boundary fields one must use in addition \erf{eq:bnd-act}.

Via equation \erf{eq:or-morita-eq} we define when we consider two CFTs to be
equivalent, namely iff there exist isomorphisms between the spaces of
bulk fields, boundary fields and boundary conditions of the two CFTs
such that the correlator for a world sheet $\X$ for the first CFT is
equal to the correlator for the corresponding world sheet $\X'$ of the
second CFT up to an overall constant that only depends on the Euler character
of $\X$ (which is equal to that of $\X'$).

In the case $A\eq B$ we are dealing with a group-like defect. The
discussion above then reduces to the one in section \ref{sec:group-sym}.


\subsection{Equivalence of CFTs on unoriented world sheets}
\label{sec:morita-unor}

As seen in the previous section, two Morita equivalent simple symmetric special 
Frobenius algebras result in equivalent CFTs on oriented world sheets.
To obtain a CFT that is well defined also on unoriented world sheets we 
need a Jandl algebra, see \cite[def.\,2.1]{tft2}. In this section we present an 
equivalence relation between simple Jandl algebras, such that
two equivalent Jandl algebras yield equivalent CFTs on unoriented surfaces. 

Let $A$ and $B$ be two Jandl algebras, and denote by $\sigma_A$ and
$\sigma_B$ their reversions. Let us for convenience repeat
part of the statement of \cite[proposition 2.10]{tft2}.
\\[-2.2em]

\dtl{Proposition}{pr:bimod-conj}
Let $X\eq(\dot X, \rho , \tilde \rho)$ be an $A$-$B$-bimodule. Then
$X^s \eq (\dot X, \rho^s, \tilde \rho^s)$ with
  \be
  \rho^s := \tilde\rho \circ c_{B,X}^{} \circ (\sigma_{\!B} \oti \id_X) 
  \qquad {\rm and} \qquad
  \tilde\rho^s := \rho \circ c_{X,A}^{} \circ (\id_X\oti\sigma_{\!A}) 
  \ee
is a $B$-$A$-bimodule.

\medskip

Pictorially, the left/right action on $X^s$ is as follows.
  \bea \begin{picture}(130,96)(0,23)
  \put(45,0)     {\includeourbeautifulpicture o1}
  \put(0,53)       {$ X^s \; :=$}
  \put(47,31.7)    {\scriptsize$\sigma_{\!B}^{}$}
  \put(48,-9.2)    {\scriptsize$B$}
  \put(71.3,-9.2)  {\scriptsize$X$}
  \put(72,122)     {\scriptsize$X$}
  \put(96.7,20.4)  {\scriptsize$\sigma_{\!A}^{}$}
  \put(97,-9.2)    {\scriptsize$A$}
  \epicture12 \labl{eq:Xsvs-pics}

\dtl{Definition}{def:Jandl-Morita}
Two simple Jandl algebras $A$ and $B$ are called {\em Jandl-Morita 
equivalent\/} iff there exists a $B$-$A$-bimodule $Y$ such that
$Y^\vee \OtB Y \Cong A$ as $A$-$A$-bimodules
and $Y^s \Cong Y^\vee$ as $A$-$B$-bimodules.

\dt{Remark}
(i)~\,As before, by lemma \ref{lem:XAB-simple}(ii) the condition
$Y^\vee \OtB Y \Cong A$ implies that also $Y \OtA Y^\vee \Cong B$.
\\[.2em]
(ii) If $Y$ generates a Jandl-Morita equivalence, one can define an analogue of 
the Frobenius-Schur indicator (see e.g.\ definition 3.10 of \cite{fuSc16} or 
equation (2.19) of \cite{tft1} for a definition). This can be done by picking an
isomorphism $g \iN \HomAB(Y^s,Y^\vee)$ and defining the constant $\nu_Y \iN 
\complex$ via (see also \cite[section 2.4]{tft2} and \cite[section 11]{fuRs11})
  \bea \begin{picture}(120,45)(0,33)
  \put(0,0)   {\includeourbeautifulpicture 68}
  \put(3,75.2)     {\scriptsize$Y^\vee$}
  \put(4,-8.5)     {\scriptsize$Y^s$}
  \put(6,36.8)     {\scriptsize$g$}
  \put(33,35)      {$ =\; \nu_Y^{} $}
  \put(71.5,-8.5)  {\scriptsize$Y^s$}
  \put(91.5,36.8)  {\scriptsize$g$}
  \put(106.5,75.2) {\scriptsize$Y^\vee$}
  \epicture18  \labl{eq:bimod-FS}
Since $Y^\vee$ and $Y^s$ are simple as $A$-$B$-bimodules, the morphism space 
$\HomAB(X^s,X^\vee)$ is one-dimensional, so that $\nu_Y$ exists and is 
independent of the choice of $g$. Also, applying \erf{eq:bimod-FS} twice 
shows immediately that $\nu_Y \iN \{\pm 1\}$.
\\[.2em]
(iii) By writing out the definitions, one can convince oneself that for simple 
Jandl algebras, the notion of `Jandl-Morita equivalence' from definition
\ref{def:Jandl-Morita} is the same as the notion of equivalence of two Jandl 
algebras in \cite[Definition 13]{fuRs11}. To this end one turns $Y \eq M^\vee$,
with $M$ the left $A$-module used in \cite{fuRs11}, into a $B$-$A$-bimodule in 
the obvious way. The isomorphism $g \iN \HomA(M,M^\sigma)$ of \cite{fuRs11} is 
related to the isomorphism $f \iN \HomAB(Y^\vee,Y^s)$ required in definition 
\ref{def:Jandl-Morita} via $f \eq \theta_M \cir g$.

\medskip

Let $A$, $B$ and $Y$ be as in definition \ref{def:Jandl-Morita}. In order to 
show that \CFTA\ and \CFTB\ are equivalent for two simple Jandl algebras 
$A$ and $B$, we essentially repeat the calculation done in section 
\ref{sec:morita}. The only new aspect is that the world sheet $\X$ can
now contain insertions of cross caps. The effect of taking the defect $Y$ past 
a cross cap is illustrated in the following sequence of deformations:
  \begin{eqnarray}\begin{picture}(400,110)(13,0)
  \put(-10,0)   {\Includeourtinynicepicture 69a }
  \put(160,0)   {\Includeourtinynicepicture 69b }
  \put(340,0)   {\Includeourtinynicepicture 69c }
  \put(78,88)   {\scriptsize$\cfta$}
  \put(1,88)    {\scriptsize$\cftb$}
  \put(65,85)   {\scriptsize$Y$}
  \put(119,50)  {$\dsty \overset{(1)}{=}$}
 \put(-10,0){
  \put(190,95)  {\scriptsize$\cftb$}
  \put(265,95)  {\scriptsize$\cfta$}
  \put(165,66)  {\scriptsize$Y$}
  \put(260,4)   {\scriptsize$Y$}
  \put(250,60)  {\cftb \put(-38,4.5){\line(4,0){20}} }
 }
  \put(440,95)  {\scriptsize$\cfta$}
  \put(355,95)  {\scriptsize$\cftb$}
  \put(334,59)  {\scriptsize$Y$}
  \put(297,50)  {$\dsty \overset{(2)}{=}$}
  \put(380,80)  {\scriptsize \begin{turn}{-80}$g$\end{turn}}
  \put(406,82)  {\scriptsize \begin{turn}{-100}$g^{\!-\!1}$\end{turn}}
  \put(429,3)   {\scriptsize$Y$}
  \end{picture}
  \nonumber\\\begin{picture}(400,120)(13,0)
  \put(-15,50)  {$\dsty \overset{(3)}{=} \nu_Y\,\frac{\dim B}{\dim Y}$}
  \put(155,50)  {$\dsty\overset{(4)}{=} \nu_Y\,\frac{\dim B}{\dim Y}$}
  \put(323,50)  {$\dsty\overset{(5)}{=} \nu_Y\,\frac{\dim B}{\dim Y}$}
  \put(50,0){
    \put(0,0)     {\Includeourtinynicepicture 69d }
    \put(96,88)   {\scriptsize$\cfta$}
    \put(-5,90)   {\scriptsize$\cftb$}
    \put(92,3)    {\scriptsize$Y$}
    \put(25,94)   {\scriptsize \begin{turn}{-80}$g$\end{turn}}
    \put(60,77)   {\scriptsize \begin{turn}{-110}$g^{\!-\!1}$\end{turn}}
    }
  \put(50,0){
    \put(170,0)   {\Includeourtinynicepicture 69e }
    \put(261,88)  {\scriptsize$\cfta$}
    \put(185,85)  {\scriptsize$\cftb$}
    \put(224,77)  {\scriptsize \begin{turn}{-110}$g^{\!-\!1}$\end{turn}}
    \put(231,43)  {\scriptsize \begin{turn}{110}$g$\end{turn}}
    \put(255,3)   {\scriptsize$Y$}
    }
  \put(53,0){
    \put(340,0)   {\Includeourtinynicepicture 69f }
    \put(395,45)  {\scriptsize$\cfta$}
    \put(355,85)  {\scriptsize$\cftb$}
    \put(402,91)  {\scriptsize$Y$}
    }
  \end{picture}
  \nonumber\\[-2pt]~\label{pic-ffrs5-69}
  \end{eqnarray}
The dashed line indicates that the local
orientation around the defect is thereby reversed.\,\footnote{~%
  For conventions regarding the labelling of defect lines on unoriented
  surfaces see section 3.8 of \cite{tft2} and section 3.4 of \cite{tft4}.}
In step (2) an isomorphism $g \iN \HomAB(Y^s,Y^\vee)$ is chosen and the 
identity morphism in the form $g^{-1} \cir g$ is inserted. We also indicate 
the half-twists of the corresponding ribbon in the 3dTFT representation, see 
section 3 of \cite{tft2}. In step (3) we first use \erf{eq:morita-move} 
and then \erf{eq:bimod-FS}. In step (4) the remaining section
of the $Y$-defect is dragged through the cross cap, and in step (5) the 
half-twist are removed and $g$ is cancelled against $g^{-1}$.  

Altogether we arrive at
  \be
  \Cf_A(\X) = \gamma^{-\chi(X)}\, (\nu_Y)^{c}\,\Cf_B(\X') \,.
  \labl{eq:unor-morita-eq}
Here $\X$, $\X'$ and $\gamma$ are as in \erf{eq:or-morita-eq}, $c$ is the 
number of cross caps, and $\chi(\X) \eq 2-2h-b-c$ is the Euler character of \X.
Recall that three cross caps can be traded for one cross cap plus one handle, 
so that the total number of cross caps is only defined modulo two. 
Since $\nu_Y^{} \eq \pm 1$, the prefactor in \erf{eq:unor-morita-eq} 
is nonetheless well-defined.

\dt{Remark}
A similar sign factor has been found in a geometric approach to WZW theories
on unoriented surfaces \cite{scsW} based on hermitian bundle gerbes with
additional structure. Such structures -- called Jandl structures in \cite{scsW}
-- actually come in pairs whose monodromies on an unoriented surface with $c$
crosscaps differ by a factor of $(-1)^c$.


\subsection{Action of duality defects on fields}

In the Ising lattice model, order-disorder duality
is at the same time a high-low temperature duality. A similar effect
occurs for defect-induced order-disorder dualities of CFTs.
To exhibit this phenomenon we need to study the behaviour of bulk
fields under the duality. As a preparation, we introduce a map $\phi_Y$ 
which is a generalisation of the basis independent $6j$-symbols 
studied in section 4.1 of \cite{tft3} to the categories of bimodules.
\\[-2em]

\dtl{Definition}{def:phi_Y}
For $Y$ a simple $A$-$B$-bimodule, the map 
$\phi_Y{:}\ \mathcal{S}^l(Y) \Times \mathcal{S}^r(Y) \To \complex^\times$ 
is defined via
  \bea \begin{picture}(240,59)(0,37)
  \put(20,0)     {\Includeourbeautifulpicture 39a }
  \put(150,0)    {\Includeourbeautifulpicture 39b }
  \put(18,-8.8)  {\scriptsize$X_g$}
  \put(35.5,-8.8){\scriptsize$Y$}
  \put(52,-8.8)  {\scriptsize$X_h$}
  \put(27,65)    {\scriptsize$\alpha$}
  \put(30,38)    {\scriptsize$\beta$}
  \put(44,55)    {\scriptsize$Y$}
  \put(37,97)    {\scriptsize$Y$}
  \put(148,-8.8) {\scriptsize$X_g$}
  \put(167.5,-8.8){\scriptsize$Y$}
  \put(183,-8.8) {\scriptsize$X_h$}
  \put(175,39)   {\scriptsize$\alpha$}
  \put(178,65)   {\scriptsize$\beta$}
  \put(160,55)   {\scriptsize$Y$}
  \put(169,97)   {\scriptsize$Y$}
  \put(80,38)    {$=~~\phi_Y(g,h)$}
  \epicture20 
  \labl{eq:phi_Y}
for $g \iN \mathcal{S}^l(Y)$ and $h \iN \mathcal{S}^r(Y)$. 

\medskip

The numbers $\phi_Y(g,h)$ do not depend on the choice of nonzero elements 
$\alpha\iN\Hom(X_g \otA Y,Y)$ and $\beta\iN\Hom(Y \otB X_h,Y)$: these two 
morphism spaces are one-dimensional, and hence choosing different 
elements changes both sides of \erf{eq:phi_Y} by the same factor.
The map $\phi_Y$ tells us how to commute two group-like 
defects attached from opposite sides to a simple $A$-$B$-defect:
  \bea \begin{picture}(420,141)(0,0)
  \put(0,0)      {\Includeourbeautifulpicture 41a }
  \put(185,0)    {\Includeourbeautifulpicture 41b }
  \put(70,20)    {\scriptsize$\cfta$}
  \put(70,120)   {\scriptsize$\cfta$}
  \put(10,20)    {\scriptsize$\cftb$}
  \put(10,120)   {\scriptsize$\cftb$}
  \put(38,4)     {\scriptsize$Y$}
  \put(49.5,64)  {\scriptsize$Y$}
  \put(38,114)   {\scriptsize$Y$}
  \put(20,65)    {\scriptsize$X_g$}
  \put(70,103)   {\scriptsize$X_h$}
  \put(250,20)   {\scriptsize$\cfta$}
  \put(250,120)  {\scriptsize$\cfta$}
  \put(190,20)   {\scriptsize$\cftb$}
  \put(190,120)  {\scriptsize$\cftb$}
  \put(236,4)    {\scriptsize$Y$}
  \put(221,64)   {\scriptsize$Y$}
  \put(236,114)  {\scriptsize$Y$}
  \put(256,64)   {\scriptsize$X_h$}
  \put(205,103)  {\scriptsize$X_g$}
  \put(111,62)   {$=~~\phi_Y(g,h)$}
  \put(299,3)    {where $\quad g\iN\mathcal{S}^l(Y)\,,~ h\iN\mathcal{S}^r(Y)~.$}
  \epicture-7 
  \labl{eq:phiY-def-def}
The following properties of the map $\phi_Y$ will be important:
\\[-2.2em]

\dtl{Proposition}{prop:phi-sep}
Let $Y$ be a simple $A$-$B$-bimodule.
\\[.3em]
(i)~\,The map $\phi_Y{:}\ \mathcal{S}^l(Y) \Times \mathcal{S}^r(Y) \To 
      \complex^\times$ is a bihomomorphism.
\\[.3em]
(ii)~Let $Y$ be in addition a duality bimodule. Then the bihomomorphism $\phi_Y$ 
is non-degenerate in the first argument, i.e.\ if for some $g,g'\iN\mathcal{S}^l
(Y)$ one has $\phi_Y(g,h) \eq \phi_Y(g',h)$ for all $h\iN\mathcal{S}^r(Y)$, then 
$g \eq g'$. In particular, $|\mathcal{S}^r(Y)| \,{\ge}\, |\mathcal{S}^l(Y)|$.

\medskip\noindent
Proof:
\\[.2em]
(i)~The proof proceeds similar to the one of proposition 4.2 of \cite{tft3}. 
Choose basis vectors $\alpha_{g,h} \iN \HomAA(X_g \otA X_h,X_{gh})$ and 
$\beta_g \iN \HomAB(X_g \otA Y,Y)$. For any $g_1,g_2 \iN \mathcal{S}^l(Y)$ there
are nonzero constants $\psi_Y(g_1,g_2)$ such that
(we implicitly use the isomorphisms \erf{eq:HomP-HomTens})
  \bea \begin{picture}(280,65)(0,43)
  \put(20,0)     {\Includeourbeautifulpicture 44a }
  \put(170,0)    {\Includeourbeautifulpicture 44b }
  \put(18.5,-8)  {\scriptsize$X_{g_1}$}
  \put(55,-8)    {\scriptsize$Y$}
  \put(35,-8)    {\scriptsize$X_{g_2}$}
  \put(60.5,69)  {\scriptsize$\beta_{g_1}$}
  \put(60.5,38)  {\scriptsize$\beta_{g_2}$}
  \put(55,107)   {\scriptsize$Y$}
  \put(169,-8)   {\scriptsize$X_{g_1}$}
  \put(218.5,-8) {\scriptsize$Y$}
  \put(193,-8)   {\scriptsize$X_{g_2}$}
  \put(192,38)   {\scriptsize$\alpha_{g_1,g_2}$}
  \put(225,69)   {\scriptsize$\beta_{g_1,g_2}$}
  \put(218.5,107){\scriptsize$Y$}
  \put(185,63)   {\scriptsize$X_{g_1,g_2}$}
  \put(87,47)    {$=~~\psi_Y(g_1,g_2)$}
  \epicture28 
  \labl{eq:psi-move}
Then on the one hand one has (we omit the labels for the couplings)
  \bea \begin{picture}(330,95)(0,41)
  \put(20,0)     {\Includeourbeautifulpicture 45a }
  \put(220,0)    {\Includeourbeautifulpicture 45b }
  \put(18,-8)    {\scriptsize$X_{g_1}$}
  \put(52,-8)    {\scriptsize$Y$}
  \put(35,-8)    {\scriptsize$X_{g_2}$}
  \put(65,-8)    {\scriptsize$X_{h}$}
  \put(54,134)   {\scriptsize$Y$}
  \put(218,-8)   {\scriptsize$X_{g_1}$}
  \put(255,-8)   {\scriptsize$Y$}
  \put(236,-8)   {\scriptsize$X_{g_2}$}
  \put(270,-8)   {\scriptsize$X_{h}$}
  \put(255,134)  {\scriptsize$Y$}
  \put(94,58)    {$=~~\phi_Y(g_1,h)\, \phi_Y(g_2,h)$}
  \epicture24 
  \labl{eq:phi-hom-proof1}
and on the other hand
  \bea \begin{picture}(400,96)(16,42)
  \put(0,0)      {\Includeourbeautifulpicture 46a }
  \put(148,0)    {\Includeourbeautifulpicture 46b }
  \put(356,0)    {\Includeourbeautifulpicture 46c }
  \put(-1,-8)    {\scriptsize$X_{g_1}$}
  \put(33,-8)    {\scriptsize$Y$}
  \put(16,-8)    {\scriptsize$X_{g_2}$}
  \put(46,-8)    {\scriptsize$X_{h}$}
  \put(34,134)   {\scriptsize$Y$}
  \put(147,-8)   {\scriptsize$X_{g_1}$}
  \put(195,-8)   {\scriptsize$Y$}
  \put(171,-8)   {\scriptsize$X_{g_2}$}
  \put(209,-8)   {\scriptsize$X_{h}$}
  \put(197,134)  {\scriptsize$Y$}
  \put(355,-8)   {\scriptsize$X_{g_1}$}
  \put(400,-8)   {\scriptsize$Y$}
  \put(379,-8)   {\scriptsize$X_{g_2}$}
  \put(416,-8)   {\scriptsize$X_{h}$}
  \put(400,134)  {\scriptsize$Y$}
  \put(160,79)   {\scriptsize$X_{g_1,g_2}$}
  \put(370,58)   {\scriptsize$X_{g_1,g_2}$}
  \put(70,61)    {$=~~\psi_Y(g_1,g_2)$}
  \put(227,61)   {$=~~\phi_Y(g_1g_2,h)\,\psi_Y(g_1,g_2)$}
  \epicture28 
  \labl{pic46}
Applying relation \erf{eq:psi-move} to the right hand side of \erf{pic46}
removes the factor $\psi_Y(g_1,g_2)$ again, and thus comparison with 
\erf{eq:phi-hom-proof1} yields
$\phi_Y(g_1,h)\, \phi_Y(g_2,h) \eq \phi_Y(g_1 g_2,h)$. The 
unit property $\phi_Y(e,h) \eq 1$ is immediate. The homomorphism
property in the second argument can be checked analogously.
\\[.2em]
(ii)~Consider the equalities
  \bea \begin{picture}(360,45)(0,44)
  \put(20,0)     {\Includeourbeautifulpicture 48a }
  \put(160,0)    {\Includeourbeautifulpicture 48b }
  \put(44,-8)    {\scriptsize$X_{g}$}
  \put(28,90)    {\scriptsize$A$}
  \put(81,33)    {\scriptsize$Y$}
  \put(34,42)    {\scriptsize$A$}
  \put(174,36)   {\scriptsize$A$}
  \put(168,90)   {\scriptsize$A$}
  \put(221,52)   {\scriptsize$Y$}
  \put(168,-8)   {\scriptsize$A$}
  \put(105,44)   {$\dsty =~~\delta_{g,e}$}
  \put(240,44)   {$\dsty =~~\delta_{g,e}\,\displaystyle\frac{\dim(Y)}{\dim(A)}\,\id_A~.$}
  \epicture24 
  \labl{eq:phi-sep-proof1}
The first equality holds because $\HomAA(X_g,A)$ is zero unless $g\eq e$ 
(i.e., unless $X_g \Cong A$), and the second follows by
lemma \ref{lem:X-insert} below. Next consider the following series of equalities:
  \begin{eqnarray}\begin{picture}(400,155)(13,0)
  \put(60,0)    {\Includeourbeautifulpicture 49a }
  \put(155,0)   {\Includeourbeautifulpicture 49b }
  \put(350,0)   {\Includeourbeautifulpicture 49c }
  \put(59,-8)   {\scriptsize$A$}
  \put(82,148)  {\scriptsize$Y$}
  \put(82,-8)   {\scriptsize$Y$}
  \put(152,-8)  {\scriptsize$X_g$}
  \put(177,43)  {\scriptsize$A$}
  \put(162,97)  {\scriptsize$Y$}
  \put(199,148) {\scriptsize$Y$}
  \put(199,-8)  {\scriptsize$Y$}
  \put(372,35)  {\scriptsize$A$}
  \put(348,-8)  {\scriptsize$X_g$}
  \put(396,-8)  {\scriptsize$Y$}
  \put(402,80)  {\scriptsize$X_h$}
  \put(396,148) {\scriptsize$Y$}
  \put(370,80)  {\scriptsize$Y$}
  \put(346,94)  {\scriptsize$Y$}
  \put(-10,63)  {$\delta_{g,e}\,\displaystyle\frac{\dim(Y)}{\dim(A)}$}
  \put(113,63)  {$\overset{(1)}{=}$}
  \put(230,63)  {$\overset{(2)}{=}~ \displaystyle
                  \sum_{h \in \mathcal{S}^r(Y)} \frac{\dim(X_h)}{\dim(Y)}$}
  \end{picture}
  \nonumber\\\begin{picture}(400,168)(13,0)
  \put(135,0)   {\Includeourbeautifulpicture 49d }
  \put(367,0)   {\Includeourbeautifulpicture 49e }
  \put(130,-8)  {\scriptsize$X_g$}
  \put(185,66)  {\scriptsize$X_h$}
  \put(157.5,-8){\scriptsize$Y$}
  \put(159,148) {\scriptsize$Y$}
  \put(363,-8)  {\scriptsize$X_g$}
  \put(387,-8)  {\scriptsize$Y$}
  \put(388.5,148) {\scriptsize$Y$}
  \put(30,63)   {$\overset{(3)}{=}~ \displaystyle
                  \frac{\dim(B)}{\dim(Y)} \sum_{h \in \mathcal{S}^r(Y)}$}
  \put(210,63)  {$\overset{(4)}{=}~ \displaystyle \frac{\dim(B)}{\dim(Y)}
                  \sum_{h \in \mathcal{S}^r(Y)}\,\phi_Y(g,h)^{-1}$}
  \end{picture}
  \nonumber\\[-11pt]~\label{eq:phi-sep-proof2} 
  \end{eqnarray}
Equality (1) is obtained by applying the identity that results from composing
both sides of \erf{eq:phi-sep-proof1} with the counit.
Step (2) amounts to \erf{eq:2-opp-sum}.
In step (3) the ribbon graph is deformed and the $A$-ribbons are removed 
(which is possible owing to the properties of $A$ and the fact that the various 
intertwiners commute with the action of $A$); also, it is used that 
$\dim(X_h) \eq \dim(B)$ (see proposition \ref{prop:eps(g)=1}). Finally, (4) 
uses the definition \erf{eq:phi_Y} of $\phi_Y(g,h)$ and that the basis elements
in $\HomAB(Y \otB X_h,Y)$ and $\HomAB(Y,Y \otB X_h)$ are dual to each other. 
\\[2pt]
Altogether, the result of \erf{eq:phi-sep-proof2} shows that
  \be
  \sum_{h \in \mathcal{S}^r(Y)} \phi_Y(g,h^{-1})
  = \frac{\delta_{g,e} \dim(Y)^2}{\dim(A)\dim(B)}
  = |\mathcal{S}^r(Y)|\, \delta_{g,e} \,,
  \labl{eq:phi-sep-proof3}
where the last equality holds by \erf{YY=ABS}.  
Now $\phi_Y(g,h) \eq \phi_Y(g',h)$ for all $h \in \mathcal{S}^r(Y)$ implies that
$\sum_{h \in \mathcal{S}^r(Y)} \phi_Y(g (g')^{-1},h) \eq |\mathcal{S}^r(Y)|$, 
which by \erf{eq:phi-sep-proof3} is the case only if $g\eq g'$.
\qed

\dtl{Remark}{rem:dont-know}
In the remainder of this section we will assume that both the simple 
$B$-$A$-defect $Y$ and the simple $A$-$B$-defect $Y^\vee$ are duality defects. 
This allows us to make stronger statements, but it is not the generic situation. 
A counter example is provided by a phase-boundary between the tetracritical 
Ising and the critical three-states Potts model, see section 
\ref{sec:IsingPotts} for details. In fact, if both $Y$ and $Y^\vee$ are simple 
duality defects, then by theorem \ref{thm:stab-ab} below their stabilisers are 
abelian. So if a simple duality defect $Y$ has a nonabelian stabiliser, then
$Y^\vee$ cannot be a duality defect. This is precisely the situation
in the example treated in section \ref{sec:IsingPotts}.

\dtl{Theorem}{thm:stab-ab}
Let $Y$ be a simple $B$-$A$-bimodule such that both $Y$ and $Y^\vee$ are 
duality bimodules. Then $\mathcal{S}^l(Y) \,{\subseteq}\, \mathcal{G}_B$ and 
$\mathcal{S}^r(Y) \,{\subseteq}\, \mathcal{G}_A$ are abelian and are
isomorphic as groups.

\medskip\noindent
Proof:\\
Since $Y$ and $Y^\vee$ are duality bimodules, by proposition 
\ref{prop:phi-sep}\,(ii) we have $|\mathcal{S}^r(Y)|\,{\ge}\,|\mathcal{S}^l(Y)|$
and $|\mathcal{S}^r(Y^\vee)| \,{\ge}\, |\mathcal{S}^l(Y^\vee)|$.
Using $\mathcal{S}^r(Y^\vee) \eq \mathcal{S}^l(Y)$ and
$\mathcal{S}^l(Y^\vee) \eq \mathcal{S}^r(Y)$, this implies
$|\mathcal{S}^r(Y)| \eq |\mathcal{S}^l(Y)|$. Denote by $G^*$ the character 
group of a group $G$. Since $Y$ is a duality bimodule, from proposition 
\ref{prop:phi-sep}\,(ii) it follows that the map 
$\varphi_Y{:}\ g \,{\mapsto}\, \phi_Y(g,\cdot)$ is an injective
group homomorphism from $\mathcal{S}^l(Y)$ to $\mathcal{S}^r(Y)^*$.
Because of $|\mathcal{S}^r(Y)| \eq |\mathcal{S}^l(Y)|$ this shows that there 
are at least $|\mathcal{S}^r(Y)|$ different one-dimensional representations
of $\mathcal{S}^r(Y)$. Since the number of inequivalent representations of a 
finite group is equal to the number of its conjugacy classes, this means that 
every conjugacy class of $\mathcal{S}^r(Y)$ must consist of a single element, 
i.e.\ $\mathcal{S}^r(Y)$ is abelian.  But a finite abelian group is isomorphic 
to its character group, and combining this isomorphism with $\varphi_Y$ we 
obtain an isomorphism of groups from $\mathcal{S}^l(Y)$ to $\mathcal{S}^r(Y)$.
Since $\mathcal{S}^r(Y)$ is abelian, so is $\mathcal{S}^l(Y)$.
\qed

\medskip

Let $Y$ be a simple $B$-$A$-bimodule such that $Y$ and $Y^\vee$ are duality 
bimodules. We now investigate what happens to bulk fields when we inflate the 
duality defect $Y$ in a world sheet. According to \erf{eq:defect-repn} the bulk 
fields of \CFTA\ carry a representation $\phi \,{\mapsto}\, D_g(\phi)$ of the 
group $\mathcal{G}_\AA$ (and the bulk fields of \CFTB\ a representation of 
$\mathcal{G}_B$). In particular, the bulk fields of \CFTA\ furnish a 
representation of the stabiliser $\mathcal{S}^r(Y)$. Let us define a map 
$F^r_{Y,UV}$ which assigns to an element $g \iN \mathcal{S}^l(Y)$ the subspace 
of bulk fields with chiral/anti-chiral labels $U$, $V$ in representation 
$\phi_Y(g,\cdot)$ of $\mathcal{S}^r(Y)$:
  \be\begin{array}{lrl}
  F^r_{Y,UV} : & \mathcal{S}^l(Y) & \!\!\longrightarrow\
    {\rm set~of~subspaces~of~}\HomAA(U \otiP A \otim V , A) 
  \\{}\\[-.9em]
  & g & \!\!\longmapsto\
  \big\{ \, v \,\big|\, D_h(v) \eq \phi_Y(g,h) v {\rm~for~all~} 
  h \iN \mathcal{S}^r(Y) \, \big\} \,.
  \eear\ee
By theorem \ref{thm:stab-ab} the stabilisers are abelian, and their
irreducible representations are one-di\-men\-sional. Together with
proposition \ref{prop:phi-sep} it follows that each irreducible
representation of $\mathcal{S}^r(Y)$ is of the form 
$h \,{\mapsto}\, \phi_Y(g,h)$ for some $g \iN \mathcal{S}^l(Y)$. 
Thus we get the direct sum decomposition
  \be
  \HomAA(U \otiP A \otim V , A)
  \cong \bigoplus_{g \in \mathcal{S}^l(Y)} F^r_{Y,UV}(g) \,.
  \ee
Analogously we set
  \be\begin{array}{lrl}
  F^l_{Y,UV} : & \mathcal{S}^r(Y) & \!\!\longrightarrow\
    {\rm set~of~subspaces~of~}\HomBB(U \otiP B \otim V , B)
  \\{}\\[-.8em]
  & h & \!\!\longmapsto\
  \big\{ \, v \,\big|\, D_g(v) \eq \phi_Y(g,h) v {\rm~for~all~} 
  g \iN \mathcal{S}^l(Y) \, \big\} 
  \eear\ee
for bulk fields of \CFTB. The following result shows that in an appropriate 
basis for the bulk fields, taking a duality defect $Y$ past a bulk field results
in a defect field sitting at the end of a single group-like defect, rather than 
a superposition thereof as one might expect from the relation
$Y^\vee \OtB Y \Cong \bigoplus_{h \in \mathcal{S}^r(Y)} X_h$.
\\[-2.2em]

\dtl{Proposition}{prop:bulk-dual}
Let $Y$ be a simple $B$-$A$-defect such that both $Y$ and $Y^\vee$ are 
duality defects, and let $g \iN \mathcal{S}^l(Y)$
and $h \iN \mathcal{S}^r(Y)$. For $\phi \iN F^r_{Y,UV}(g)$ a bulk field 
of \CFTA\ and $\phi' \iN F^l_{Y,UV}(h)$ a bulk field of \CFTB\ we have
  \begin{eqnarray}&&\begin{picture}(320,85)(0,0)
  \put(-5,0)     {\Includeourbeautifulpicture 40a }
  \put(140,0)    {\Includeourbeautifulpicture 40b }
  \put(70,40)    {$\dsty = ~ \frac{\dim(B)}{\dim(Y)}$}
  \put(47,1)     {\scriptsize$Y$}
  \put(14,40)    {\scriptsize$\phi$}
  \put(55,70)    {\scriptsize$\cfta$}
  \put(17,70)    {\scriptsize$\cftb$}
 \put(25,0){
  \put(133,10)   {\scriptsize$Y$}
  \put(137,30)   {\scriptsize$\phi$}
  \put(140,70)   {\scriptsize$\cftb$}
  \put(170,40)   {\scriptsize$X_{g^{-1}}$}
  \put(206,1)    {\scriptsize$Y$}
  \put(205,65)   {\scriptsize$Y$}
 }
  \end{picture} \nonumber\\
                {\rm and} \nonumber\\
  &&\begin{picture}(200,95)(0,0)
  \put(25,0)     {\Includeourbeautifulpicture 40c }
  \put(183,0)    {\Includeourbeautifulpicture 40d }
 \put(-231,0){
  \put(262,1)    {\scriptsize$Y$}
  \put(283,40)   {\scriptsize$\phi'$}
  \put(270,70)   {\scriptsize$\cfta$}
  \put(231,70)   {\scriptsize$\cftb$}
 }
 \put(-202,0){
  \put(389,1)    {\scriptsize$Y$}
  \put(393,64)   {\scriptsize$Y$}
  \put(465,12)   {\scriptsize$Y$}
  \put(448,22)   {\scriptsize$\phi'$}
  \put(404,70)   {\scriptsize$\cfta$}
  \put(362,70)   {\scriptsize$\cftb$}
  \put(405,40)   {\scriptsize$X_{h^{-1}}$}
 } 
  \put(100,40)   {$\dsty =~\frac{\dim(A)}{\dim(Y)}$}
  \end{picture}
  \nonumber\\[-13pt]~ \label{eq:bulk-dual} 
  \end{eqnarray}

\noindent 
Proof:\\
We establish the first equality in \erf{eq:bulk-dual}; the
second equality can be seen analogously.
Let $u \iN \mathcal{S}^l(Y)$ and $\phi \iN F^r_{Y,UV}(u)$. Note that
  \begin{eqnarray}\begin{picture}(420,85)(20,0)
  \put(0,0)      {\Includeourbeautifulpicture 42a }
  \put(179,0)    {\Includeourbeautifulpicture 42b }
  \put(116,48)   {\scriptsize$X_g$}
  \put(299,48)   {\scriptsize$X_g$}
  \put(10,67)    {\scriptsize$Y$}
  \put(223,77)   {\scriptsize$Y$}
  \put(215,50)   {\scriptsize$h$}
  \put(40,33)    {\scriptsize$\phi$}
  \put(223,33)   {\scriptsize$\phi$}
  \put(10,40)    {\scriptsize$\cfta$}
  \put(76,65)    {\scriptsize$\cftb$}
  \put(205,5)    {\scriptsize$\cfta$}
  \put(265,65)   {\scriptsize$\cftb$}
  \put(147,35)   {$=$}
  \end{picture}
  \nonumber\\ \begin{picture}(420,90)(20,0)
  \put(83,0)     {\Includeourbeautifulpicture 42c }
  \put(353,3)    {\Includeourbeautifulpicture 42a }
  \put(202,50)   {\scriptsize$X_g$}
  \put(469,50)   {\scriptsize$X_g$}
  \put(92,3)     {\scriptsize$Y$}
  \put(113,15)   {\scriptsize$h$}
  \put(99,76)    {\scriptsize$Y$}
  \put(365,70)   {\scriptsize$Y$}
  \put(127,40)   {\scriptsize$\phi$}
  \put(392,35)   {\scriptsize$\phi$}
  \put(363,40)   {\scriptsize$\cfta$}
  \put(445,66)   {\scriptsize$\cftb$}
  \put(5,35)     {$=~ \phi_Y(g,h^{-1})$}
  \put(223,35)   {$=~ \phi_Y(g,h^{-1})\,\phi_Y(u,h^{-1})$}
  \end{picture}
  \nonumber\\[-20pt]~ \label{eq:phi-commute-g} 
  \end{eqnarray}
where we also used that by remark \ref{rem:M35}\,(iii), $\eps(g)\eq1$. In order 
for \erf{eq:phi-commute-g} to be nonzero, we thus need $\phi_Y(g u,h) \eq 1$ 
for all $h \iN \mathcal{S}^l$, which by proposition \ref{prop:phi-sep}\,(ii) 
implies $g \eq u^{-1}$. But then, using also \erf{eq:2-opp-sum}, we have
  \bea \begin{picture}(360,44)(0,40)
  \put(20,0)     {\Includeourbeautifulpicture 43a }
  \put(220,0)    {\Includeourbeautifulpicture 43b }
  \put(272,24)   {\scriptsize$X_g$}
  \put(73,72)    {\scriptsize$Y$}
  \put(308,78)   {\scriptsize$Y$}
  \put(311,2)    {\scriptsize$Y$}
  \put(212,30)   {\scriptsize$Y$}
  \put(40,38)    {\scriptsize$\phi$}
  \put(242,28)   {\scriptsize$\phi$}
  \put(85,67)    {\scriptsize$\cfta$}
  \put(15,67)    {\scriptsize$\cftb$}
  \put(325,67)   {\scriptsize$\cfta$}
  \put(235,67)   {\scriptsize$\cftb$}
  \put(232,-7)   {\cfta \put(0,3){\line(0,1){29}}\put(0.5,3){\line(0,1){29}} }
  \put(111,38)   {$=\displaystyle\sum_{g\in\mathcal{S}^l(Y)}
                   \frac{\dim(B)}{\dim(Y)}$}
  \epicture19 \labl{pic-ffrs5-43}
Moreover, in the sum on the right hand side, only the term
$g\eq u^{-1}$ can be nonzero.
\qed

\dt{Remark}
In \cite{ruel'5}, based on previous results in \cite{ruve,ruel'3}, order-disorder
dualities were investigated via the symmetry properties of boundary states.
The boundary states were defined to also include sectors twisted by a symmetry 
of the CFT, and dualities can be found by checking if one can find an invertible
transformation on the boundary states that exchanges some periodic sectors with 
twisted ones. Using defect lines we can recover this relation by considering
one-point functions of a bulk field on a disk. These one-point functions give 
the coefficients of the Ishibashi states in the expansion of a boundary state. 
Consider a disk in phase \CFTA\ with boundary condition $M$, and let $Y$ be a 
$B$-$A$-duality defect such that also $Y^\vee$ is a duality defect. Then the 
manipulations
  \begin{eqnarray}\begin{picture}(420,131)(20,0)
  \put(0,0)      {\Includeourbeautifulpicture 62a }
  \put(230,0)    {\Includeourbeautifulpicture 62b }
  \put(116,85)   {\scriptsize$M$}
  \put(347,85)   {\scriptsize$M$}
  \put(65,62)    {\scriptsize$\phi$}
  \put(294,67)   {\scriptsize$\phi$}
  \put(286,23)   {\scriptsize$Y$}
  \put(25,75)    {\scriptsize$\cfta$}
  \put(255,75)   {\scriptsize$\cfta$}
  \put(158,60)   {$=~\displaystyle\frac{\dim(A)}{\dim(Y)}$}
  \end{picture}
  \nonumber\\ \begin{picture}(420,137)(20,0)
  \put(120,0)    {\Includeourbeautifulpicture 62c }
  \put(307,0)    {\Includeourbeautifulpicture 62d }
  \put(150,79)   {\scriptsize$\cftb$}
  \put(335,75)   {\scriptsize$\cftb$}
  \put(232,95)   {\scriptsize$M$}
  \put(382,95)   {\scriptsize$Y\OtA M$}
  \put(185,63)   {\scriptsize$\phi$}
  \put(360,64)   {\scriptsize$\theta$}
  \put(187,102)  {\scriptsize$Y$}
  \put(379,70)   {\scriptsize$X_g$}
  \put(50,60)    {$=~\displaystyle\frac{\dim(A)}{\dim(Y)}$}
  \put(277,60)   {$=$}
  \end{picture}
  \nonumber\\[-16pt]~ \label{eq:ishi-sym} 
  \end{eqnarray}
show that the coefficient of the periodic Ishibashi state belonging to the bulk 
field labelled $\phi$ in the expansion of the boundary state for $M$ is the 
same as the coefficient of the $g$-twisted Ishibashi state belonging to the 
disorder field labelled $\theta$ in the boundary state for $Y \otA M$ (in an 
appropriate normalisation of twisted Ishibashi states and disorder fields). We 
also used proposition \ref{prop:bulk-dual} to choose $\phi$ such that only a 
single group-like defect contributes on the \rhs\ of \erf{eq:ishi-sym}.
Note that the above argument only shows that the relation \erf{eq:ishi-sym} 
is a necessary condition for the existence of a two-sided duality.


\subsection{High-low temperature duality}\label{sec:highlow}

Consider a bulk field labelled $\phi$ of \CFTA\ that is invariant under the 
action of $\mathcal{S}^r(Y)$, i.e.\ $\phi\iN F^r_{Y,UV}(e)$. After acting with
the defect $Y$, it becomes the bulk field $\tfrac{\dim(B)}{\dim(Y)}\, D_Y(\phi)$
(in the notation \erf{eq:D_nu-def}) of \CFTB. Since $\mathcal{S}^l(Y)$ is the 
left-stabiliser of $Y$, the field $D_Y(\phi)$, in turn, lies in $F^l_{Y,UV}(e)$. 
According to \erf{YY=ABS}, inflating the dual defect $Y^\vee$ 
in the world sheet thus takes $\tfrac{\dim(B)}{\dim(Y)}\,D_Y(\phi)$ to 
  \be
  \frac{\dim(A)\dim(B)}{\dim(Y)^2}\, D_{Y^\vee} \cir D_Y(\phi) = 
  \frac{1}{|\mathcal{S}^r(Y)|}
  \sum_{h \in \mathcal{S}^r(Y)} D_h(\phi) = \phi \,.
  \ee
In fact, the map $\varphi \,{\mapsto}\, |\mathcal{S}^r(Y)|^{-1}
\sum_{h \in \mathcal{S}^r(Y)} D_h(\varphi)$ is a projector to the
subspace of bulk fields invariant under $\mathcal{S}^r(Y)$.

The subspaces $F^r_{Y,UV}(e)$ of the multiplicity spaces of bulk fields
are of particular interest, because they are related to the high/low temperature
dualities which turn into the order-disorder duality induced by $Y$ at
the critical point. To see that, let us study the situation 
that \CFTA\ is perturbed by the field $\phi \iN F^r_{Y,UV}(e)$. 
Applying the duality induced by $Y$ relates an order correlator of the
perturbed \CFTA\ to a disorder correlator in \CFTB\ perturbed by 
$\phi' \,{:=}\, \tfrac{\dim(B)}{\dim(Y)}\, D_Y(\phi)$. Schematically,
  \be
  \Big\langle ({\rm fields~in~}\CFTA)\,
  \eE^{ \lambda \int\!\phi(z) \,{\rm d}^2z } \Big\rangle
  \,=\, \Big\langle ({\rm dual~fields~in~}\CFTB)\, \eE^{ \lambda \int\!\phi'(z)
  \,{\rm d}^2z } \Big\rangle \,.
  \labl{eq:high-low}
The precise form of this relation is obtained by expanding the left hand side
of \erf{eq:high-low} in a perturbation series in $\lambda$ and 
applying the defect $Y$ at each order. In \cite{ffrs3}
it was noted that the high-low temperature duality of the Ising model 
can be found in this way upon choosing $\phi$ to be the energy field $\eps$.


\sect{TFT formulation of defect correlators}\label{sec:tft-form}

To prove the rules for manipulating defects laid out in section
\ref{sec:calc-def}, and to establish the non-degeneracy of the defect 
two-point function on the sphere which is instrumental for our purposes, we 
employ the formulation of RCFT correlators in terms of three-dimensional 
topological field theory (3-d TFT) that was developed in \cite{fffs2} and 
\citeOtoV.  In this approach the chiral CFT is realised by the boundary degrees 
of freedom of an appropriate 3-d TFT \cite{witt27,frki}. One can then use the 
geometry of a three-manifold together with a certain network of Wilson lines to
combine left and right moving chiral degrees of freedom in the correct manner.


\subsection{TFT derivation of the rules of section 
            \ref{sec:calc-def}}\label{sec:calc-tft}

A 3-d TFT can be constructed from any modular tensor category \C\
\cite{retu,TUra}. The modular tensor category relevant for the application
to CFT is $\C \eq \Rep(\mathcal{V})$ (or more precisely, an equivalent strict 
ribbon category), but as already mentioned in remark 
\ref{rem:modular}, for the calculations in this paper it is irrelevant 
whether \C\ can be realised as the representation category of a vertex algebra
or not. Given an oriented three-manifold $\M$ with embedded ribbon
graph, the 3-d TFT assigns to the boundary $\partial \M$ a vector space
$\mathcal{H}(\partial \M)$ and to $\M$ itself a vector $Z(\M)$ in 
$\mathcal{H}(\partial \M)$.  For references and more details on our conventions 
regarding the 3-d TFT constructed from $\calc$ we refer to 
\cite[sect.\,2]{tft1} and \cite[sect.\,3.1]{tft4}.

To obtain the correlator for an oriented world sheet $\X$, one considers 
the {\em connecting manifold\/} $\MX$, defined as
  \be
  \MX =: \X \times [-1,1] /{\sim} \,,
  \quad \text{where}~(x,t) \,{\sim}\, (x,-t)~~\text{for all}~x \iN \partial\X
  \,,\,t\iN[-1,1]~.
  \ee
This amounts to a `fattening' of the world sheet. Note that
$\iota {:}\ x \,{\mapsto}\, (x,0)$ gives an embedding of $\X$ into $\MX$.
The relevant ribbon graph -- or framed Wilson graph -- 
in $\MX$ is obtained by choosing a dual triangulation\,\footnote{~%
  That is, a covering of $\X$ with a 2-complex
  whose vertices are three-valent and whose faces can be arbitrary polygons.}
with directed edges on $\X$ and inserting ribbons in $\MX$ along the images of 
these edges under the embedding $\iota$. A ribbon has an orientation as a 
surface and a direction, and it is labelled by an object of \C. The 
ribbons are to be embedded in $\iota(\X)$ such that their orientation is 
opposite to that of $\iota(\X)$ and their direction is opposite to that of 
the edge of the dual triangulation\,\footnote{~%
  That orientation and direction are chosen opposite to one another
  is just a convention (see section 3.1 of \cite{tft4} for the reasoning
  behind it), and does not have any deeper meaning.}.
If the edge of the dual triangulation lies in a region of the world sheet
in phase $\CFTA$, then the corresponding ribbon is labelled by the object $A$ 
of \C.

Close to defect lines, world sheet boundaries and field insertions, special
pieces of ribbon graph have to be inserted. Specifically, close to a 
defect labelled by an $A$-$B$-bimodule $Y$, the ribbon graph looks like
  \bea \begin{picture}(420,198)(9,0)
  \put(0,40)    {\Includeourtinynicepicture 74a }
  \put(0,40){
     \setlength{\unitlength}{.90pt}\put(-7,-19){
     \put( 33,102)    {\cfta}
     \put( 93, 49)    {\cftb}
     \put( 68, 86) {\scriptsize$ Y $}
     }\setlength{\unitlength}{1pt}}
  \put(180,0)   {\Includeourtinynicepicture 74b }
  \put(180,0){
     \setlength{\unitlength}{.90pt}\put(-6,-4){
     \put(176, 23) {\tiny$ 1 $}
     \put(165, 34) {\tiny$ 2 $}
     \put(155, 46) {\tiny$ 3 $}
     \put(246,144) {\tiny$ 1 $}
     \put(237,153) {\tiny$ 2 $}
     \put( 56,103) {\scriptsize \begin{turn}{-40}$\Delta$\end{turn}}
     \put(138,114) {\scriptsize$ \rho $}
     \put(183,143) {\scriptsize$ \rho $}
     \put( 75,125) {\scriptsize$ A $}
     \put( 41, 85) {\scriptsize$ A $}
     \put( 99,110) {\scriptsize$ A $}
     \put(205,120) {\scriptsize$ B $}
     \put(156,130) {\scriptsize$ Y $}
     \put(138, 97) {\scriptsize$ Y $}
     }\setlength{\unitlength}{1pt}}
  \put(130,70)  {$ \longmapsto $}
  \epicture-9 
  \labl{eq:defect-cobord}
In this picture the orientations and a possible choice of dual triangulation are
also shown. All ribbons are showing their `black side', i.e.\ their surface
orientation is opposite to the one indicated on the embedded world sheet.
In a neighbourhood of a world sheet boundary we have
  \bea \begin{picture}(420,172)(5,0)
  \put(0,15)    {\Includeourtinynicepicture 75a }
  \put(0,15){
     \setlength{\unitlength}{.90pt}\put(-16,-12){
     \put( 74,140)    {\cfta}
     \put(115,123) {\scriptsize$ M $}
     }\setlength{\unitlength}{1pt}}
  \put(180,0)   {\Includeourtinynicepicture 75b }
  \put(180,0){
     \setlength{\unitlength}{.90pt}\put(-11,-18){
     \put( 92, 77) {\tiny$ 1 $}
     \put( 82, 86) {\tiny$ 2 $}
     \put( 85,109) {\scriptsize \begin{turn}{-40}$\Delta$\end{turn}}
     \put(163,100) {\scriptsize \begin{turn}{160}$\rho$\end{turn}}
     \put( 58,103) {\scriptsize$ A $}
     \put(125,103) {\scriptsize$ A $}
     \put( 90,141) {\scriptsize$ A $}
     \put(189,134) {\scriptsize$ M $}
     }\setlength{\unitlength}{1pt}}
  \put(130,70)  {$ \longmapsto $}
  \epicture-9 \labl{pic-ffrs5-75}
For a defect field insertion in left/right representation $U_i \Times U_j$, 
there are additional
ribbons, labelled by the simple objects $U_i$ and $U_j$, which connect the 
embedded world sheet $\iota(\X)$ to the boundary of the connecting manifold:
  \bea \begin{picture}(420,184)(0,45)
  \put(-20,105)  {\includeourbeautifulpicture 76 }
  \put(-20,105)  {
     \put(30,5)     {\scriptsize$X$}
     \put(55,8)     {\scriptsize$\theta_{ij}$}
     \put(90,0)     {\scriptsize$Y$}
     \put(70,25)    {\cfta}
     \put(20,-30)   {\cftb}
    }
  \put(120,0)    {\Includeournicelargepicture 85{} }
 \put(-10,0)  {
  \put(350,227)    {$\partial\MX$}
  \put(342,17)     {$\partial\MX$}
  \put(291,234)    {\tiny $1$}
  \put(284,221)    {\tiny $2$}
  \put(316,192)    {\scriptsize$ U_j$}
  \put(310,41)     {\scriptsize$ U_i$}
  \put(237,128)    {\scriptsize$X$}
  \put(349,128)    {\scriptsize$Y$}
  \put(295,120)    {\scriptsize $\theta_{ij}$}
  \put(330,168.1)  {$t\eq0$-plane}
  \put(220,62.3)   {\tiny $1$}
  \put(198,77)     {\tiny $2$}
  \put(274,19)     {\tiny $2$}
  \put(297.8,4)    {\tiny $1$}
  \put(390,30)     {\tiny $1$}
  \put(374.3,50)   {\tiny $3$}
  \put(363,46)     {\tiny $2$}
    }
  \put(110,110)  {$ \longmapsto $}
  \epicture25 
  \labl{eq:defect-field-cobord}
The special case of the insertion of a local bulk field is obtained for
$X \eq Y \eq A$. A more detailed description of the TFT construction is 
given in appendix A of \cite{tft5} and in section 3.3 of \cite{tft4}.

We now derive some of the rules given in section \ref{sec:calc-def}; the 
remaining ones can be verified along similar lines.  Let us start by explaining 
why the fusion of defects corresponds to the tensor product of bimodules, 
see \erf{pic-ffrs5-01}. We have the equality
  \begin{eqnarray}\begin{picture}(420,300)(23,0)
  \put(0,90)    {\Includeourtinynicepicture 77a }
  \put(0,90){
     \setlength{\unitlength}{.90pt}\put(-5,-4){
     \put( 83,121) {\scriptsize$ A $}
     \put(159,132) {\scriptsize$ B $}
     \put(174,102) {\scriptsize$ C $}
     \put(100,100) {\scriptsize$ X $}
     \put(147, 93) {\scriptsize$ Y $}
     \put( 21, 88) {\tiny $1$}
     \put( 46, 96) {\tiny $2$}
     }\setlength{\unitlength}{1pt}}
  \put(242,0)   {\Includeourtinynicepicture 77b }
  \put(242,0){
     \setlength{\unitlength}{.90pt}\put(-5,-4){
     \put( 83,121) {\scriptsize$ A $}
     \put(168,100) {\scriptsize$ C $}
     \put( 87, 88) {\scriptsize$ X $}
     \put(136, 84) {\scriptsize$ Y $}
     \put(145,146) {\scriptsize$ X $}
     \put(202,149) {\scriptsize$ Y $}
     \put(151,117) {\scriptsize$ X{\otimes_B}Y $}
     \put(149,131.5){\scriptsize$ e_{X,Y} $}
     \put(117, 98) {\scriptsize$ r_{X,Y} $}
     \put( 21, 88) {\tiny $1$}
     \put( 46, 96) {\tiny $2$}
     }\setlength{\unitlength}{1pt}}
  \put(203,65)  {$=$}
  \end{picture}
  \nonumber\\[-18pt]~ \label{eq:fusion-derive} 
  \end{eqnarray}
Here we show only the relevant fragments of the complete cobordisms,
and it is understood that $Z(\cdot)$ is applied to each side of the equality.
We also passed to the `blackboard framing' convention for the ribbons, in which
a solid line means that the ribbon is showing its `white' side to the reader. 
Note that the orientation of the embedded world sheet is reversed
with respect to \erf{eq:defect-cobord}\,--\,\erf{eq:defect-field-cobord},
which means that, for convenience, we are drawing the cobordism viewed from a
different angle, so that we face the white side of the embedded ribbons.
The equality in \erf{eq:fusion-derive} follows from the corresponding equality 
for morphisms, namely 
$e_{X,Y} \cir r_{X,Y} \eq P_{X,Y}$, see section \ref{sec:chiral}.

Next, the second rule in \erf{pic-ffrs5-03} amounts to the following identity
for invariants of cobordisms:
  \begin{eqnarray}\begin{picture}(420,300)(23,0)
  \put(0,90)    {\Includeourtinynicepicture 77a }
  \put(0,90){
     \setlength{\unitlength}{.90pt}\put(-5,-4){
     \put( 83,121) {\scriptsize$ A $}
     \put(159,131) {\scriptsize$ B $}
     \put(173,102) {\scriptsize$ C $}
     \put( 93,100) {\scriptsize$ X_\rho $}
     \put(146, 93) {\scriptsize$ X_\sigma $}
     \put( 21, 88) {\tiny $1$}
     \put( 46, 96) {\tiny $2$}
     }\setlength{\unitlength}{1pt}}
  \put(242,0)   {\Includeourtinynicepicture 77b }
  \put(242,0){
     \setlength{\unitlength}{.90pt}\put(-5,-4){
     \put( 83,121) {\scriptsize$ A $}
     \put(167,100) {\scriptsize$ C $}
     \put( 83, 90) {\scriptsize$ X_\rho $}
     \put(136, 82) {\scriptsize$ X_\sigma $}
     \put(139,146) {\scriptsize$ X_\rho $}
     \put(201,150) {\scriptsize$ X_\sigma $}
     \put(151,117) {\scriptsize$ X_\mu $}
     \put(155,130) {\scriptsize$ \overline\alpha $}
     \put(123, 97) {\scriptsize$ \alpha $}
     \put( 21, 88) {\tiny $1$}
     \put( 46, 96) {\tiny $2$}
     }\setlength{\unitlength}{1pt}}
  \put(188,65)  {$\dsty =~\sum_{\mu,\alpha}$}
  \end{picture}
  \nonumber\\[-18pt]~ \label{eq:fusion2-derive} 
  \end{eqnarray}
Here the morphisms are the basis morphisms as chosen in 
\erf{eq:Xmu-XY-basis-HomP}. Equation \erf{eq:fusion2-derive} follows from 
semisimplicity of the bimodule category $\clc AC$; we have
  \be
  \sum_{\mu,\alpha} {\overline\Lambda}_{(\rho\sigma)\mu}^{\alpha}
  \circ {\Lambda}_{(\rho\sigma)\mu}^{\alpha}
  = P_{X_\rho,X_\sigma} \,.
  \ee

The ribbon graph for the identity \erf{eq:2-opp-sum} is similar to 
\erf{eq:fusion2-derive}. The corresponding equality for morphisms of \C\ reads
  \bea \begin{picture}(290,66)(0,54)
  \put(20,0)     {\Includeourbeautifulpicture 78a }
  \put(20,0){
     \setlength{\unitlength}{.38pt}\put(0,0){
     \put( 60,180) {\scriptsize$ B $}
     \put( -5,-19) {\scriptsize$ X_\rho $}
     \put(125,-19) {\scriptsize$ X_\sigma^\vee $}
     \put( -5,318) {\scriptsize$ X_\rho $}
     \put(125,318) {\scriptsize$ X_\sigma^\vee $}
     }\setlength{\unitlength}{1pt}}
  \put(220,0)    {\Includeourbeautifulpicture 78b }
  \put(220,0){
     \setlength{\unitlength}{.38pt}\put(0,0){
     \put( 31,232) {\scriptsize$ \Lambda^\gamma_{(\mu\sigma)\rho} $}
     \put( 31, 62) {\scriptsize$ \overline\Lambda^\gamma_{(\mu\sigma)\rho} $}
     \put( 18,150) {\scriptsize$ X_\mu $}
     \put(  1,-19) {\scriptsize$ X_\rho $}
     \put( 95,-19) {\scriptsize$ X_\sigma^\vee $}
     \put(  1,318) {\scriptsize$ X_\rho $}
     \put( 95,318) {\scriptsize$ X_\sigma^\vee $}
     }\setlength{\unitlength}{1pt}}
  \put(111, 60) {$\dsty=~~\sum_{\mu,\gamma}~\frac{\dim(X_\mu)}{\dim(X_\rho)}$}
  \epicture35 
  \labl{eq:dual-fusion-derive}
To obtain this equality, first note that
  \bea
  L^\gamma_{(\rho\sigma)\mu} = 
  (\id_{X_\mu} \oti \tilde d_{X_\sigma}) \circ
    ({\overline \Lambda}_{(\mu\sigma)\rho}^\gamma \oti \id_{X_\sigma^\vee})
    \circ e_{X_\rho,X_\sigma^\vee}
  \,\in \HomAC(X_\rho \otB X_\sigma^\vee,X_\mu)~\quad{\rm and}
  \\{}\\[-.5em]
  \overline L^\gamma_{(\rho\sigma)\mu} = 
    r_{X_\rho,X_\sigma^\vee} \circ
    (\Lambda_{(\mu\sigma)\rho}^\gamma \oti \id_{X_\sigma^\vee}) \circ
    (\id_{X_\mu} \oti b_{X_\sigma}) 
   \,\in \HomAC(X_\mu, X_\rho \otB X_\sigma^\vee)
  \eear\ee
provide bases of the two morphism spaces, respectively. Since the bimodule 
$X_\mu$ is simple, we have $L^\gamma_{(\rho\sigma)\mu} {\circ}\, 
\overline L^\eps_{(\rho\sigma)\mu} \eq c_{\gamma\eps}\, \id_{X_\mu}$ for some 
constants $c_{\gamma\eps}$ (which can also depend on $\rho,\sigma,\mu$). To 
determine the value of the constant, one takes the trace of both sides and
uses the defining properties of the morphisms $ \Lambda$ and
$\overline \Lambda$ introduced in \erf{eq:Xmu-XY-basis-HomP}. This results in 
$c_{\gamma\eps} \dim(X_\mu)\eq \delta_{\gamma,\eps}\, \dim(X_\rho)$. Thus 
  \be
  L^\gamma_{(\rho\sigma)\mu} \circ \overline L^\eps_{(\rho\sigma)\mu}
  = \delta_{\gamma,\eps}\, \frac{\dim(X_\rho)}{\dim(X_\mu)}\, \id_{X_\mu} \,.
  \labl{eq:fusion2-aux1}
Next, by semisimplicity
of $\clc AC$ there are numbers $C_{\mu,\gamma\eps}$ such that
  \be
  \id_{X_\rho \OtB X_\sigma^\vee} = 
  \sum_{\mu\in\KK_{AC}} \sum_{\gamma,\eps}
  C_{\mu,\gamma\eps}\, \overline L^\eps_{(\rho\sigma)\mu}
  \cir L^\gamma_{(\rho\sigma)\mu} \,.
  \labl{eq:dual-fusion-derive-aux1}
These constants can be determined by composing both sides with
$\overline L^\beta_{(\rho\sigma)\nu}$ from the right, which yields
  \be
  \overline L^\beta_{(\rho\sigma)\nu} = 
  \sum_{\eps} C_{\nu,\beta\eps}\, \frac{\dim(X_\rho)}{\dim(X_\nu)}\,
  \overline L^\eps_{(\rho\sigma)\nu} \,.
  \ee
Since the morphisms $\overline L^\beta_{(\rho\sigma)\nu}$ form a basis, this
forces $C_{\nu,\beta\eps} \eq \delta_{\beta,\eps}\, \dim(X_\nu)/{\dim}(X_\rho)$.
This shows that the constants in \erf{eq:dual-fusion-derive-aux1}
indeed have the value used in \erf{eq:dual-fusion-derive}.

Next consider the equality \erf{eq:def-bulk-act} for wrapping a defect line
around a bulk insertion. In terms of cobordisms, it amounts to the
following defining equality for $D_{\mu\nu\alpha}(\phi)$:
  \bea \begin{picture}(420,165)(20,0)
  \put(0,0)     {\Includeourtinynicepicture 79a }
  \put(0,0){
     \setlength{\unitlength}{.90pt}\put(-5,-4){
     \put(116,148) {\scriptsize$ U_i $}
     \put( 98, 39) {\scriptsize$ U_j $}
     \put(132, 86) {\scriptsize$ \phi $}
     \put(168, 89) {\scriptsize$ B $}
     \put( 94, 83) {\scriptsize$ B $}
     \put(233,120) {\scriptsize$ A $}
     \put( 87,120) {\scriptsize$ X_\mu $}
     \put( 67, 75) {\scriptsize$ X_\nu $}
     \put(108,120) {\scriptsize$ \alpha $}
     \put(165.3,61){\tiny $1$}
     \put(200, 74) {\tiny $2$}
     }\setlength{\unitlength}{1pt}}
  \put(240,0)   {\Includeourtinynicepicture 79b }
  \put(240,0){
     \setlength{\unitlength}{.90pt}\put(-5,-4){
     \put(116,148) {\scriptsize$ U_i $}
     \put( 98, 39) {\scriptsize$ U_j $}
     \put(130,105) {\scriptsize$ D_{\mu\nu\alpha}(\phi) $}
     \put(168, 89) {\scriptsize$ A $}
     \put( 87,120) {\scriptsize$ X_\mu $}
     \put(141.3,61){\tiny $1$}
     \put(176, 74) {\tiny $2$}
     }\setlength{\unitlength}{1pt}}
  \put(234,80)  {$=$}
  \epicture-9 \labl{pic-ffrs5-79}
This enables us to deduce the explicit form of the linear map $D_{\mu\nu\alpha}$
that we introduced in \erf{eq:D-def}:
  \bea \begin{picture}(120,155)(0,58)
  \put(20,0)     {\includeourbeautifulpicture 80 }
  \put(20,0){
     \setlength{\unitlength}{.38pt}\put(0,0){
     \put( 30,-20) {\scriptsize$ U_i $}
     \put(238,-20) {\scriptsize$ U_j $}
     \put(171,418) {\scriptsize$ \phi $}
     \put(165,282) {\scriptsize$ B $}
     \put(176,460) {\scriptsize$ B $}
     \put( 30,570) {\scriptsize$ A $}
     \put( 85,-20) {\scriptsize$ X_\mu $}
     \put( 72,453) {\scriptsize$ X_\nu $}
     \put( 94,153) {\scriptsize$ \alpha $}
     }\setlength{\unitlength}{1pt}}
  \put(20,0){
     \put(-80,100){$D_{\mu\nu\alpha}(\phi) ~=$}
     }
  \epicture36 \labl{pic-ffrs5-80}

As a final rule we establish the one for inserting a little
defect loop as in equation \erf{eq:insertloop}. The corresponding identity
for invariants of cobordisms is
  \begin{eqnarray}\begin{picture}(420,324)(23,0)
  \put(0,110)   {\Includeourtinynicepicture 81a }
  \put(240,0)   {\Includeourtinynicepicture 81b }
  \put(79,196)  {\scriptsize$ A $}
  \put(154,191) {\scriptsize$ A $}
  \put(170,221) {\scriptsize$ A $}
  \put(166,235) {\tiny $1$}
  \put(194.3,247) {\tiny $2$}
  \put(307,93)  {\scriptsize$ A $}
  \put(352,90)  {\scriptsize$ A $}
  \put(415,101) {\scriptsize$ A $}
  \put(406,125) {\tiny $1$}
  \put(436,136) {\tiny $2$}
  \put(387.5,117) {\scriptsize$ Y $}
  \put(165,65)  {$=~\displaystyle\frac{\dim(A)}{\dim(Y)}$}
  \end{picture}
  \nonumber\\[-18pt]~ \label{pic-ffrs5-81}
  \end{eqnarray}
That this is indeed valid is a consequence of the following result (note that
it requires the algebra $A$ to be simple):
\\[-2.2em]

\dtl{Lemma}{lem:X-insert}
For any left $A$-module $M$ over a simple symmetric special Frobenius algebra
$A$ one has
  \bea \begin{picture}(200,43)(0,39)
  \put(20,0)     {\includeourbeautifulpicture 47 }
  \put(32,65)    {\scriptsize$ A $}
  \put(33,19)    {\scriptsize$ A $}
  \put(80,51)    {\scriptsize$ M $}
  \put(102,40)   {$=~ \displaystyle\frac{\dim(M)}{\dim(A)}\, \id_A~.$}
  \epicture16 
  \labl{eq:X-insert}
Proof:\\
Denote the morphism on the left hand side of \erf{eq:X-insert} by $f$. Using 
that $A$ is symmetric Frobenius, one verifies that $f \iN \HomAA(A,A)$. Since 
$A$ is simple, this space is one-dimensional, so that $f \eq\lambda\, \id_A$ 
for some $\lambda \iN \complex$. Composing with unit and counit determines 
this coefficient to be $\lambda \eq \dim(M)/{\dim}(A)$.
\qed


\subsection{Non-degeneracy of defect two-point correlators}\label{sec:proof}

In this section we prove a non-degeneracy result for the two-point correlator
for disorder fields on $S^2$ , i.e.\ for
  \bea \begin{picture}(180,44)(0,38)
  \put(43,0)     {\includeourbeautifulpicture 31 }
  \put(39,63)  {\scriptsize$ X_{\tau} $}
  \put(100,78) {\scriptsize$ \gamma $}
  \put(121,49) {\scriptsize$ X_{\sigma}$}
  \put(95,-7)  {\scriptsize$\delta $}
  \put(97.5,49){\scriptsize$ X_{\nu}$}
  \put(143,39) {\scriptsize$ X_{\mu}$}
  \put(55,40)  {\scriptsize$ \cfta$}
  \put(130,70) {\scriptsize$ \cftb$}
  \put(170,25) {\scriptsize$ \theta_{\bar\imath\bar\jmath,\beta} $}
  \put(77,30)  {\scriptsize$ \theta_{ij,\alpha} $}
  \put(0,36)     {$C~:=$}
  \epicture23 
  \labl{eq:gen-def-2pt}
and derive a number of consequences.
In \erf{eq:gen-def-2pt}, $X_\mu$ is a simple $B$-$B$-defect, $X_\nu$ a simple
$A$-$A$-defect, $X_\tau$ and $X_\sigma$ are simple $B$-$A$-defects, and 
$\gamma, \delta$ label basis elements in the relevant morphism spaces.
As in \erf{eq:A-def} we can write the correlator \erf{eq:gen-def-2pt}
as a multiple of a product of two-point blocks:
  \be
  C = \GammA(\mu,\nu)_{ij\alpha\beta}^{\tau\sigma\gamma\delta}\,
  \beta[i,\bar\imath](z,w)\, \beta[j,\bar\jmath](z^*\!,w^*)\,.
  \labl{eq:def-2pt-corr}
The coefficients $\GammA(\mu,\nu)_{ij\alpha\beta}^{\tau\sigma\gamma\delta}$ can 
be computed by applying the TFT construction for defect correlators outlined in 
section \ref{sec:calc-tft} to the correlator \erf{eq:def-2pt-corr}. This 
results in the following ribbon invariant in $S^3$:
  \begin{eqnarray}\begin{picture}(400,166)(12,0)
  \put(84,0)     {\Includeourtinynicepicture 23a }
  \put(303,0)    {\Includeourtinynicepicture 23b }
  \put(240,150)  {\scriptsize \fbox{$S^3$}}
  \put(442,150)  {\scriptsize \fbox{$S^3$}}
  \put(-19,74)   {$\dsty\GammA(\mu,\nu)_{ij\alpha\beta}^{\tau\sigma\gamma\delta}
                   ~=~ \frac{1}{S_{00}^{\,2}}$}
  \put(237,79)   {\scriptsize$ B$}
  \put(209,61)   {\scriptsize$ \beta$}
  \put(211,41)   {\scriptsize$ \bar\imath$}
  \put(200,-5)   {\scriptsize$ i$}
  \put(135,39)   {\scriptsize$ X_{\mu}$}
  \put(209,106)  {\scriptsize$ \bar\jmath$}
  \put(162,151)  {\scriptsize$ j$}
  \put(107,47)   {\scriptsize$ \delta$}
  \put(130,61)   {\scriptsize$ X_{\sigma}$}
  \put(146,51)   {\scriptsize$ \gamma$}
  \put(148,70)   {\scriptsize$ X_{\nu}$}
  \put(144,82)   {\scriptsize$ \alpha$}
  \put(112,79)   {\scriptsize$ A$}
  \put(175,95)   {\scriptsize$ X_{\tau}$}
  \put(389,153)  {\scriptsize$ B$}
  \put(398,132)  {\scriptsize$ \beta$}
  \put(355,123)  {\scriptsize$ \bar\imath$}
  \put(308,55)   {\scriptsize$ i$}
  \put(332,82)   {\scriptsize$ X_{\mu}$}
  \put(425,102)  {\scriptsize$ \bar\jmath$}
  \put(452,58)   {\scriptsize$ j$}
  \put(349,45)   {\scriptsize$ \delta$}
  \put(362,54)   {\scriptsize$ X_{\sigma}$}
  \put(368,80)   {\scriptsize$ \gamma$}
  \put(379,69)   {\scriptsize$ X_{\nu}$}
  \put(397,53)   {\scriptsize$ \alpha$}
  \put(376,19)   {\scriptsize$ A$}
  \put(415,70)   {\scriptsize$ X_{\tau}$}
  \put(261,74)   {$\dsty=~ \frac{1}{S_{00}^{\,2}}$}          
  \end{picture}
  \nonumber\\[1pt]~ \label{eq:A-rib} 
  \end{eqnarray}
Let $R_1$ denote the set of labels $(ij\alpha\beta)$ and $R_2$ the set of 
labels $(\tau\sigma\gamma\delta)$. By definition, their cardinalities are
  \bea
  |R_1| = \sum_{i,j \in \II} 
  \dimc\HomAA(U_i \otiP A \otim U_j,X_\nu)\cdot
  \dimc\HomAA(U_{\bar\imath} \otip X_\mu \otim U_{\bar\jmath},B) \,, 
  \\{}\\[-.7em]
  |R_2| = \sum_{\sigma,\tau\in\KK_{BA}}
  \dimc \HomBA(X_\sigma \otA X_\nu,X_\tau)\cdot
  \dimc \HomBA(X_\tau,X_\mu \otB X_\sigma)\,.
  \eear\ee
Re-expressing $|R_1|$ via \erf{eq:ZijXY} and performing
the sum over $\tau$ in $|R_2|$ we can alternatively write
  \be
  |R_1| = \sum_{i,j \in \II} 
  Z(A)_{ij}^{A|X_\nu} Z(B)_{\bar\imath\bar\jmath}^{X_\mu|B} \,, \quad~
  |R_2| = \sum_{\sigma\in\KK_{BA}}\!\!
  \dimc\HomBA(X_\sigma \otA X_\nu,X_\mu \otB X_\sigma)\,.
  \labl{eq:R1R2-gen}

\dtl{Theorem}{thm:gen-non-deg}
Let $A$ and $B$ be simple symmetric special Frobenius algebras and let 
$\mu \iN \KK_{BB}$ and $\nu \iN \KK_{AA}$. Then the label sets $R_1$ and $R_2$ 
either obey $|R_1|\eq|R_2|\eq0$, or else the $|R_1| \Times |R_2|$\,-matrix 
$\GammA(\mu,\nu)$ with entries given by \erf{eq:A-rib} is non-degenerate.

\medskip

The idea of the proof is to relate the assertion (which is a statement about
conformal blocks on surfaces of genus zero) to properties of conformal blocks 
on surfaces of genus one. It relies on a number of ingredients. The first is a 
certain projector $\rm P$ on the vector space $\calh({\rm T}_{\!X_\mu,X_\nu})$,
with ${\rm T}_{\!X_\mu,X_\nu}$ the extended surface consisting of a
2-torus with two marked points with labels $(X_\mu,+)$ and $(X_\nu,-)$
(here the signs $(\cdot,\pm)$ refer to different orientations of the core of a 
ribbon, see the conventions in \cite[sect.\,2.4]{tft1} and 
\cite[sect.\,3.1]{tft4}),
  \bea \begin{picture}(405,230)(0,49)
  \put(80,0)          {\includeourbeautifulpicture 24 }
  \put(80,0){
     \setlength{\unitlength}{.38pt}\put(0,0){
     \put(359,544)    {\scriptsize $B$}
     \put(369,287)    {\scriptsize $A$}
     \put(381,360)    {\scriptsize $B$}
     \put(467,106)    {\scriptsize $A$}
     \put(532,225)    {\scriptsize $X_\nu$}
     \put(532,483)    {\scriptsize $X_\mu$}
     }\setlength{\unitlength}{1pt}}
  \put(98,254)        {\tiny $1$}
  \put(83,249)        {\tiny $2$}
  \put(315.6,243)     {\tiny $1$}
  \put(302.8,265)     {\tiny $2$}
  \put(82.5,162.2)    {\scriptsize $(X_\mu,+)$}
  \put(82.5,64)       {\scriptsize $(X_\nu,-)$}
  \put(332,183)       {\scriptsize $(X_\mu,-)$}
  \put(337,82)        {\scriptsize $(X_\nu,+)$}
  \put(23,138)        {$ {\rm P} \;:= $}
  \epicture25 \labl{eq:P-def}
This is just a slight generalisation of the projector given in (5.15) of 
\cite{tft5}, with the $A$-ribbons running from left to right replaced by an
$X_\mu$- and an $X_\nu$-ribbon, respectively. That \erf{eq:P-def}
indeed defines a projector is seen in the same way as in \cite{tft5}.

Let ${\rm Im}({\rm P})\,{\subseteq}\,{\rm End}(\calh({\rm T}_{\!X_\mu,X_\nu}))$
be the image of the projector $\rm P$. 
We will introduce two bases for the vector space ${\rm Im}({\rm P})$
and show that the matrix transforming one basis into the other is given by 
$\GammA(\mu,\nu)$, thus establishing that $\GammA(\mu,\nu)$ is nondegenerate.

For $\alpha \iN \HomAA(U_i \otiP A \otim U_j,X_\nu)$ and 
$\beta \iN \HomAA(U_{\bar\imath} \otip X_\mu \otim U_{\bar\jmath},B)$, consider 
the vectors ${\rm M}^{\mu\nu}_{ij,\alpha\beta} \iN \calh({\rm T}_{X_\mu,X_\nu})$
given by the invariant
  \bea \begin{picture}(120,160)(0,68)
  \put(0,0)        {\includeourbeautifulpicture 25 }
  \put(-68,121)    {$ {\rm M}_{ij,\alpha\beta}^{\mu\nu} \;:= $}
  \put(5,194)      {\tiny $2$}
  \put(18,200)     {\tiny $1$}
  \put(47,139)     {\scriptsize $X_\mu$}
  \put(47,87)      {\scriptsize $X_\nu$}
  \put(89,91.5)    {\small $\alpha$}
  \put(109.5,128)  {\scriptsize $\ib$}
  \put(118.1,108)  {\scriptsize $i$}
  \put(124,86)     {\scriptsize $A$}
  \put(126,145)    {\small $\beta$}
  \put(131,188)    {\scriptsize $\jb$}
  \put(137,213)    {\scriptsize $j$}
  \put(140,140)    {\scriptsize $B$}
  \epicture44 \labl{eq:M_mu-def}
The three-manifold in this figure is a solid torus in ``wedge presentation''. 
The boundary of the solid torus is the vertical face containing the two marked 
points, which is the 2-torus ${\rm T}_{\!X_\mu,X_\nu}$. The other two vertical 
faces are to be identified, as are the horizontal faces at the top and bottom.
We refer to section 5.1 of \cite{tft5} for more details on the
wedge presentation of three-manifolds.

\dtl{Lemma}{lem:M_mu-basis}
The invariants ${\rm M}^{\mu\nu}_{ij,\alpha\beta}$, 
with $(ij\alpha\beta) \iN R_1$, form a basis of $\,{\rm Im}({\rm P})$.

\medskip\noindent
Proof:\\
The invariants \erf{eq:M_mu-def} constitute a generalisation
of the ones displayed in (5.10) of \cite{tft5}. The proof
works in the same way as the one of lemma 5.2\,(ii) of \cite{tft5}.
\qed

To find the second basis, consider the cobordisms ${\rm K}^{\mu\nu}
_{\sigma\tau\gamma\delta}{:}\  \emptyset \To {\rm T}_{X_\mu,X_\nu}$ and 
$\overline{\rm K}^{\mu\nu}_{\sigma\tau\gamma\delta}{:}\ 
{\rm T}_{X_\mu,X_\nu} \To \emptyset$ given by
  \begin{eqnarray}\begin{picture}(420,153)(18,0)
  \put(74,0)     {\Includeourtinynicepicture 26{} }
  \put(322,0)    {\Includeourtinynicepicture 27{} }
  \put(170,70)   {\scriptsize $X_{\mu}$}
  \put(177.5,54) {\scriptsize $\overline{\delta}$}
  \put(160,40)   {\scriptsize $X_{\tau}$}
  \put(144,54.5) {\scriptsize $X_{\sigma}$}
  \put(151,69)   {\scriptsize $\overline{\gamma}$}
  \put(122,43)   {\scriptsize $X_{\nu}$}
  \put(410,120)  {\scriptsize $X_{\mu}$}
  \put(375,75)   {\scriptsize ${\delta}$}
  \put(398,60)   {\scriptsize $X_{\tau}$}
  \put(390,86)   {\scriptsize ${\sigma}$}
  \put(393,95)   {\scriptsize ${\gamma}$}
  \put(432,47)   {\scriptsize $X_{\nu}$}
  \put(0,78)     {${\rm K}^{\mu\nu}_{\sigma\tau\gamma\delta}~:=$}
  \put(264,78)   {$\overline{\rm K}^{\mu\nu}_{\sigma\tau\gamma\delta}~:=$}
  \end{picture}
  \nonumber\\[2pt]~ \label{eq:K_mu-bas} 
  \end{eqnarray}
Here $\gamma$ runs over a basis of the space
$\HomAB(X_\sigma,X_\tau \otA X_\nu)$ and $\bar\gamma$ over the corresponding 
dual basis of $\HomAB(X_\tau \otA X_\nu,X_\sigma)$; the meaning of $\delta$ 
and $\bar\delta$ is analogous. Again, we implicitly use the isomorphisms 
\erf{eq:HomP-HomTens}. To establish that \erf{eq:K_mu-bas} are dual 
bases, the following result is helpful:
\\[-2.2em]

\dtl{Lemma}{lem:S2xI-id}
Let $A$ be a simple symmetric special Frobenius algebra and $X_\mu$ a simple 
$A$-$A$-bimodule. Then
  \bea \begin{picture}(300,59)(0,44)
  \put(20,0)     {\Includeourtinynicepicture 33a }
  \put(200,0)    {\Includeourtinynicepicture 33b }
  \put(19,53)    {\scriptsize $A$}
  \put(57,26)    {\scriptsize $X_{\mu}$}
  \put(98,95)    {\scriptsize \fbox{$S^2\times I$}}
  \put(240,70)   {\scriptsize $A$}
  \put(240,30)   {\scriptsize $A$}
  \put(278,95)   {\scriptsize \fbox{$S^2\times I$}}
  \put(133,51)   {$=~\displaystyle\frac{\delta_{\mu,0}}{\dim(A)}$}
  \epicture19 
  \labl{eq:S2xI-id}
Proof:\\
Let us start from the ribbon graph invariant on the left hand side of 
\erf{eq:S2xI-id}, to be denoted by $Q$. By the same reasoning as in the proof 
of lemma 5.2 in \cite{tft1}, we see that $Q \iN \End(\mathcal{H}(S^2;X_\mu))$ 
is an idempotent. Next consider the equalities
  \begin{eqnarray} \begin{picture}(420,175)(5,0)
  \put(80,0)     {\Includeourtinynicepicture 34a }
  \put(245,0)    {\Includeourtinynicepicture 34b }
  \put(79,110)   {\scriptsize $A$}
  \put(119,27)   {\scriptsize $X_{\mu}$}
  \put(158,165)  {\scriptsize \fbox{$S^2\times I$}}
  \put(119,52)   {\scriptsize $k$}
  \put(119,137)  {\scriptsize $X_{\mu}$}
  \put(103,38)   {\scriptsize $\overline{\alpha}$}
  \put(103,67)   {\scriptsize $\alpha$}
 \put(25,0){
  \put(230,105)  {\scriptsize $A$}
  \put(256,65)   {\scriptsize $A$}
  \put(268,83)   {\scriptsize $A$}
  \put(259,27)   {\scriptsize $X_{\mu}$}
  \put(298,165)  {\scriptsize \fbox{$S^2\times I$}}
  \put(259,135)  {\scriptsize $X_{\mu}$}
  \put(243,38)   {\scriptsize $\overline{\alpha}$}
  \put(243,93)   {\scriptsize $\alpha$}
  \put(170,88)   {$=~\displaystyle\sum_\alpha$}
 }
  \put(0,88)     {$Q~=~\displaystyle\sum_{k\in\II}\sum_\alpha$}
  \put(365,88)   {$=~0\quad{\rm if}~~\mu\ne0~.$}
  \end{picture}
  \nonumber\\[-23pt]~ \label{pic-ffrs5-34}
  \end{eqnarray}
Here in the first step the bimodule $X_\mu$ is regarded as an object of \C\ and 
decomposed into simple objects of \C. The 
second step follows because $\mathcal{H}(S^2;U_k)$ has nonzero dimension only 
for $U_k\eq\one$, so that the sum over $k$ restricts to $k\eq0$. We also 
deformed the $A$-ribbon and used that $A$ is symmetric. Finally, the morphism 
in the dashed box constitutes an element in $\HomAA(A,X_\mu)$, which 
is non-zero only for $\mu\eq0$. If $\mu\eq0$, i.e.\ if $X_\mu \eq A$, then 
the morphism from $A$ to $A$ on the left hand side of \erf{eq:S2xI-id}
is just the projector on the left center $C_l(A)$ of $A$, see section 2.4 of 
\cite{ffrs} for details and references. Since $A$ is simple we have 
$\dimc\Hom(\one,C_l(A)) \eq 1$ (this follows from the case $U\eq V\eq \one$
of proposition 2.36 of \cite{ffrs}). The unit morphism $\eta$ of $A$ lies in 
the left center, so that the left and the right hand side of \erf{eq:S2xI-id}
must be proportional. The constant is
determined by demanding the right hand side to be an idempotent.
\qed 

\dtl{Lemma}{lem:K_mu-basis}
(i)~~${\rm K}^{\mu\nu}_{\sigma\tau\gamma\delta} 
    \iN {\rm Im}({\rm P})$.
\\[.3em]
(ii)~\,$\overline{\rm K}^{\mu\nu}_{\sigma'\tau'\gamma'\delta'} \cir 
    {\rm K}^{\mu\nu}_{\sigma\tau\gamma\delta} 
    = \delta_{\sigma,\sigma'} \, \delta_{\tau,\tau'} \,
    \delta_{\gamma,\gamma'} \, \delta_{\delta,\delta'}$. 
\\[.2em]
    In particular, the vectors
    $\{ {\rm K}^{\mu\nu}_{\sigma\tau\gamma\delta} 
    \}_{\sigma,\tau,\gamma,\delta}$ are linearly independent.
\\[.3em]
(iii) The vectors $\{ {\rm K}^{\mu\nu}_{\sigma\tau\gamma\delta} 
    \}_{\sigma,\tau,\gamma,\delta}$ span ${\rm Im}({\rm P})$.

\medskip\noindent
Proof:\\
(i) can be seen by using the property that $\bar\gamma$ and $\bar\delta$ are 
bimodule intertwiners -- this allows one to remove the $A$- and $B$-ribbons 
in the composition ${\rm P} \cir {\rm K}^{\mu\nu}_{\sigma\tau\gamma\delta}$ by 
moves similar to the ones used in the proof of lemma 5.2(iii) of \cite{tft5}.
\\[.3em]
For (ii) note that composing $\overline{\rm K}^{\mu\nu}
_{\sigma'\tau'\gamma'\delta'} \cir {\rm K}^{\mu\nu}_{\sigma\tau\gamma\delta}$
yields the following ribbon graph in $S^2 \Times S^1$:
  \begin{eqnarray}\begin{picture}(420,356)(0,0)
  \put(111,94)   {\Includeourtinynicepicture 32a }
  \put(302,0)    {\Includeourtinynicepicture 32b }
  \put(210,145)  {\scriptsize $\overline{\delta}$}
  \put(203,191)  {\scriptsize $X_{\sigma}$}
  \put(205,214)  {\scriptsize $\overline{\gamma}$}
  \put(188,236)  {\scriptsize $X_{\tau}$}
  \put(175,160)  {\scriptsize $X_{\mu}$}
  \put(175,208)  {\scriptsize $X_{\nu}$}
  \put(137,145)  {\scriptsize ${\delta'}$}
  \put(135,191)  {\scriptsize $X_{\sigma'}$}
  \put(140,214)  {\scriptsize ${\gamma'}$}
  \put(153,236)  {\scriptsize $X_{\tau'}$}
  \put(220,230)  {\scriptsize \fbox{$S^2\times S^1$}}
  \put(398,215)  {\scriptsize $\overline{\gamma}$}
  \put(396,187)  {\scriptsize $X_{\sigma}$}
  \put(402,46)   {\scriptsize $\overline{\delta}$}
  \put(396,35)   {\scriptsize $X_{\tau}$}
  \put(332,215)  {\scriptsize ${\gamma'}$}
  \put(327,187)  {\scriptsize $X_{\sigma'}$}
  \put(329,46)   {\scriptsize ${\delta'}$}
  \put(327,35)   {\scriptsize $X_{\tau'}$}
  \put(307,305)  {\scriptsize $B$}
  \put(375,279)  {\scriptsize $A$}
  \put(357,170)  {\scriptsize $B$}
  \put(366,208)  {\scriptsize $X_{\nu}$}
  \put(375,115)  {\scriptsize $A$}
  \put(366,62)   {\scriptsize $X_{\mu}$}
  \put(420,320)  {\scriptsize \fbox{$S^2\times S^1$}}
  \put(0,178)    {$\overline{\rm K}^{\mu\nu}_{\sigma'\tau'\gamma'\delta'} \circ
                   {\rm K}^{\mu\nu}_{\sigma\tau\gamma\delta}~\overset{(1)}{=}$}
  \put(274,178)  {$\overset{(2)}{=}$}
  \end{picture}
  \nonumber\end{eqnarray}
  \begin{eqnarray}\begin{picture}(420,362)(0,0)
  \put(210,0)    {\Includeourtinynicepicture 32c }
  \put(304,320)  {\scriptsize $X_{\tau}$}
  \put(280,248){\scriptsize $X_{\tau}$}
  \put(304,209)  {\scriptsize $\overline{\gamma}$}
  \put(280,155)  {\scriptsize $X_{\sigma}$}
  \put(280,108)  {\scriptsize $X_{\sigma}$}
  \put(311,48)   {\scriptsize $\overline{\delta}$}
  \put(304,25)   {\scriptsize $X_{\tau}$}
  \put(235,220)  {\scriptsize $X_{\tau'}$}
  \put(239,210)  {\scriptsize $\gamma'$}
  \put(235,188)  {\scriptsize $X_{\sigma'}$}
  \put(235,81)   {\scriptsize $X_{\sigma'}$}
  \put(235,48)   {\scriptsize ${\delta'}$}
  \put(235,25)   {\scriptsize $X_{\tau'}$}
  \put(240,90)   {\scriptsize $\overline{\varepsilon}_2$}
  \put(236,162)  {\scriptsize $X_{\zeta}$}  
  \put(240,175)  {\scriptsize ${\varepsilon}_2$}
  \put(240,230)  {\scriptsize $\overline{\varepsilon}_1$}
  \put(240,314)  {\scriptsize ${\varepsilon}_1$}
  \put(236,302)  {\scriptsize $X_{\xi}$}
  \put(340,320)  {\scriptsize \fbox{$S^2\times S^1$}}
  \put(215,139)  {\scriptsize $B$}
  \put(280,202)  {\scriptsize $X_{\nu}$}
  \put(215,274)  {\scriptsize $B$}
  \put(280,64)   {\scriptsize $X_{\mu}$}
  \put(20,178)   {$\overset{(3)}{=}~\displaystyle\sum_{\zeta,\xi\in\KK_{BB}}
                   \sum_{\eps_1,\eps_2}~\frac{\dim(X_\zeta)}{\dim(X_{\sigma'})}
                   \,\frac{\dim(X_\xi)}{\dim(X_{\tau'})}$}
  \end{picture}
  \nonumber\\\begin{picture}(420,264)(0,0)
  \put(133,0)    {\Includeourtinynicepicture 32d }
  \put(234,50)   {\scriptsize $\overline{\delta}$}
  \put(159,50)   {\scriptsize ${\delta'}$}
  \put(205,37)   {\scriptsize $X_{\tau}$}
  \put(205,67)   {\scriptsize $X_{\sigma}$}
  \put(205,98)   {\scriptsize $X_{\mu}$}
  \put(205,124)  {\scriptsize $X_{\sigma}$}
  \put(205,205)  {\scriptsize $X_{\tau}$}
  \put(204,158)  {\scriptsize $X_{\nu}$}
  \put(255,200)  {\scriptsize \fbox{$S^2\times S^1$}}
  \put(227,166)  {\scriptsize $\overline{\gamma}$}
  \put(163,166)  {\scriptsize ${\gamma'}$}
  \put(20,128)   {$\overset{(4)}{=}~\displaystyle\frac{\delta_{\sigma,\sigma'}
                   \delta_{\tau,\tau'}}{\dim(X_\sigma)\dim(X_\tau)}$}
  \put(311,128)  {$\overset{(5)}{=}~\delta_{\sigma,\sigma'}\,\delta_{\tau,\tau'}
                   \,\delta_{\gamma,\gamma'}\,\delta_{\delta,\delta'}~.$}
  \end{picture}
  \nonumber\\[-19pt]~ \label{pic-ffrs5-32}
  \end{eqnarray}
The manipulations leading to the individual equalities are as follows: 
(1) consists in composing the two cobordisms \erf{eq:K_mu-bas}. 
For (2) one first wraps the $X_\mu$-ribbon around the `horizontal' $S^2$ so 
that it runs behind the vertical ribbons, and then uses the presence of the 
$X_\mu$- and $X_\nu$-ribbons to insert the $A$- and $B$-ribbons (this 
equality is more easily checked in the opposite direction, using the presence 
of the two bimodule ribbons to remove $A$ and $B$). In (3) the identity 
\erf{eq:dual-fusion-derive} is applied to the morphisms in the two 
dashed boxes. In (4) lemma \ref{lem:S2xI-id} is applied to restrict the sums 
to $\zeta \eq \xi \eq 0$. Finally, in (5) one uses that $\gamma'$ and 
$\bar\gamma$, as well as $\delta'$ and $\bar\delta$, label dual bases.
\\[.3em]
To show (iii), we start from the ribbon invariants 
  \bea \begin{picture}(300,91)(0,68)
  \put(90,0)     {\Includeourtinynicepicture 35{} }
  \put(138,49)   {\scriptsize $X_{\nu}$}
  \put(123,101)  {\scriptsize $X_{\mu}$}
  \put(152,36)   {\scriptsize ${\beta}$}
  \put(155,101)  {\scriptsize ${\alpha}$}
  \put(184,48)   {\scriptsize $n$}
  \put(204,100)  {\scriptsize $m$}
  \put(206,66)   {\scriptsize $t$}
  \put(223,50)   {\scriptsize $s$}
  \put(219,58)   {\scriptsize $\varphi$}
  \put(197,76)   {\scriptsize $\overline{\varepsilon}$}
  \put(0,82)     {$b_{nmts,\alpha\beta\eps\varphi} ~:=$}
  \epicture40 \labl{pic-ffrs5-35}
which form a basis of $\calh({\rm T}_{\!X_\mu,X_\nu})$. Here $n, m, s, t$ label 
simple objects of $\calc$ and $\alpha, \beta, \eps, \sigma$ label basis elements
in the appropriate morphism spaces. Applying the idempotent
$\rm P$ and deforming the resulting ribbon graph suitably gives
  \begin{eqnarray}\begin{picture}(420,347)(20,0)
  \put(16,0)     {\Includeourtinynicepicture 36{} }
  \put(105,162)  {\scriptsize $X_{\nu}$}
  \put(220,293)  {\scriptsize $X_{\mu}$}
  \put(173,242)  {\scriptsize ${\beta}$}
  \put(296,177)  {\scriptsize ${\alpha}$}
  \put(199,250)  {\scriptsize $\overline{\varepsilon}$}
  \put(271,168)  {\scriptsize $\varphi$}
  \put(195,241)  {\scriptsize $n$}
  \put(283,187)  {\scriptsize $m$}
  \put(232,205)  {\scriptsize $t$}
  \put(324,131)  {\scriptsize $s$}
  \put(202,200)  {\scriptsize $A$}
  \put(255,209)  {\scriptsize $B$}
  \put(299,127)  {\scriptsize $A$}
  \put(349,136)  {\scriptsize $B$}
  \put(0,331)    {${\rm P} \circ b_{nmts,\alpha\beta\eps\varphi} ~=$}
  \end{picture}
  \nonumber\\[-22pt]~ \label{eq:Pob} 
  \end{eqnarray}
Inside the dashed boxes we deal with the induced $B$-$A$-bimodules 
$B\oti U_s\oti A$ and $B \oti U_t \oti A$, which we can decompose in simple 
$B$-$A$-bimodules $X_\sigma$ and $X_\tau$, respectively. In this decomposition, 
the morphisms inside the dashed circles give elements in $\HomBA(X_\tau,X_\sigma
\otA X_\nu)$ and $\HomBA(X_\mu \otB X_\sigma,X_\tau)$, respectively. It follows 
that the vector \erf{eq:Pob} can be written as a linear combination of the 
elements ${\rm K}^{\mu\nu}_{\sigma\tau\gamma\delta}$ in 
\erf{eq:K_mu-bas}. Thus indeed ${\rm Im}({\rm P})$ is
spanned by the vectors ${\rm K}^{\mu\nu}_{\sigma\tau\gamma\delta}$.
\qed

\medskip\noindent
{\bf Proof of theorem \ref{thm:gen-non-deg}:}
\\[.2em]
By lemmas \ref{lem:M_mu-basis} and \ref{lem:K_mu-basis} we know that 
${\rm M}^{\mu\nu}_{ij,\alpha\beta}$ and ${\rm K}^{\mu\nu}_{\sigma\tau
\gamma\delta}$ are bases of ${\rm Im}({\rm P})$. Thus we can write
  \be
  {\rm M}^{\mu\nu}_{ij,\alpha\beta}
  = \sum_{\sigma',\tau',\gamma',\delta'} 
  \GammB(\mu,\nu)_{ij\alpha\beta}^{\sigma'\tau'\gamma'\delta'}
  {\rm K}^{\mu\nu}_{\sigma'\tau'\gamma'\delta'}
  \ee
for some nondegenerate matrix $\GammB(\mu,\nu)$. Composing both sides
with  $\overline{\rm K}^{\mu\nu}_{\sigma\tau\gamma\delta}$ 
from the left and using lemma \ref{lem:K_mu-basis}\,(ii) yields
  \bea \begin{picture}(370,98)(0,58)
  \put(180,0)    {\Includeourtinynicepicture 28{} }
  \put(344,154)  {\scriptsize \fbox{$S^3$}}
  \put(250,59)   {\scriptsize $X_{\nu}$}
  \put(248,110)  {\scriptsize $X_{\mu}$}
  \put(287,117)  {\scriptsize ${\beta}$}
  \put(287,38)   {\scriptsize ${\alpha}$}
  \put(176,127)  {\scriptsize $j$}
  \put(275,125)  {\scriptsize $\bar\jmath$}
  \put(305,108)  {\scriptsize $B$}
  \put(306,58)   {\scriptsize $A$}
  \put(295,72)   {\scriptsize $i$}
  \put(280,84)   {\scriptsize $\bar\imath$}
  \put(222,64)   {\scriptsize $\sigma$}
  \put(233,47)   {\scriptsize $\delta$}
  \put(236,74)   {\scriptsize $\gamma$}
  \put(347,92)   {\scriptsize $X_{\tau}$}
  \put(0,78)     {$\GammB(\mu,\nu)_{ij\alpha\beta}^{\sigma\tau\gamma\delta}
                   ~=~\overline{\rm K}^{\mu\nu}_{\sigma\tau\gamma\delta}\circ
                   {\rm M}^{\mu\nu}_{ij,\alpha\beta} ~=$}
  \epicture33 \labl{pic-ffrs5-28}
The ribbon graph on the right hand side can be deformed into the one shown 
in \erf{eq:A-rib}, so that in fact
$\GammA(\mu,\nu) \eq \GammB(\mu,\nu)/S_{00}^{\,2}$.
\qed

Lemma \ref{lem:non-deg} is a consequence of theorem \ref{thm:gen-non-deg}, 
obtained by specialising to $\nu \eq 0$.  Since for $\GammA(\mu,\nu)$ to be 
nondegenerate the cardinalities \erf{eq:R1R2-gen} of the index sets $R_1$ and 
$R_2$ must be equal, another immediate consequence is the following 
\\[-2.3em]

\dtl{Corollary}{cor:R1=R2}
For $A, B$ simple symmetric special Frobenius algebras, $X_\nu$ a simple 
$A$-$A$-bimodule and $X_\mu$ a simple $B$-$B$-bimodule, we have
  \bea\displaystyle\hspace*{-1em}
  \sum_{i,j \in \II} \dimc\HomAA(U_i \otiP A \otim U_j, X_\nu)\,
  \dimc\HomBB(U_{\bar\imath} \otip X_\mu \otim U_{\bar\jmath}, B)
  \\[-.2em] \hspace*{11.3em}\displaystyle
  = \sum_{\tau\in\KK_{BA}} \dimc\HomBA(X_\tau\otA X_\nu,X_\mu\otB X_\tau)\,.
  \eear\labl{eq:R1=R2}

Recall that the space $\HomAA(U_i \otiP A \otim U_j, X_\nu)$ labels disorder
fields which turn a defect $A$ into a defect $X_\nu$ and which are in in 
left/right representation $(i,j)$. Thinking of the CFT as the scaling limit of 
a lattice model, one would expect that for every defect $X_\nu$ there is at 
least one disorder field on which that defect can start, i.e.\ at least one 
pair $(i,j)$ for which $\HomAA(U_i \otiP A \otim U_j, X_\nu)$ is nonzero.
In \rep\ theoretic terms, this is established by the following 
\\[-2.3em]

\dtl{Proposition}{prop:UiAUj-submod}
Every simple $A$-$A$-bimodule of a simple symmetric special Frobenius algebra
\A\ in a modular tensor category $\calc$ is a sub-bimodule of $U \otiP A \otim V$ 
for suitable simple objects $U, V$ of $\calc$.

\medskip\noindent
Proof:\\
It suffices to show that $\dimc\HomAA(U_i \otiP A \otim U_j, X_\nu)$ is nonzero
for a suitable choice of $(i,j) \iN \II\Times\II$. Suppose the contrary. Then the 
left hand side of \erf{eq:R1=R2}, evaluated for $A\eq B$ and $\mu \eq \nu$, is 
zero. The sum on the right hand side, on the other hand, contains the 
contribution $\dimc\HomAA(A \otA X_\mu,X_\mu \otA A)$ for $\tau\eq0$, which is 
equal to one. This is a contradiction; hence 
$\dimc\HomAA(U_i \otiP A \otim U_j, X_\nu)$ cannot be zero for all $i,j$.
\qed

\dtl{Remark}{rem:alpha-ind}
(i)~\,By taking suitable direct sums, proposition \ref{prop:UiAUj-submod} 
implies in particular that every $A$-$A$-bimodule is a sub-bimodule of 
$U \otiP A \otim V$ for suitable objects $U$ and $V$.
\\[.2em]
(ii)~It is instructive to reformulate proposition \ref{prop:UiAUj-submod} in 
terms of $\alpha$-induction. Recall that, given an algebra $A$ in a braided 
tensor category \C, there are two tensor functors $\alpha_\AA^\pm$ from \C\ to 
the category of $A$-$A$-bimodules, called $\alpha$-induction, with the right 
$A$-action on $\alpha_\AA^\pm(U)$ involving the braiding, see e.g.\ section 5.1 
of \cite{ostr}. Comparing the definitions (see equations (2.18) of \cite{tft4} 
and (2.33) of \cite{ffrs}) one finds
  \be
  \alpha^+_\AA(U) = A \otip U \qquad {\rm and} \qquad
  \alpha^-_\AA(U) = A \otim U \,.
  \ee
This leads to the isomorphisms
  \be
  \alpha^+_\AA(U)  \otA \alpha^-_\AA(V)
  \cong ( A \otip U ) \otA (A \otim V) \cong U \otiP A \otim V \,,
  \ee
where the last isomorphism is again easily seen by writing
out the definitions. Proposition \ref{prop:UiAUj-submod} 
then implies that for simple $A$, every $A$-$A$-bimodule
is a submodule of the tensor product $\alpha^+_\AA(U) \otA \alpha^-_\AA(V)$
of two $\alpha$-induced bimodules. This result has been obtained previously in 
the framework of subfactor theory in \cite[Theorem\,5.10]{boek} and has been 
conjectured in the category theoretic framework in Claim 2 of \cite{ostr}.

\medskip

When analysing the fusion rules of defect lines, it is useful to know that 
defects labelled by $\alpha$-induced bimodules always lie in the center of the 
fusion algebra. This is established in the following 
\\[-2.3em]

\dtl{Lemma}{lem:alpha-fuse}
Let $A$ be a symmetric special Frobenius algebra in an idempotent complete 
ribbon category $\calc$. For every object $U$ of \C\ and for every 
$A$-$A$-bimodule $Y$ we have, for $\nu \iN \{\pm\}$, an isomorphism 
  \be
  \alpha^\nu_\AA(U) \otA Y  \cong Y \otA \alpha^\nu_\AA(U) 
  \ee
of $A$-$A$-bimodules.

\medskip\noindent
Proof:\\
An isomorphism can be constructed using the braiding of $\calc$. For example, 
for $\nu \eq {+}$ consider the two morphisms
  \bea \begin{picture}(380,165)(0,62)
  \put(45,0)     {\Includeourbeautifulpicture 73a }
  \put(260,0)    {\Includeourbeautifulpicture 73b }
  \put(50,42)    {\scriptsize $A$}
  \put(88,160)   {\scriptsize $A$}
  \put(113,160)  {\scriptsize $U$}
  \put(51,160)   {\scriptsize $Y$}
  \put(120,181)  {$r_{Y,\alpha^+_\AA(U)}$}
  \put(68,223)   {\scriptsize $Y\OtA\,\alpha^+_{\!A}(U)$}
  \put(120,31)   {$e_{\alpha^+_{\!A}(U),Y}$}
  \put(66,-8)    {\scriptsize $\alpha^+_\AA(U)\OtA Y$}
  \put(315,42)   {\scriptsize $A$}
  \put(277,160)  {\scriptsize $A$}
  \put(315,160)  {\scriptsize $U$}
  \put(340,160)  {\scriptsize $Y$}
  \put(346,181)  {$r_{\!\alpha^+_{\!A}(U),Y}$}
  \put(292,223)  {\scriptsize $\alpha^+_{\!A}(U)\OtA Y$}
  \put(346,31)   {$e_{Y,\alpha^+_{\!A}(U)}$}
  \put(292,-8)   {\scriptsize $Y\OtA\,\alpha^+_{\!A}(U)$}
  \put(0,107)    {$f~:=$}
  \put(175,107)  {and}
  \put(215,107)  {$g~:=$}
  \epicture44 \labl{pic-ffrs5-73}
One verifies that $f \iN \HomAA(\alpha^+_\AA(U) \otA Y,Y \otA \alpha^+_\AA(U))$ and 
that $g \iN \HomAA(Y\otA \alpha^+_\AA(U),\alpha^+_\AA(U)
      $\linebreak[0]$
\otA Y)$. It is furthermore 
straightforward to check that $g$ is a left and right inverse to $f$.
\qed


\sect{Defects for simple current theories}\label{sec:sicu}

A class of symmetric special Frobenius algebras that play an important role in 
applications and are at the same time particularly tractable are 
Schellekens algebras. By definition, a {\em Schellekens algebra\/} is a 
symmetric special Frobenius algebra which is simple as a left-module over 
itself and all of whose simple subobjects are invertible objects of \C,
or, in other words \cite{scya,scya6}, simple currents.

The structure and representation theory of Schellekens algebras has been 
developed in \cite{tft3}.  As an object of \C, a Schellekens algebra is of 
the form 
  \be
  A \cong \bigoplus _{h \in H} L_h
  \ee
with $H$ a subgroup of the group $\PicC$
of isomorphism classes of invertible objects (simple currents) of \C. 
The subgroup $H\eq H(A)$ 
is called the {\em support\/} of $A$. Since $\calc$ is braided, its Picard group
$\PicC$ is abelian and, as a consequence, the support $H$ is abelian as well.
Indeed, the simple objects appearing in $A$ are even further restricted:
\\[-2.2em]

\dtl{Lemma}{lem:schel-dim}
For any Schellekens algebra $A \Cong \bigoplus_{h \in H} L_h$ in
an additive ribbon category \C\ we have:
   \def\leftmargini{1.8em}\begin{itemize}\addtolength\itemsep{-3pt}
\item[(i)] 
$\dim(L_h)\eq 1$ (rather than ${-}1$) for all $h \iN H$.
\item[(ii)] 
The twist $\theta_h\,{:=}\, \theta_{U_h}$ obeys 
$(\theta_h)^{N_h} \eq {\rm id}_{L_h}$ for $N_h$ the order of $L_h$.
\\[2pt]
(Invertible objects obeying this relation are said to be in the
`effective center' of $\calc$.)
\end{itemize}

\medskip\noindent
Proof:\\
(i)~\,Since by definition $A$ is simple as a left module over itself, it is in 
particular simple as an algebra, i.e.\ is a simple $A$-$A$-bimodule. Also, it 
follows from the defining properties of a symmetric special Frobenius algebra
that $\dim(A) \,{\ne}\, 0$, see section \ref{sec:chiral}. On the other hand, 
$\dim(\cdot)$ is a group homomorphism from $H$ to $\zet_2$. Thus $\sum_{h\in H} 
\dim(L_h) \eq \dim(A) \,{\ne}\, 0$ implies that $\dim(\cdot)$ is trivial on $H$.
\\[3pt]
(ii) is Proposition 3.14 of \cite{tft3}.
\qed

One and the same object of the form $A \Cong \bigoplus_{h \in H} L_h$ can, 
in general,
be endowed with several non-isomorphic structures of a symmetric special
Frobenius algebra. A convenient concept for classifying non-isomorphic
algebra structures is the Kreuzer-Schellekens bihomomorphism:
\\[-2.2em]

\dtl{Definition}{def:ksb}
Given a subgroup $H$ of $\PicC$ that is contained in the effective center,
a {\em Kreuzer-Schel\-le\-kens bihomomorphism\/} (or {\em KSB\/}, for short) 
on $H$ is a bihomomorphism
  \be
  \Xi:\quad H\Times H \to \complex^\times
  \ee
which on the diagonal coincides with the quadratic form on $H$ that is given by
the twist, i.e.\ $\theta_{L_g} \eq \Xi(g,g)\, {\rm id}_{L_g}$
for all $g\iN H$.

\medskip

Let us fix monomorphisms $e_g{:}\ L_g \To A$ as auxiliary data. One can show 
(proposition 3.20 of \cite{tft3}) that for any Schellekens algebra $A$ with 
multiplication $m$ the equation
  \be
  m\circ c_{L_g,L_h} \circ (e_g\oti e_h) =: \Xi_A(h,g)\, m \circ (e_g\oti e_h) 
  \labl{Xidef}
defines a KSB $\Xi_A$ on the support $H$ of $A$. This KSB is independent of the 
choice of morphisms $e_h$ and characterises a Schellekens algebra up to 
isomorphism. Conversely, every 
KSB appears in the description of some Schellekens algebra. In 
brief, the KSB plays the role for the classification of Schellekens algebras 
that is played by alternating bihomomorphisms for the classification of 
twisted group algebras of abelian groups.

\medskip

The main goal of this section is to extract as much information as possible 
about the bimodules of a Schellekens algebra $A$ from the support $H(A)$ and the 
KSB $\Xi_A$. As it turns out, to this end we need to know the automorphism 
group of $A$; this group is determined in the next subsection.


\subsection{Automorphisms of Schellekens algebras}

The endomorphisms of $A$ as an object of $\calc$ are of the form
  \be
  t_\psi := \bigoplus_{h\in H} \psi(h)\, \id_{L_h} \,,
  \labl{eq:schellekens-auto}
where $\psi{:}\ H\To \complex$ is an arbitrary function. Obviously,
  \be
  t_{\psi_1} \circ t_{\psi_2} = t_{\psi_1\cdot\psi_2} \, ,
  \ee
where the product of functions on $H$ is defined by pointwise multiplication.

\dtl{Lemma}{lem:schellekens-auto}
The endomorphism $t_\psi$ of $A$ is an algebra automorphism if and only
if $\psi$ is a character of $H$. This identifies canonically the
character group of $H$ with $\Aut(A)$.

\medskip\noindent
Proof:\\
The statement follows from the fact that for a Schellekens algebra the function
$\mu$ defined by
$m\cir (e_g\oti e_h)\eq \mu(g,h)\, e_{gh}$ is nonzero for all $g,h\iN H$.
\qed

\medskip

For any simple object of \C\ we set
  \be
  \chii_U(h) := \frac{s_{U,L_h}^{}}{s_{U,\one}^{}} \,.
  \ee
As shown in \cite{tft3} (see definition 3.24 and proposition 3.26 of 
\cite{tft3}), this furnishes a character $\chii_U$ on $H$, and it is 
just the exponentiated negative {\em monodromy charge\/} of $U$, 
  \be
  \chii_U(h) 
  = \frac{\theta_{L_h \oti U}}{\theta_{L_h}\, \theta_{U}} \,. 
  \ee
Furthermore,
  \be
  c_{U,L_g}^{} = \chii_U(g)\, c_{L_g,U}^{\,-1} \qquad \text{and} \qquad
  \chii_{U^\vee}^{}(g) = \chii_U(g^{-1}) \,.
  \labl{c=chi.c-1}
For brevity, we denote the algebra automorphism of $A$ corresponding to the 
character $\chii_U$ by $t_U$, i.e.\ write $t_U^{}\,{:=}\,t_{\chi_U^{}}$.
 
These considerations immediately imply the following result (which
justifies the term monodromy charge).
\\[-2.2em]

\dtl{Lemma}{c+dim.c-1}
Let $A$ be a Schellekens algebra and $U$ be a simple object of $\calc$. Then
\\[.3em]
(i)~~$c_{U,A}^{} = c_{A,U}^{\,-1} \cir (\id_U \oti t_U)$\,.
\\[.3em]
(ii)~\,$c_{A,U}^{} = c_{U,A}^{\,-1} \cir (t_U \oti \id_U)$\,.

\medskip \noindent
Proof: \\
For (i) note that
expanding the braiding morphisms in direct sums, \erf{c=chi.c-1} implies that
  \be
  c_{U,A}^{} = \bigoplus_{h\in H} c_{U,L_h}^{}
  = \bigoplus_{h\in H} \chii_U(h)\, c_{L_h,U}^{-1}
  = c_{U,A}^{-1} \cir (t_U \oti \id_U) \,.
  \ee
(ii) is seen similarly.  
\qed

\medskip

We close this subsection with the
\\[-2.3em]

\dtl{Lemma}{lem:schel-psi-bubble}
Let $\calc$ be a ribbon category, $A$ a Schellekens algebra in $\calc$,
and $t_\psi \iN \Aut(A)$. Then
  \be
  m \cir (\id_A \oti t_\psi) \cir \Delta = \delta_{\psi,e} \,\id_A
  = m \cir (t_\psi \oti \id_A) \cir \Delta \,.
  \labl{eq:tr-psi}
Proof:\\
First note that $\tr(t_\psi) \eq \sum_{h \in H} \psi(h) \dim(L_h)$. Now 
according to lemma \ref{lem:schel-dim} we have $\dim(L_h)\eq1$, so that
$\tr(t_\psi) \eq \sum_{h \in H} \psi(h) \eq |H|\, \delta_{\psi,e}$.
Together with the fact that $A$ is a symmetric special Frobenius algebra
it follows that
  \bea \begin{picture}(420,80)(0,58)
  \put(0,0)      {\Includeourtinynicepicture 52a }
  \put(95,0)     {\Includeourtinynicepicture 52b }
  \put(240,0)    {\Includeourtinynicepicture 52c }
  \put(392,0)    {\Includeourtinynicepicture 52d }
  \put(28,61.8)  {\scriptsize$t_\psi$}
  \put(140.6,48.8){\scriptsize$t_\psi$}
  \put(284.4,48.8){\scriptsize$t_\psi$}
  \put(63,62)    {$=$}
  \put(173,62)   {$=~\displaystyle\frac1{\dim(A)}$}
  \put(317,62)   {$=~\displaystyle\frac{{\rm tr}(t_\psi)}{\dim(A)}$}
  \epicture33 \labl{pic-ffrs5-52}
Because of $\dim(A) \eq |H|$ this implies the first
equality in \erf{eq:tr-psi}. The second equality can be seen in the same way.
\qed


\subsection{Bimodules of Schellekens algebras}\label{sec:bischel}

To enter the discussion of bimodules, we observe that both the left and the 
right action of $A$ on an $A$-$A$-bimodule can be twisted by the action of 
automorphisms of $A$ (see section 9 of \cite{fuRs11}):
\\[-2.2em]

\dtl{Definition}{def:schel-ind-bimod}
Let $\calc$ be a tensor category and $A$ an algebra 
in \C. For $X \eq (\dot X,\rho_l, \rho_r)$ an $A$-$A$-bimodule and
$t,t' \iN  \Aut(A)$ we denote by ${}_t X_{t'}$ the $A$-$A$-bimodule 
  \be
  \big(\dot X, \rho_l \cir( t \oti \id_{\dot X} ) ,
  \rho_r \cir( \id_{\dot X} \oti t' ) \big) \,.
  \ee

In the case of a Schellekens algebra $A$ with support $H$ we write
${}_\varphi X_\psi \,{:=}\, {}_{t_\varphi}X_{t_\psi}$ for 
$\varphi,\psi \iN H^*$, i.e.
  \be
  {}_\varphi X_\psi =
  \big(\dot X, \rho_l \cir( t_\varphi \oti \id_{\dot X} ) ,
  \rho_r \cir( \id_{\dot X} \oti t_\psi ) \big) \,.
  \ee
We also use the abbreviation $X_\psi \equiv {}_{\idsmall} X_\psi$. 

\medskip

Next we assume that \C\ is modular. We can then
show that every bimodule over a Schellekens algebra
can be recovered as a submodule of a twisted $\alpha$-induced bimodule:
\\[-2.2em]

\dtl{Proposition}{prop:schel-submod}
Let $A$ be a Schellekens algebra in modular tensor category \C. Then every 
simple $A$-$A$-bimodule is a submodule of $\alpha_\AA^+(U)_\psi$ for some 
simple object $U$ of \C\ and some automorphism $t_\psi \iN \Aut(A)$.

\medskip\noindent
Proof:\\
Let $X$ be a simple $A$-$A$-bimodule. Since a Schellekens algebra is
in particular simple, by proposition \ref{prop:UiAUj-submod} there are 
simple objects $U$ and $V$ of $\calc$ such that $X$ is a sub-bimodule
of $U \otiP A \otim V$. We claim that for a Schellekens algebra one has
  \be
  U \otiP A \otim V \cong\, \alpha_\AA^+(U \Oti V)_{\psi} 
  \ee
as $A$-$A$-bimodules,
with $\psi\iN H^*$ the character for which $t_\psi$
is the automorphism \erf{eq:schellekens-auto}, $t_\psi \eq t_V^{\,-1}$. 
An isomorphism is provided by
  \be
  f = c_{U,A}^{} \otimes \id_V .
  \ee
To see that $f$ is an isomorphism of bimodules, note that
  \bea \begin{picture}(420,98)(9,57)
  \put(0,0)      {\Includeourbeautifulpicture 72a }
  \put(93,0)     {\Includeourbeautifulpicture 72b }
  \put(227,0)    {\Includeourbeautifulpicture 72c }
  \put(360,0)    {\Includeourbeautifulpicture 72d }
  \put(-1,-8)    {\scriptsize$A$}
  \put(11,-7)    {\scriptsize$U{\otimes^+}\!A$}
  \put(17.5,-13.7){\scriptsize${\otimes^-}V$}
  \put(44,-8)    {\scriptsize$A$}
  \put(7,149)    {\scriptsize$\alpha_{\!A}^+(U{\otimes}V)_{\psi}$}
  \put(31,90)    {\scriptsize$f$}
  \put(90,-8)    {\scriptsize$A$}
  \put(115,-8)   {\scriptsize$U$}
  \put(115,149)  {\scriptsize$A$}
  \put(137,-8)   {\scriptsize$A$}
  \put(137,149)  {\scriptsize$U$}
  \put(161,-8)   {\scriptsize$V$}
  \put(161,149)  {\scriptsize$V$}
  \put(183,-8)   {\scriptsize$A$}
  \put(225,-8)   {\scriptsize$A$}
  \put(250,-8)   {\scriptsize$U$}
  \put(249,149)  {\scriptsize$A$}
  \put(271,-8)   {\scriptsize$A$}
  \put(272,149)  {\scriptsize$U$}
  \put(295,-8)   {\scriptsize$V$}
  \put(295.7,149){\scriptsize$V$}
  \put(318,-8)   {\scriptsize$A$}
  \put(316.4,30) {\tiny$t_V^{-1}$}
  \put(359,-8)   {\scriptsize$A$}
  \put(365,149)  {\scriptsize$\alpha_{\!A}^+(U{\otimes}V)_{\psi}$}
  \put(390,42)   {\scriptsize$f$}
  \put(371,-7)    {\scriptsize$U{\otimes^+}\!A$}
  \put(377.5,-13.7)  {\scriptsize${\otimes^-}V$}
  \put(404,-8)   {\scriptsize$A$}
  \put(72,71)    {$=$}
  \put(206,71)   {$=$}
  \put(339,71)   {$=$}
  \epicture42 \labl{pic-ffrs5-72}
where in the second step lemma \ref{c=chi.c-1}\,(i) is used.
Thus $X$ is also a sub-bimodule of $\alpha_\AA^+(U \Oti V)_{\psi}$.
If $U \Oti V$ is not simple, then this bimodule is a direct sum
of bimodules of the form $\alpha_\AA^+(U_i)_{\psi}$ with $U_i$ simple,
and $X$ is contained in at least one of the summands.
\qed

\medskip

This result implies immediately that any simple $A$-$A$-bimodule, considered
as an object of \C, is a direct sum of simple objects of $\calc$ whose classes
in $K_0(\calc)$ are on the same $H$-orbit. Thus the problem to classify simple 
$A$-$A$-bimodules of a Schellekens algebra decomposes into separate 
problems for each $H$-orbit.

\medskip

The situation simplifies when one considers $H$-orbits of simple objects in
$\calc$ on whose class in $K_0(\calc)$ the support $H$ of $A$ acts freely by 
fusion. Objects with this property have been termed non-fixed points in the 
physics literature. We call such a simple object $H$-{\em torsorial\/}, and 
the orbit of its 
class in $K_0(\calc)$ a torsorial $H$-orbit. Note that invertible objects are 
in particular $H$-torsorial for any subgroup $H$ of the Picard group \PicC.

As shown in section 5 of \cite{tft3}, the algebraic structure that controls 
the decomposition of the $A$-$A$-bimodule $A\oti U\oti A$ is a twisted group
algebra. If $U$ is $H$-torsorial, then the underlying group is given by
  \be
  {\mathcal S}_{AA}(U) = \{(h,h^{-1}) \,|\, h\iN H \} \cong H \,.
  \ee
The commutator two-cocycle describing the twist is, according to proposition 
5.4 of \cite{tft3},
  \be\bearll
  \eps(g,g^{-1},h,h^{-1}) \!\! &= \phi_U(1,1)\, \beta(h,g^{-1})\, \Xi_A(h,g)\, 
  \Xi_A(g^{-1}\!,h^{-1})
  \\{}\\[-.8em] &= \beta(h,g^{-1})\, \Xi_A(h,g)\, \Xi_A(g,h) = 1 \,,
  \eear\ee
with $\beta$ defined by $c_{L_g,L_h} \eq \beta(g,h)\,c_{L_h,L_g}^{-1}$.
Thus the group algebra is in fact untwisted. 

We conclude that there are exactly $|H|$ simple $A$-$A$-bimodules that are
direct sums of simple elements on the $H$-orbit of a given torsorial simple 
object $U$. On the other hand, proposition \ref{prop:iso-bimod-2} below provides 
$|H|$ mutually non-isomorphic $A$-$A$-bimodules on this orbit, namely the 
bimodules $\alpha_\AA^+(U)_\varphi$ for $\varphi \iN H^*$.  
Furthermore, applying Frobenius reciprocity in the form
  \be
  \HomA({\rm Ind}_\AA(U),{\rm Ind}_\AA(U)) \cong \Hom_\calc(U,A\oti U) 
  = \bigoplus_{g\in H} \Hom(U,L_g\oti U) = \Hom(U,U)
  \labl{short-frobenius}
shows that the bimodules $\alpha^\pm_\AA(U)$ induced from $H$-torsorial simple
objects $U$ are simple as left $A$-modules and thus, a fortiori, simple as 
$A$-$A$-bimodules. Thus the $\alpha_\AA^+(U)_\varphi$ represent the $|H|$ 
isomorphism classes of simple $A$-$A$-bimodules that can be constructed 
using objects on the $H$-orbit of $U$. Note that for this argument it is not 
required that the category is modular.
\\[-2.2em]

\dtl{Proposition}{prop:iso-bimod-1}
For $A$ an algebra in a braided tensor category \C, 
twists by automorphisms of $A$ can be shifted from the 
left to the right action of $A$ on $\alpha$-induced bimodules, and vice versa:
  \be
  {}_{\varphi^{-1}\psi_1} \alpha_\AA^\pm(U)_{\psi_2} \Cong 
  {}_{\psi_1} \alpha_\AA^\pm(U)_{\varphi\psi_2}
  \ee
for any object $U$ of \C.

\medskip\noindent
Proof: \\[2pt]
The claim follows by verifying, similarly as in the proof of
lemma 6\,(i) of \cite{fuRs11}, that $t_\varphi\oti\id_U$ furnishes
an isomorphism between the two bimodules.
\qed

\dtl{Proposition}{prop:iso-bimod-2}
Let $A$ be a Schellekens algebra in an additive
ribbon category \C\ and $U$ a simple object of \C.
   \\[-1.5em]~
   \def\leftmargini{1.7em}\begin{itemize}\addtolength\itemsep{-3pt}
\item[(i)] 
One can work with one type of $\alpha$-induction only:
$_{\psi_1} \alpha_\AA^+(U)_{\psi_2} \Cong 
{}_{\psi_1} \alpha_\AA^-(U)_{\chi_U^{}\psi_2} $. 
\item[(ii)] 
If $U$ is $H$-torsorial, then one has $\alpha_\AA^+(U)_\varphi \Cong 
\alpha_\AA^+(U)_\psi$ as bimodules if and only if $\varphi \eq \psi$.
\end{itemize} 

\medskip\noindent
Proof: \\[2pt]
(i)~~follows immediately from lemma \ref{c+dim.c-1}.
\\[2pt]
(ii)~Because of proposition \ref{prop:iso-bimod-1} it suffices to establish 
the claim for $\varphi \,{\equiv}\, 1$. By the definition of $\alpha$-induction
(see formula \erf{eq:UxXxV}, remark \ref{rem:alpha-ind}\,(ii) and
definition \ref{def:schel-ind-bimod}),
we have $\alpha_\AA^+(U)_\psi \eq A_\psi \otip U$. Suppose now there exists
an isomorphism $f \iN \HomAA(A \otip U,A_\psi \otip U)$, and consider 
its partial trace
  \be
  g := (\id_A \oti \tilde d_U) \cir (f \oti \id_{U^\vee}) \cir (\id_A \oti b_U)
  \in \Hom(A,A) \,.
  \labl{g-from-f}
Since $U$ is $H$-torsorial, $f$ can be written as $f \eq \sum_{h \in H} f_h\, 
\id_{L_h} \oti \id_U$, with all of the coefficients $f_h \iN \complex$ being 
nonzero. Substituting this decomposition of $f$ gives
$g \eq \dim(U) \sum_{h\in H}f_h\,\id_{L_h}$. Since $U$ is simple, $\dim(U)\nE0$,
and since the $f_h$ are nonzero, we have $g \nE 0$.
It is also easy to check that in fact $g \iN \HomAA(A,A_\psi)$. Since
$A$ and $A_\psi$ are simple and $g$ is nonzero, $g$ is an isomorphism.
Lemma~6\,(iv) of \cite{fuRs11} now implies that $t_\psi$ is an
inner automorphism of $A$, $t_\psi \iN \text{Inn}(A)$ (for the definition of 
the notion of inner automorphism, see section 8 of \cite{fuRs11}). However, for a 
Schellekens algebra one has $\text{Inn}(A) \eq \{ \id_A \}$, and 
thus $t_\psi \eq \id_A$. 
\\[.2em]
The converse direction of the claim is trivial.
\qed

By a calculation generalising \erf{short-frobenius} one shows that 
the bimodules $_{\psi_1} \alpha_\AA^+(U\oti L_h)_{\psi_2}$ and 
$_{\psi_1} \alpha_\AA^+(U)_{\psi_2'}$ are isomorphic as left $A$-modules for 
any $h\iN H$ and for every choice of $\psi_2,\psi_2'\iN H^*$. The following 
result establishes equivalences of $A$-$A$-bimodules between bimodules of 
this type.
\\[-2.3em]

\dtl{Proposition}{prop:schel-iso-subtle}
Let $A$ be a Schellekens algebra in an additive ribbon 
category, let $H$ be the support of $A$, and let $U$ be an $H$-torsorial 
simple object. Then for any $\varphi,\psi\iN H^*$ we have the isomorphism 
  \be
  {}_{\psi} \alpha_\AA^+(L_h\oti U)_{\varphi} \cong
  {}_{\psi} \alpha_\AA^+(U)_{\Xi_A(\cdot,h)\varphi} 
  \ee
of $A$-$A$-bimodules.

\medskip\noindent
Proof: \\
First note that, owing to the fact that $\alpha$-induction is a tensor functor, 
the assertion can be reduced to the case $U\eq\one$:
  \be\bearll
  {}_\psi\alpha_\AA^+(L_h\oti U)_\varphi \!\!
  &\cong {}_\psi\alpha_\AA^+(L_h) \ota \alpha_\AA^+(U)_\varphi
  \\{}\\[-.7em]
  &\cong {}_\psi\alpha_\AA^+(\one)_{\Xi_A(\cdot,h)} \ota \alpha_\AA^+(U)_\varphi
  \\{}\\[-.7em]
  &\cong {}_{\psi\Xi_A(\cdot,h)^{-1}}\alpha_\AA^+(\one) \ota \alpha_\AA^+(U)_\varphi
  \cong {}_\psi\alpha_\AA^+(U)_{\Xi_A(\cdot,h)\varphi} \,.
  \eear\ee 
To establish the isomorphism for $U\eq\one$, we first show that the
two morphisms
  \be\bearll
  s_h := (\id_A\oti r_h) \cir \Delta \in \Hom(A,A\oti L_h) & {\rm and}
  \\{}\\[-.6em]
  f_h := \dim(A)\, m \cir (\id_A \oti e_h) \in \Hom(A\oti L_h,A) \,,
  \eear\labl{sh,fh}
with $e_h$ and $r_h$ embedding and restriction morphisms for $L_h$ as a
retract of $A$, are each other's inverse. That $s_h$ is left-inverse to $f_h$
follows by noting that after applying the Frobenius property to $s_h\cir f_h$ 
one can insert an idempotent $p_\one$ in the intermediate $A$-line, and then 
using that
  \be
  p_\one \equiv r_\one\circ e_\one = \frac1{\dim(A)}\, \eta\cir\eps \,,
  \ee
which, in turn, follows by using that $\Hom(\one,A) \Cong \complex$ and 
composing with the counit $\eps$. That $s_h$ is also a right-inverse follows 
by noting that $f_h \cir s_h$ is $\dim(A)$ times the left hand side of
\erf{pic-ffrs5-52} with $t_\psi$ replaced by $p_h \eq r_h \cir e_h$, 
and using also $\tr(p_h)\eq1$.
\\[2pt]
Next we observe that by combining \erf{Xidef} with the Frobenius property
it follows that
  \begin{eqnarray}\begin{picture}(415,96)
  \put(-20,0)    {\Includeourbeautifulpicture 87a }
  \put(141,0)    {\Includeourbeautifulpicture 87b }
  \put(297,0)    {\Includeourbeautifulpicture 87c }
  \put(426,0)    {\Includeourbeautifulpicture 87d }
  \put(-15.5,-8) {\scriptsize$L_g$}
  \put(-13.5,94) {\scriptsize$A$}
  \put(3.4,-8)   {\scriptsize$L_{g'}$}
  \put(3,94)     {\scriptsize$L_h$}
 \put(161,0) {
  \put(-15.5,-8) {\scriptsize$L_g$}
  \put(-13.5,94) {\scriptsize$A$}
  \put(3.4,-8)   {\scriptsize$L_{g'}$}
  \put(3,94)     {\scriptsize$L_h$}
 }
 \put(313,0) {
  \put(-15.5,-8) {\scriptsize$L_g$}
  \put(-13.5,94) {\scriptsize$A$}
  \put(3.4,-8)   {\scriptsize$L_{g'}$}
  \put(3,94)     {\scriptsize$L_h$}
 }
 \put(442,0) {
  \put(-15.5,-8) {\scriptsize$L_g$}
  \put(-13.5,94) {\scriptsize$A$}
  \put(3.4,-8)   {\scriptsize$L_{g'}$}
  \put(3,94)     {\scriptsize$L_h$}
 }
  \put(35,45)    {$=~~\Xi_A(g',gh^{-1})_{}^{-1}$}
  \put(195,45)   {$=~~\Xi_A(g',gh^{-1})_{}^{-1}$}
  \put(348,45)   {$=~~\Xi_A(g',h)$}
  \end{picture}
  \nonumber\\[8pt]~\label{pic-ffrs5-87}
  \end{eqnarray}
where the factor on the right hand side arises as the product
$\Xi_A(g',gh^{-1})_{}^{-1}\,\Xi_A(g',g)$. This shows that
$s_h$ is an isomorphism of bimodules from 
${}_\psi A_{\Xi_A(\cdot,h)\varphi} \eq 
{}_\psi\alpha_\AA^+(\one)_{\Xi_A(\cdot,h)\varphi}$ to
${}_\psi\alpha_\AA^+(L_h)_\varphi$.
\qed

\medskip

Another special situation of interest in which we can make more specific
statements are Schellekens algebras that are braided commutative, i.e.\
obey $m\cir c_{A,A} \eq m$. This class is also of particular practical interest,
since it describes extensions of the chiral algebra and is therefore used to 
implement projections on conformal field theories, e.g.\ \cite{fusw} in string 
theory the GSO-projection and the alignment of fermionic boundary conditions in 
various sectors. In this case the following results hold also for simple 
objects which are not $H$-torsorial.
\\[-2.2em]

\dtl{Theorem}{thm:schel-comm-bimod}
Let $A$ be a commutative Schellekens algebra in a modular tensor category \C, 
and let $M_\kappa$, $\kappa \iN \JJ_A$, be 
representatives of the isomorphism classes of simple left $A$-modules. 
Then the isomorphism classes of simple $A$-$A$-bimodules can be labelled by 
pairs consisting of an element of $\JJ_A$ and an algebra automorphism:
  \be
  \KK_{AA} = \{ (\kappa,\psi) \,|\, \kappa \iN \JJ_A,\, \psi \iN \Aut(A) \} \,.
  \ee
A representative $M_{\kappa,\psi}$ of $(\kappa,\psi)$ is given by the simple 
left module $M_\kappa$ with right action
  \bea \begin{picture}(80,60)(0,42)
  \put(20,0)     {\Includeourtinynicepicture 51{} }
  \put(50,-8)  {\scriptsize$A$}
  \put(51,26.5){\scriptsize$\psi$}
  \put(24,-9)  {\scriptsize$\dot M_\kappa$}
  \put(24,99.7){\scriptsize$\dot M_\kappa$}
  \epicture24 \labl{pic-ffrs5-51}
In particular, all simple $A$-$A$-bimodules are already simple as left 
$A$-modules, and the equality $|\KK_{AA}| \eq \dim(A) \, |\JJ_A|$ holds.

\medskip\noindent
Proof:\\
First note that in order to verify that $M_{\kappa,\psi}$ is indeed a bimodule 
one uses commutativity of $A$. All $M_{\kappa,\psi}$ are simple as bimodules 
because they are already simple as left modules. Let now $X$ be a simple 
$A$-$A$-bimodule. Then by proposition \ref{prop:schel-submod}, $X$ is a
sub-bimodule of $\alpha^+_\AA(U)_\psi$ for some $U\iN\objc$ and some 
$\psi\iN\Aut(A)$; we denote by $e\iN \Hom(X,\alpha^+_\AA(U)_\psi)$ the 
corresponding embedding morphism.
\\[.2em]
Let further $f_\kappa\iN\HomA(\inda(U),M_\kappa)$ be a morphism of $A$-modules. 
Because of
  \bea \begin{picture}(340,77)(0,50)
  \put(20,0)     {\Includeourtinynicepicture 57a }
  \put(140,0)    {\Includeourtinynicepicture 57b }
  \put(265,0)    {\Includeourtinynicepicture 57c }
  \put(20,-8)    {\scriptsize$A$}
  \put(40.5,-8)  {\scriptsize$U$}
  \put(60,-8)    {\scriptsize$A$}
  \put(61,20.5)  {\scriptsize$\psi$}
  \put(30,78.8)  {\scriptsize$f_\kappa$}
  \put(30,125)   {\scriptsize$\dot M_\kappa$}
  \put(149,-8)   {\scriptsize$A$}
  \put(171,-8)   {\scriptsize$U$}
  \put(190,-8)   {\scriptsize$A$}
  \put(190,20.5) {\scriptsize$\psi$}
  \put(159,78.8) {\scriptsize$f_\kappa$}
  \put(159,125)  {\scriptsize$\dot M_\kappa$}
  \put(265,-8)   {\scriptsize$A$}
  \put(286,-8)   {\scriptsize$U$}
  \put(305,-8)   {\scriptsize$A$}
  \put(306,52.5) {\scriptsize$\psi$}
  \put(274,28)   {\scriptsize$f_\kappa$}
  \put(274,125)  {\scriptsize$\dot M_\kappa$}
  \put(100,56)   {$=$}
  \put(228,56)   {$=$}
  \epicture31 \labl{pic-ffrs5-57}
$f_\kappa$ is also a morphism of $A$-$A$-bimodules from
$\alpha^+_\AA(U)_\psi$ to $M_{\kappa,\psi}$; here it is also used that $A$ is 
commutative. Since $e$ is nonzero and since $\inda(U)$ can be written as a 
direct sum of simple modules, there exists an $f_\kappa$ such that $e \cir 
f_\kappa$ is nonzero. Thus $e \cir f_\kappa\iN \Hom(X,M_{\kappa,\psi})$ 
is a nonzero morphism of bimodules. Since both $X$ and
$M_{\kappa,\psi}$ are simple, it is thus an isomorphism. Hence every 
simple $A$-$A$-bimodule is isomorphic to one of the $M_{\kappa,\psi}$.
\\[.2em] 
Suppose now that for some choice of $\kappa,\mu \iN \KK_{AA}$ and of
$\psi,\varphi \iN \Aut(A)$ there is an isomorphism $f\iN \HomAA(M_{\kappa,\psi}, 
M_{\mu,\varphi})$. Then $f$ is in particular an isomorphism of $A$-left modules,
so that $\kappa\eq\mu$. Furthermore,
  \begin{eqnarray}
  \begin{picture}(420,170)(0,0)
  \put(20,0)     {\Includeourtinynicepicture 53a }
  \put(110,0)    {\Includeourtinynicepicture 53b }
  \put(220,0)    {\Includeourtinynicepicture 53c }
  \put(330,0)    {\Includeourtinynicepicture 53d }
  \put(22,-8)    {\scriptsize$\dot M_{\kappa,\psi}$}
  \put(22,168)   {\scriptsize$\dot M_{\kappa,\varphi}$}
  \put(27,88)    {\scriptsize$f$}
  \put(112,-8)   {\scriptsize$\dot M_{\kappa,\psi}$}
  \put(112,168)  {\scriptsize$\dot M_{\kappa,\varphi}$}
  \put(117,127)  {\scriptsize$f$}
  \put(151.6,67) {\scriptsize$A$}
  \put(222,-8)   {\scriptsize$\dot M_{\kappa,\psi}$}
  \put(222,168)  {\scriptsize$\dot M_{\kappa,\varphi}$}
  \put(227,84)   {\scriptsize$f$}
  \put(261.6,67) {\scriptsize$A$}
  \put(332,-8)   {\scriptsize$\dot M_{\kappa,\psi}$}
  \put(332,168)  {\scriptsize$\dot M_{\kappa,\varphi}$}
  \put(337,110)  {\scriptsize$f$}
  \put(379.5,77) {\scriptsize$A$}
  \put(353,53.5) {\scriptsize$\psi$}
  \put(374,105)  {\scriptsize$\varphi$}
  \put(70,82)    {$=$}
  \put(181,82)   {$=$}
  \put(291,82)   {$=$}
  \end{picture}
  \nonumber\end{eqnarray}
  \begin{eqnarray}
  \begin{picture}(420,179)(0,0)
  \put(60,0)     {\Includeourtinynicepicture 53e }
  \put(200,0)    {\Includeourtinynicepicture 53f }
  \put(370,0)    {\Includeourtinynicepicture 53a }
  \put(62,-8)    {\scriptsize$\dot M_{\kappa,\psi}$}
  \put(62,168)   {\scriptsize$\dot M_{\kappa,\varphi}$}
  \put(67,140)   {\scriptsize$f$}
  \put(101,54.5) {\tiny$\psi^{-1}$}
  \put(104.6,77) {\scriptsize$\varphi$}
  \put(103,38)   {\scriptsize$A$}
  \put(204,-8)   {\scriptsize$\dot M_{\kappa,\psi}$}
  \put(204,168)  {\scriptsize$\dot M_{\kappa,\varphi}$}
  \put(209,129.8){\scriptsize$f$}
  \put(248,51)   {\tiny$\varphi\,\psi^{-1}$}
  \put(252,34.5) {\scriptsize$A$}
  \put(373,-8)   {\scriptsize$\dot M_{\kappa,\psi}$}
  \put(373,168)  {\scriptsize$\dot M_{\kappa,\varphi}$}
  \put(378,88)   {\scriptsize$f$}
  \put(20,82)    {$=$}
  \put(150,82)   {$=$}
  \put(291,82)   {$=~ \delta_{\varphi\psi^{-1},\idsmall}$}
  \end{picture}  
  \nonumber\\[-1em]~ \label{pic-ffrs5-53}
  \end{eqnarray} 
(the last step uses lemma \ref{lem:schel-psi-bubble}), showing that 
$\psi \eq \varphi$.  
\\
Thus the $M_{\kappa,\psi}$ are pairwise non-isomorphic.
\qed

\dtl{Remark}{rem:schel-comm-bimod} 
The statement does not apply to non-commutative algebras. For example, denote 
by $U_l$ the simple object given by the integrable highest weight representation
of $A_1^{(1)}$ with highest weight $l$ at level $k$. In the theory described by 
the D-type modular invariant of the WZW model based on $A_1^{(1)}$ at level 
$k \eq 2\,{\bmod}\, 4$, the object $U_{k/2}$ carries two non-isomorphic 
structures $U_{k/2}^+$ and $U_{k/2}^-$ of (simple) left module, and there is a 
simple bimodule structure defined on $U_{k/2}\,{\oplus}\, U_{k/2}$, which as a 
left module decomposes as $U_{k/2}^+\,{\oplus}\,U_{k/2}^-$.  Also the sum rule 
$|\KK_{AA}| \eq \dim(A) \, |\JJ_A|$ does not hold any longer.

\bigskip

As a generalisation of lemma 6\,(ii) of \cite{fuRs11}
the following statement will be useful,
\\[-2.3em]

\dtl{Lemma}{lem:schel-tens}
Let $A$ be an algebra in a braided tensor category \C. Then 
for $U,V \iN \objc$ and $\psi,\varphi \iN \Aut(A)$ we have
  \be
  \alpha_\AA^+(U)_\psi \ota \alpha_\AA^+(V)_\varphi
  \cong \alpha_\AA^+(U \Oti V)_{\psi\varphi}
  \ee
as $A$-$A$-bimodules.

\medskip\noindent
Proof:\\
First we use proposition \ref{prop:iso-bimod-1} to consider instead of
$\alpha_\AA^+(U)_\psi$ the isomorphic bimodule ${}_{\psi^{-1}}\alpha_\AA^+(U)$.
Then we use that, since $\alpha$-induction is a tensor functor, we have
$\alpha_\AA^+(U) \ota \alpha_\AA^+(V) \Cong \alpha_\AA^+(U \Oti V)$,
and thereby automatically also 
${}_{\psi^{-1}}\alpha_\AA^+(U) \ota \alpha_\AA^+(V)_\varphi \Cong
{}_{\psi^{-1}}\alpha_\AA^+(U \Oti V)_{\varphi}$. The claim then follows 
by invoking once more proposition \ref{prop:iso-bimod-1}.
\qed


\subsection{The generic symmetry group}\label{sec:SymGen}

For any full local \cft\ described by a Schellekens algebra $A$, the results
of the previous subsection allow us to determine a subgroup of the symmetry
group of that \cft: those given by group-like defects that are induced from 
invertible objects of \C. We call this group the {\em generic group of 
symmetries\/} and denote it by \SymGen. It can be described as follows.
\\[-2.3em]

\dtl{Proposition}{prop:gen-sym}
The generic group of symmetries of a \cft\ corresponding to
a Schellekens algebra $A$ with support $H$ and KSB $\Xi_A$ is
  \be
  \SymGen = H^* \times_H \PicC \,.
  \ee
Here $H$ acts on \PicC\ by (left) multiplication, and via $\psi^h(\cdot) \eq 
\psi(\cdot)\, \Xi_A(\cdot,h)$ for $h\iN H$ from the right on $H^*$. 

\medskip\noindent
Proof:\\
The assertion is a special case of proposition \ref{prop:all-torsorial} below, 
to which we turn after the following remarks.
 
\dtl{Remark}{rem:gen-sym}
(i)~~As a quotient of an abelian group, $\SymGen$ is abelian.
    In other words: non-abelian symmetries come from resolved fixed points. 
    This is illustrated in the next section with the example of the Potts model.
\\[2pt]
(ii)~It is instructive to consider the special example of a Schellekens algebra
    with support $H\Cong\zet_2$.
    Two cases must be distinguished: the nontrivial invertible element $U_h$ 
    can have twist $\theta_h\eq{\pm}\id_{U_h}$. If $\theta_h\eq\id_{U_h}$, then
    $\Xi_A(h,h)\eq1$ and the KSB is trivial. As a consequence, $H$ acts trivially
    on $H^*$ and we get
  \be
  \SymGen \equiv H^* {\times_H}\, \PicC \cong H^* \times \PicC/H \,.
  \ee
Note that this is an abelian group of the same order as $\PicC$, but the two
groups are, in general, not isomorphic. For $\theta_h\eq{-}\id_{U_h}$, we have
$\Xi_A(h,h)\eq{-}1$ and the KSB is nontrivial. As a consequence, the action of
$H$ on $H^*$ removes that factor, so that
  \be
  \SymGen \equiv H^* {\times_H}\, \PicC \cong \PicC \,.
  \ee
This should not come as a surprise, since in this case $A$ is an Azumaya 
algebra, which implies that $\PicCAA \Cong \PicC$.
\\[2pt]
(iii)~The assertion in proposition \ref{prop:gen-sym}, which was announced in 
\cite{scfr} with a sketch of the proof, has been refined in theorem 4.5 of
\cite{naid}, where also the associator on the
subcategory whose objects correspond to elements of \SymGen\ is discussed.

\dtl{Proposition}{prop:all-torsorial}
Let \C\ be a semisimple ribbon category and let $A$ be a Schellekens algebra 
in \C\ with support $H$. If all simple objects of \C\ are $H$-torsorial, then
there is a ring isomorphism 
  \be
  K_0(\CAA) \cong \zet H^*_{} \otimes_{\zet H}^{} K_0(\C)
  \ee
which preserves the distinguished bases.

\medskip\noindent
Proof:\\
We will construct a surjective ring homomorphism
$f{:}\ \zet H^*_{} \,{\otimes_{\zet}^{}}\, K_0(\C) \To K_0(\CAA)$ and show that 
it descends to a ring isomorphism
$\tilde f{:}\ \zet H^*_{} \,{\otimes_{\zet H}^{}}\, K_0(\C) \To K_0(\CAA)$. 
For $X$ an object of \C\ (respectively, of \CAA), denote by $[X]$ its class in 
$K_0(\C)$ (respectively, in $K_0(\CAA)$). It is enough to define $f$ on pairs 
$(\psi,[U])$ with $U$ a simple object of \C. We set
$f(\psi,[U]) \,{:=}\, [ \alpha^+(U)_\psi ]$. As observed in
section \ref{sec:bischel} (see the arguments before proposition
\ref{prop:iso-bimod-1}), every simple $A$-$A$-bimodule is isomorphic
to a bimodule of the form $\alpha^+(U)_\psi$, so $f$ is surjective.
That $f$ is a ring homomorphism follows by direct calculation
using lemma \ref{lem:schel-tens}. Define a right action of $H$ on $H^*$ by
$\psi^h(\cdot) \,{:=}\, \psi(\cdot)\, \Xi_A(\cdot,h)$ for $h\iN H$, 
and a left action on $K_0(\C)$ by ${}^h[U] \,{:=}\, [L_h \oti U]$.
It is then an immediate consequence of proposition \ref{prop:schel-iso-subtle}
that $f(\psi^h,[U]) \eq f(\psi,{}^h[U])$ for all $h\iN H$. Thus
$f$ indeed gives rise to a well-defined surjective ring homomorphism 
$\tilde f{:}\ \zet H^*_{} \,{\otimes_{\zet H}^{}}\, K_0(\C)\To K_0(\CAA)$.
It remains to show that $\tilde f$ is injective. Let us denote the image
of $(\psi,[U])$ in $\zet H^*_{} \,{\otimes_{\zet H}^{}}\, K_0(\C)$ by 
$\{\psi,[U]\}$. Suppose that $\tilde f(\varphi,[U]) \eq \tilde f(\psi,[V])$ 
for simple objects $U$ and $V$ of \C. Then by definition
$\alpha^+(U)_\varphi \Cong \alpha^+(V)_\psi$; this is only possible
if $U$ and $V$ lie on the same $H$-orbit. We can thus use the action
of $H$ to find $\psi'$ such that $\{\psi',[U] \} \eq \{\psi,[V]\}$. 
By proposition \ref{prop:iso-bimod-2}\,(ii), the resulting equality 
$\tilde f(\varphi,[U]) \eq \tilde f(\psi',[U])$ then implies $\varphi\eq\psi'$
and thus $\{\varphi,[U]\} \eq \{\psi',[U]\} \eq \{\psi,[V]\}$.
Hence $\tilde f$ is injective.
\qed

\medskip

Let us add a few comments on the situation when the symmetry group \PicCAA\ 
contains non-generic elements, which is a prerequisite for having a nonabelian 
symmetry group. A non-generic element must appear as the class of a proper 
subobject of a twisted $\alpha$-induced bimodule $\alpha_\AA^+(U)_\psi$ for 
some simple object $U$ of \C. Now $\dim_{A|A}(\alpha^+(U)_\psi)\eq\dim(U)$, and
if $\alpha^+(U)_\psi$ is a direct sum of $N_U$ simple bimodules, then 
\cite{tft3} $N_U$ must be a divisor of $|H|$ and each of the simple 
sub-bimodules $S$ has the same dimension $\dim_{A|A}(S) \eq \dim(U)/{N_U}$. 
Since a group-like bimodule has dimension 1, this implies that non-generic 
symmetries can only come from non-$H$-torsorial simple objects of \C\ with 
low integral dimension. In all known classes of models such objects are rare 
(see e.g.\ \cite{jf18,garw}). Moreover, in many models that come in series, 
e.g.\ labelled by a ``level'', such as minimal or WZW models, the dimension of 
non-$H$-torsorial objects grows with the level. In all such cases non-abelian 
symmetries will be a low-level phenomenon. 
Also, of course, independently of whether there are non-$H$-torsorial objects or
not, the symmetry is abelian whenever the algebra $A$ is Azumaya, since then 
$\calc$ and \CAA\ are equivalent tensor categories.  


\sect{Tetracritical Ising and three-states Potts model}\label{sec:IsingPotts}

In this section we investigate topological defects and phase boundaries for 
the tetracritical Ising and for the critical three-states Potts model. For 
both models the chiral symmetry algebra contains the Virasoro vertex algebra 
of central charge $c\eq 4/5$, which is rational. Thus we only need to require 
preservation of the Virasoro symmetry, and the analysis below gives all 
topological defects and phase boundaries consistent with conformal symmetry.

\subsection{Chiral data of the minimal model M(5,6)}

The M(5,6) minimal model has central charge $c\eq 4/5$. The first of the
following two tables gives the lowest conformal weight of the representation 
corresponding to a given entry of the Kac table. The second table shows our 
choice of representatives and the names we will use for these representations.

\begin{center}
\begin{tabular}{c||c|c|c|c|c|}
\hline & & & & & \\[-.6em]
4 & $3$ & $\tfrac{13}{8}$ & $\tfrac{2}{3}$ & $\tfrac{1}{8}$ & $0$ \\[.3em]
\hline & & & & & \\[-.6em]
3 & $\tfrac{7}{5}$ & $\tfrac{21}{40}$ & $\tfrac{1}{15}$ & 
    $\tfrac{1}{40}$ & $\tfrac{2}{5}$ \\[.3em]
\hline & & & & & \\[-.6em]
3 & $\tfrac{2}{5}$ & $\tfrac{1}{40}$ & $\tfrac{1}{15}$ & 
    $\tfrac{21}{40}$ & $\tfrac{7}{5}$ \\[.3em]
\hline & & & & & \\[-.6em]
1 & $0$ & $\tfrac{1}{8}$ & $\tfrac{2}{3}$ & $\tfrac{13}{8}$ & $3$ \\[.3em]
\hline\hline & & & & & \\[-.6em]
 & 1 & 2 & 3 & 4 & 5 
\end{tabular}
\hspace{5em}
\begin{tabular}{c||c|c|c|c|c|}
\hline & & & & & \\[-.6em]
4 & & & & & \\[.3em]
\hline & & & & & \\[-.6em]
3 & $\hat\one$ & $\hat u$ & $\hat f$ & $\hat v$ & $\hat w$  \\[.3em]
\hline & & & & & \\[-.6em]
2 & & & & & \\[.3em]
\hline & & & & & \\[-.6em]
1 & $\one$ & $u$ & $f$ & $v$ & $w$  \\[.3em]
\hline\hline & & & & & \\[-.6em]
 & 1 & 2 & 3 & 4 & 5 
\end{tabular}
\end{center}

\noindent
Denote by $\calc_{5,6}$ the representation category of the vertex algebra
for the M(5,6) model. The index set $\II$ of representative simple objects in 
$\calc_{5,6}$ will be taken in the order
  \be
  \II = \{ \one, u, f, v, w, \hat\one, \hat u, \hat f, \hat v, \hat w \} \,. 
  \labl{eq:M56-I}
All objects of $\calc_{5,6}$ are isomorphic to their duals, so that 
$\bar k \eq k$ for all $k\iN\II$. The fusion rules of
$\calc_{5,6}$ are of the form $\mathfrak{su}(2)_4 \times (\text{Lee-Yang})$.
In more detail, the fusion ring of $\{ \one, u, f, v, w \}$
is that of $\mathfrak{su}(2)_4$, i.e.\ $u \cdot u \eq 1 + f$,~
$u \cdot f \eq u + v$, etc.\
Multiplication with hatted fields is determined by the rules
  \be
  \hat \one \cdot x \eq \hat x \quad{\rm for}\quad x \iN \{ \one, u, f, v, w \}
  \qquad \text{and} \qquad \hat \one \cdot \hat \one \eq \one + \hat \one 
  \ee
together with associativity and commutativity.

The modular $S$-matrix is given by (see e.g.\ section 10.6 of \cite{DIms}) 
  \be
  S = \begin{pmatrix} \zeta M & ~~\,\xi M \\ \xi M   & -\zeta M \end{pmatrix}
  \,, ~~~{\rm where}~~~    
  \zeta = \sqrt{\tfrac12(1\,{-}\,\tfrac{1}{\sqrt{5}})} \,,~~~
  \xi = \sqrt{\tfrac12(1\,{+}\,\tfrac{1}{\sqrt{5}})}
  \labl{eq:M56-S-mat}
and $M$ is the matrix
  \be
  M = \begin{pmatrix} 
  \tfrac{1}{2\sqrt{3}} & \tfrac12 & \tfrac{1}{\sqrt{3}} &
      \tfrac12 & \tfrac{1}{2\sqrt{3}} \\[.4em]
  \tfrac12 & -\tfrac12 & 0 & \tfrac12 & -\tfrac12 \\[.4em]
  \tfrac{1}{\sqrt{3}} & 0 & -\tfrac{1}{\sqrt{3}} &
      0 & \tfrac{1}{\sqrt{3}} \\[.4em]
  \tfrac12 & \tfrac12 & 0 & -\tfrac12 & -\tfrac12 \\[.4em]
  \tfrac{1}{2\sqrt{3}} & -\tfrac12 & \tfrac{1}{\sqrt{3}} &
      -\tfrac12 & \tfrac{1}{2\sqrt{3}} 
      \end{pmatrix} \,.
  \ee
Note that $\xi/\zeta \eq \tfrac12\,(1{+}\sqrt5)$. Rows and columns of the matrix 
\erf{eq:M56-S-mat} are ordered according to \erf{eq:M56-I}.
The invariant $s_{i,j}$ of the Hopf link is the ratio
  \be
  s_{i,j} = S_{i,j} / S_{0,0}
  \labl{eq:M56-smat}
and thus in particular the quantum dimension of the simple objects are
  \be
\begin{tabular}{c|cccccccccc}
$k$ & ~$\one$ & $u$ & $f$ & $v$ & $w$ & 
      $\hat\one$ & $\hat u$ & $\hat f$ & $\hat v$ & $\hat w$ 
\\[.3em] \hline &&\\[-.7em]       
$\dim(U_k)$ & ~$1$ & $\sqrt{3}$ & 2 & $\sqrt{3}$ & $1$ & $\tfrac{\xi}{\zeta}$ &
$\sqrt{3}\,\tfrac{\xi}{\zeta}$ & $2\,\tfrac{\xi}{\zeta}$ &
$\sqrt{3}\,\tfrac{\xi}{\zeta}$ & $\tfrac{\xi}{\zeta}$  
\end{tabular}
  \ee

\medskip

The entries of the fusion matrices $\FF$ are not really needed in the sequel.
For the sake of concreteness, we will nonetheless use the
following entry, obtained in the conventions of (A.6) in \cite{grrw2}:
  \be
  \Fs wwfff\one = \tfrac{81}{26} \,.
  \labl{eq:M56-fixed-F}


\subsection{The tetracritical Ising model}

The tetracritical Ising model is the A-series modular invariant. It is 
described by the Morita class of the symmetric special Frobenius algebra 
$A\eq\one$. Using the ribbon invariant (5.30) of \cite{tft1} one
finds that the torus partition function is diagonal, as it should be
($k\eq\bar k$ since all representations are self-conjugate):
  \be
  Z(A)_{ij} = \delta_{ij} \,.
  \ee
Further, $A\eq\one$ implies that the simple objects $U_k$, $k \iN \II$, are also
representatives of
all simple $A$-left modules as well as $A$-bimodules, and thus
  \be
  \JJ_A = \II \, , \qquad \KK_{AA} = \II \,.
  \ee
Accordingly, the fusion algebra of $A$-$A$-defects is just the fusion algebra 
of $\calc_{5,6}$. We see that there are two group-like $A$-$A$-defects $X_\one$ 
and $X_w$, which form a group isomorphic to $\zet_2$, and that there are no 
$A$-$A$-duality defects that are not already group-like.

The bulk fields of \CFTA\ are elements of $\HomAA(U_i \otiP A \otim U_j,A) 
\eq \Hom(U_i \oti U_j, \one)$, which has dimension $\delta_{ij}$. Consider the 
bulk field with left/right representation index $i$ labelled by the basis 
morphism $\phi_i\,{:=}\,\lambda_{(ii)0}$ (compare formula (2.29) of \cite{tft1}). 
The action \erf{eq:D_nu-def} on bulk fields amounts to multiplication with a 
ratio of $s$-matrix elements (not assuming $k \eq \bar k$ for the moment),
  \bea \begin{picture}(420,46)(0,46)
  \put(0,31)     {\Includeourtinynicepicture 50a }
  \put(0,31){
     \setlength{\unitlength}{.90pt}\put(-4,-4){
     \put(  1, 48) {\scriptsize$ D_\nu $}
     \put( 30, 20) {\scriptsize$ \phi_i $}
     }\setlength{\unitlength}{1pt}}
  \put(97,0)     {\Includeourtinynicepicture 50b }
  \put(97,0){
     \setlength{\unitlength}{.90pt}\put(-12,-9){
     \put( 47, 72) {\scriptsize$ i $}
     \put( 53, 31) {\scriptsize$ \overline\imath $}
     \put( 91, 82) {\scriptsize$ \nu $}
     }\setlength{\unitlength}{1pt}}
  \put(245,0)    {\Includeourtinynicepicture 50c }
  \put(245,0){
     \setlength{\unitlength}{.90pt}\put(-12,-9){
     \put( 54, 85) {\scriptsize$ i $}
     \put(45.2,51) {\scriptsize$ \overline\imath $}
     }\setlength{\unitlength}{1pt}}
  \put(65,47)    {$\hat=$}
  \put(198,47)   {$=~~\displaystyle\frac{s_{i,\overline\nu}}{s_{i,0}}$}
  \put(330,47)   {$\hat=~~\displaystyle\frac{s_{i,\overline\nu}}{s_{i,0}}$}
  \put(380,46)   {\Includeourtinynicepicture 50d } 
  \put(380,46){
     \setlength{\unitlength}{.90pt}\put(-4,-4){
     \put(9,1) {\scriptsize$ \phi_i $}
     }\setlength{\unitlength}{1pt}}
  \epicture21 \labl{pic-ffrs5-50}
{}From the explicit form of the $s$-matrix \erf{eq:M56-smat} we find that 
the group-like $A$-$A$-defect $D_w$ acts on on morphisms labelling $A$-bulk 
fields as
  \be\bearll
   D_w(\phi_i) = \phi_i \quad & {\rm for} ~ i \in \{ \one, f, w, 
     \hat\one, \hat f, \hat w \}  \,,
   \\{}\\[-.8em]
   D_w(\phi_i) = - \phi_i & {\rm for} ~ i \in \{ u,v,\hat u, \hat v \} \,.
  \eear\ee


\subsection{The three-state Potts model}\label{sec:potts-def}

The Potts model is the D-series modular invariant. It is described by the 
Morita class of the symmetric special Frobenius algebra $B\eq \one\,{\oplus}\,
U_w$. For concreteness, let us fix $m_{ww}^{~~\one}\eq 1$ in the expansion in 
\cite[eq.\,(3.7)]{tft1} for the multiplication on $B$. The automorphism group 
of $B$ is 
  \be
  \Aut(B) = \{e,\omega\} \cong \zet_2
  \ee
with
  \be
  e = \id_A \qquad \text{and} \qquad \omega = \id_\one \oplus (-\id_{U_w}) \,.
  \labl{e,omega}
To determine the simple $B$-modules we use the method presented in 
section 4.2 of \cite{tft3}. This tells us that the induced modules
  \be
  M_x = \indb(U_x)  ~\quad {\rm with}~~ x \iN \{\one,u,\hat\one,\hat u \}
  \labl{eq:potts-ind-simp}
are simple and that, owing to $\dimc\big(\HomB(\indb(U_x),\indb(U_x))\big)\eq 
\dimc\big(\Hom(A \oti U_x,U_x)\big)\eq 2$ for $x \iN \{f,\hat f\}$, we have
  \be
  \indb(U_{\!f}) \cong M_e \oplus M_\omega \qquad \text{and} \qquad
  \indb(U_{\!\hat f}) \cong M_{\hat e} \oplus M_{\hat \omega} 
  \labl{eq:Potts-ind-decomp}
for some simple $B$-modules $M_e$, $M_\omega$, $M_{\hat e}$ and 
$M_{\hat \omega}$. (While we use the same symbols $e$ and $\omega$ that label 
elements of the group $\Aut(B)$ to distinguish the simple modules, it should 
be kept in mind that this labelling is not canonical and that the two modules
only form a torsor over $\Aut(B)$.) These modules can be described as follows. 
Let $M_e \eq (U_{\!f},\rho_f)$, where $\rho_f\iN\Hom(A\oti U,U)$ is of the form
  \bea \begin{picture}(230,53)(0,40)
  \put(60,0)     {\Includeourtinynicepicture 55a }
  \put(60,0){
     \setlength{\unitlength}{.90pt}\put(-7,-12){
     \put(  8, 2)  {\scriptsize$ B $}
     \put( 35, 2)  {\scriptsize$ U_f $}
     \put( 35,111) {\scriptsize$ U_f $}
     }\setlength{\unitlength}{1pt}}
  \put(156,0)    {\Includeourtinynicepicture 55b }
  \put(156,0){
     \setlength{\unitlength}{.90pt}\put(-7,-12){
     \put(  8,  2) {\scriptsize$ B $}
     \put( 35,  2) {\scriptsize$ U_f $}
     \put( 35,111) {\scriptsize$ U_f $}
     \put(6.5, 72) {\scriptsize$ U_w $}
     }\setlength{\unitlength}{1pt}}
  \put(0,41)     {$\rho_f~=$}
  \put(114,41)   {$\oplus\quad\mu$}
  \epicture27 \labl{pic-ffrs5-55}
  for some $\mu\iN\complex$.
In order for $(U,\rho_f)$ to be a $B$-module, $\rho_f$ must obey in particular
  \bea \begin{picture}(420,66)(13,49)
  \put(0,0)      {\Includeourtinynicepicture 56a }
  \put(0,0){
     \setlength{\unitlength}{.90pt}\put(-12,-16){
     \put( 12,  7) {\scriptsize$ U_w $}
     \put( 36,  7) {\scriptsize$ U_w $}
     \put( 19, 76) {\scriptsize$ B $}
     \put( 39, 63) {\scriptsize$ B $}
     \put( 60,  7) {\scriptsize$ M_e $}
     \put( 60,140) {\scriptsize$ M_e $}
     }\setlength{\unitlength}{1pt}}
  \put(110,0)    {\Includeourtinynicepicture 56b }
  \put(110,0){
     \setlength{\unitlength}{.90pt}\put(-12,-16){
     \put( 12,  7) {\scriptsize$ U_w $}
     \put( 36,  7) {\scriptsize$ U_w $}
     \put( 25, 80) {\scriptsize$ B $}
     \put( 60,  7) {\scriptsize$ M_e $}
     \put( 60,140) {\scriptsize$ M_e $}
     }\setlength{\unitlength}{1pt}}
  \put(250,14)   {\Includeourtinynicepicture 56c }
  \put(250,14){
     \setlength{\unitlength}{.90pt}\put(-16,-16){
     \put( 14,  8) {\scriptsize$ U_w $}
     \put( 50,  8) {\scriptsize$ U_w $}
     \put( 91,  8) {\scriptsize$ U_f $}
     \put( 67, 69) {\scriptsize$ U_f $}
     \put( 52,109) {\scriptsize$ U_f $}
     }\setlength{\unitlength}{1pt}}
  \put(360,14)   {\Includeourtinynicepicture 56d }
  \put(360,14){
     \setlength{\unitlength}{.90pt}\put(-16,-16){
     \put( 14,  8) {\scriptsize$ U_w $}
     \put( 50,  8) {\scriptsize$ U_w $}
     \put( 91,  8) {\scriptsize$ U_f $}
     \put( 52,109) {\scriptsize$ U_f $}
     }\setlength{\unitlength}{1pt}}
  \put(77,51)    {$=$}
  \put(193,51)   {$\Longrightarrow \qquad \mu^2$}
  \put(335,51)   {$=$}
  \epicture31 \labl{pic-ffrs5-56}
Using the $\FF$-matrix element \erf{eq:M56-fixed-F} we see that
$\mu^2 \eq \tfrac{26}{81}$. Let us choose $\mu \eq {+}\sqrt{26}/9$.
One verifies that $(U_{\!f},\rho_f)$ is indeed a $B$-module.

For $\psi \iN \Aut(B)$ and $M \eq (\dot M,\rho)$ a $B$-module, denote by 
${}_\psi M$ the $B$-module ${}_\psi M \eq 
(\dot M, \rho \circ (\psi \oti \id_{\dot M}))$. Then we can choose
  \be
  M_\omega = {}_\omega (M_e) \,,~~~~
  M_{\hat e} = M_e \oti \hat\one \,,~~~~
  M_{\hat \omega} = {}_\omega (M_e) \oti \hat\one 
  \labl{eq:potts-ind-decomp}
with $\omega$ as defined in \erf{e,omega}.
The label set of simple $B$-modules is then
  \be
  \JJ_B = \{ \one,u,e,\omega, \hat\one,\hat u,\hat e,\hat\omega\} \,.
  \ee
According to theorem \ref{thm:schel-comm-bimod}, the
isomorphism classes of simple $B$-$B$-bimodules are labelled by pairs
  \be
  \KK_{BB} = 
  \{ (\kappa,\psi) \,|\, \kappa \iN \JJ_B,\, \psi \iN \Aut(B) \} \,.
  \ee
Thus there are $16$ isomorphism classes of
simple bimodules. Their dimensions are
  \be
  \begin{array}{c|cccccccc}
  x & ~\one & u & e & \omega & \hat\one & \hat u & \hat e & \hat\omega 
  \\[.2em] \hline \\[-.8em]
  \dim_B(M_{x,\psi})~ & 
  ~1 & \sqrt{3} & 1 & 1 &  \tfrac{\xi}{\zeta} & \sqrt{3}\,\tfrac{\xi}{\zeta} & 
  \tfrac{\xi}{\zeta} & \tfrac{\xi}{\zeta}  
  \eear \ee

\smallskip

It follows that there are six group-like bimodules, namely
  \be
  \mathcal{G}_B = \{ 
  (\one,e),\, (\one,\omega),\, (e,e),\, (e,\omega),\, (\omega,e),\,
  (\omega,\omega) \} \,.
  \ee

The fusion algebra of defects in the Potts model (or in the related 
$\mathfrak{su}(2)_4$ WZW model) has also been considered in \cite{chmp,coSc}. 
To see how it can be obtained in the present framework, we start with
the bimodule $\hat B \eq M_{\hat \one,e} \eq \alpha^+_B(\hat\one)$. From lemmas
\ref{lem:alpha-fuse} and \ref{lem:schel-tens} we see that
  \be
  \hat B \otB X \cong X \otB \hat B \qquad \text{and} \qquad
  \hat B \otB \hat B \cong B \oplus \hat B 
  \ee
for any $B$-$B$-bimodule $X$. Further, with \erf{eq:potts-ind-simp} and 
\erf{eq:potts-ind-decomp} it is straightforward to check that
  \be
  M_{x,\psi} \OtB \hat B \cong M_{\hat x,\psi}
  \qquad \text{for all} ~~ x \in \{1,u,e,\omega\}
  ~~\text{and all}~~ \psi \iN \Aut(B) \,.
  \ee
It is therefore enough to understand the tensor products of $M_{x,\psi}$ for 
$x \iN \{1,u,e,\omega\}$. Let us work in the Grothendieck ring of the tensor 
category of $B$-bimodules to simplify notation. We denote the isomorphism class 
of a bimodule $X$ by $[X]$ and abbreviate $[M_{x,\psi}] \,{=:}\, (x,\psi)$. 
Lemma \ref{lem:schel-tens} then tells us that
  \be
  [ \alpha^+_B(U_x)_\phi ] \cdot [ \alpha^+_B(U_y)_\psi ] 
  = [ \alpha^+_B( U_x \oti U_y )_{\phi\psi} ] 
  \ee
for $x,y \iN \{1,u,f\}$ and $\phi,\psi \iN \{e,\omega\}$. Using 
$(\one,\phi) \eq [B_\phi]$, $(u,\phi) \eq [\alpha^+(U_u)_\phi]$ and noting 
further the decomposition $[ \alpha^+_B(U_{\!f})_\phi ] \eq (e,\phi) + 
(\omega,\phi)$ which follows from \erf{eq:Potts-ind-decomp}, we find 
  \bea
  (\one,\phi) \cdot (\one,\psi) = (\one,\phi\psi) \,, \qquad\quad
  (\one,\phi) \cdot (u,\psi) = (u,\phi\psi) \,,
  \\{}\\[-.7em]
  (u,\phi) \cdot (u,\psi) = (\one,\phi\psi)+(e,\phi\psi)+(\omega,\phi\psi) \,,
  \\{}\\[-.7em]
  \big( (e,\phi)+(\omega,\phi) \big) \cdot (u,\psi) = 2\, (u,\phi\psi) \,.
  \eear\labl{eq:Potts-bimod-prods1}
It follows in particular that
  \be
  (\one,\phi) \cdot (u,\psi) = 
  (e,\phi)\cdot (u,\psi) = (\omega,\phi) \cdot (u,\psi) = (u,\phi\psi) \,.
  \labl{eq:Potts-u-fix}

Let us now have a closer look at the group-like bimodules. \erf{eq:Potts-u-fix} 
tells us that $(u,\psi)$ is a fixed point for the group-like bimodules labelled
by $(\one,e)$, $(e,e)$, $(\omega,e)$. Together with the third equality in 
\erf{eq:Potts-bimod-prods1} it follows that these form the stabiliser of 
$(u,\psi)$ and thus a subgroup of $\mathcal{G}_B$. A group with three elements 
is isomorphic to $\zet_3$, and so
  \be
  (e,e)\cdot(e,e) = (\omega,e) \,,~~~
  (\omega,e) \cdot (\omega,e) = (e,e) \,,~~~
  (e,e) \cdot (\omega,e) = (\omega,e) \cdot (e,e) = (\one,e) \,.
  \ee
For the multiplication by $(\one,\phi)$ we find by direct computation that 
  \be
  (\one,\phi) \cdot (\alpha,\psi) = (\alpha\phi,\phi\psi) \qquad \text{and}
  \qquad  (\alpha,\psi) \cdot (\one,\phi)  = (\alpha,\phi \psi) 
  \labl{eq:Potts-1phi-mult}
for $\alpha,\phi,\psi \iN \{e,\omega\}$. This shows that the fusion of bimodules
is nonabelian. Since there are just two non-isomorphic groups of order 6, the 
cyclic group, which is abelian, and the dihedral group, which is nonabelian and 
isomorphic to the symmetric group $S_3$, it follows that the Picard group of 
$B$-bimodules is isomorphic to $S_3$ and thus indeed coincides with the symmetry
group one expects for the three-state Potts model. In the sequel we will work 
out this group structure more explicitly (also note again that the 
identification of bimodule labels with group elements is not canonical). We 
first derive the equalities \erf{eq:Potts-1phi-mult}. For the first product we 
need to show that $M_{\one,\phi}\otB M_{\alpha,\psi} \Cong 
M_{\alpha\phi,\phi\psi}$. To this end we write the projector
$P \iN \End_{B|B}(M_{\one,\phi} \oti M_{\alpha,\psi})$, whose image defines the
tensor product, in the form $P \eq e \cir r$ such that $r \cir e \eq 
\id_{M_{\alpha\phi,\phi\psi}}$.  This is done in the following equation (for 
convenience we also indicate the left and right action of $B$):
  \begin{eqnarray}\begin{picture}(420,196)(21,0)
  \put(0,0)      {\Includeourtinynicepicture 58a }
  \put(0,0){
     \setlength{\unitlength}{.90pt}\put(-15,-11){
     \put( 13,  3) {\scriptsize$ B $}
     \put( 47,  3) {\scriptsize$ B $}
     \put(108,  3) {\scriptsize$ U_f $}
     \put( 47,224) {\scriptsize$ B $}
     \put(108,224) {\scriptsize$ U_f $}
     \put(143,  3) {\scriptsize$ B $}
     \put(145, 39) {\scriptsize$ \psi $}
     \put( 96, 82) {\scriptsize$ \alpha $}
     \put( 93,165) {\scriptsize$ \alpha $}
     \put( 65,165) {\scriptsize$ \phi $}
     \put( 23,120) {$ P $}
     }\setlength{\unitlength}{1pt}}
  \put(163,0)    {\Includeourtinynicepicture 58b }
  \put(163,0){
     \setlength{\unitlength}{.90pt}\put(-15,-11){
     \put(15,0){
     \put( 13,  3) {\scriptsize$ B $}
     \put( 47,  3) {\scriptsize$ B $}
     \put(108,  3) {\scriptsize$ U_f $}
     \put( 47,224) {\scriptsize$ B $}
     \put(108,224) {\scriptsize$ U_f $}
     \put(143,  3) {\scriptsize$ B $}
     \put( 80,131) {\scriptsize$ B $}}
     \put( 18, 42) {\scriptsize$ \alpha{\circ}\phi^{-\!1} $}
     \put( 54, 42) {\scriptsize$ \alpha{\circ}\phi^{-\!1} $}
     \put(153, 42) {\scriptsize$ \alpha \circ \psi $}
     \put( 54,188) {\scriptsize$ \phi{\circ}\alpha^{-\!1} $}
     }\setlength{\unitlength}{1pt}}
  \put(336,0)    {\Includeourtinynicepicture 58c }
  \put(336,0){
     \setlength{\unitlength}{.90pt}\put(-15,-11){
     \put(15,0){
     \put( 13,  3) {\scriptsize$ B $}
     \put( 55,  3) {\scriptsize$ B $}
     \put(108,  3) {\scriptsize$ U_f $}
     \put( 47,224) {\scriptsize$ B $}
     \put(108,224) {\scriptsize$ U_f $}
     \put(144.5,3) {\scriptsize$ B $}
     \put( 83,157) {\scriptsize$ B $}}
     \put( 18, 42) {\scriptsize$ \alpha{\circ}\phi^{-\!1} $}
     \put( 62, 42) {\scriptsize$ \alpha{\circ}\phi^{-\!1} $}
     \put(153, 37) {\scriptsize$ \phi \circ \psi $}
     \put(151, 61) {\scriptsize$ \alpha{\circ}\phi^{-\!1} $}
     \put( 54,196) {\scriptsize$ \phi{\circ}\alpha^{-\!1} $}
     \put( 42,148) {$ e $}
     \put( 56, 73) {$ r $}
     }\setlength{\unitlength}{1pt}}
  \put(144,93)   {$=$}
  \put(319,93)   {$=$}
  \end{picture}
  \nonumber\\[6pt]~ \label{pic-ffrs5-58}
  \end{eqnarray}
To make the connection with \erf{eq:Potts-1phi-mult} we also use that $\Aut(B)$
is commutative and that $\psi \eq \psi^{-1}$ for all $\psi \iN \Aut(B)$. The 
corresponding calculation for the second product in \erf{eq:Potts-1phi-mult} is
  \bea \begin{picture}(370,153)(0,43)
  \put(20,0)     {\Includeourtinynicepicture 59a }
  \put(20,0){
     \setlength{\unitlength}{.90pt}\put(-14,-11){
     \put( 18,  3) {\scriptsize$ B $}
     \put( 53,  3) {\scriptsize$ U_f $}
     \put(116,  3) {\scriptsize$ B $}
     \put(150,  3) {\scriptsize$ B $}
     \put( 53,224) {\scriptsize$ U_f $}
     \put(116,224) {\scriptsize$ B $}
     \put( 20, 43) {\scriptsize$ \alpha $}
     \put(152, 43) {\scriptsize$ \phi $}
     \put( 68,143) {\scriptsize$ \psi $}
     \put( 36,172) {\scriptsize$ \alpha $}
     \put( 23,110) {$ P $}
     }\setlength{\unitlength}{1pt}}
  \put(215,0)    {\Includeourtinynicepicture 59b }
  \put(215,0){
     \setlength{\unitlength}{.90pt}\put(-17,-11){
     \put( 22,  3) {\scriptsize$ B $}
     \put( 67,  3) {\scriptsize$ U_f $}
     \put(105,  3) {\scriptsize$ B $}
     \put(153,  3) {\scriptsize$ B $}
     \put( 68,224) {\scriptsize$ U_f $}
     \put(124,224) {\scriptsize$ B $}
     \put( 97, 27) {\scriptsize$ \alpha \circ \psi $}
     \put(146, 30) {\scriptsize$ \psi \circ \phi $}
     \put(152, 57) {\scriptsize$ \alpha $}
     \put( 24, 89) {\scriptsize$ \alpha $}
     \put(105,190) {\scriptsize$ \psi^{-\!1}{\circ}\,\alpha^{\!-\!1}$}
     \put( 46,145) {$ e $}
     \put( 57, 23) {$ r $}
     }\setlength{\unitlength}{1pt}}
  \put(177,91)   {$=$}
  \epicture27 \labl{pic-ffrs5-59}
where we also use that the algebra $B$ is commutative.
\\ 
Finally we can express $(\alpha,\psi) \eq (\alpha,e) \,{\cdot}\, (\one,\psi)
\eq (\one,\psi) \,{\cdot}\, (\alpha\psi,e)$. Thus all products are determined 
in terms of the ones already computed. 

We proceed to display the $S_3$ group structure of the fusion. We use the cycle
notation for elements of $S_n$, i.e.\ $(a\,b\,c\,\cdots\, d)$ stands for the 
permutation $a \,{\mapsto}\, b\,$, $b \,{\mapsto}\, c\,$,\,...\,, 
$d \,{\mapsto}\, a$, and consider 
the following assignment of permutations to elements of $\mathcal{G}_B$:
  \be
  \begin{array}{rlrlrl}
  \id   &\!\mapsto (\one,e)   \,,~~ &
  (123) &\!\mapsto (e,e)      \,,~~ &
  (132) &\!\mapsto (\omega,e) \,,
  \\{}\\[-.8em]
  (12) &\!\mapsto (\one,\omega)   \,,~~&
  (23) &\!\mapsto (\omega,\omega) \,,~~  &
  (13) &\!\mapsto (e,\omega) \,.
  \eear \ee
One verifies that this is a group isomorphism
$S_3 \,{\stackrel\cong\to}\, \mathcal{G}_B$. For example,
$(23)\cdot(13) \eq (123)$, as well as $(\omega,\omega)\cdot(e,\omega) \eq 
(\omega,e)\cdot(\one,\omega)\cdot(\one,\omega) \cdot(\omega,e) \eq (e,e)$.

Inspecting once more the fusion rules \erf{eq:Potts-bimod-prods1}
we see that $(u,e)$ and $(u,\omega)$ are the only
simple duality defects that are not already group-like, i.e.
  \be
  \mathcal{D}_{BB} = \mathcal{G}_B \cup \{ (u,e), (u,\omega) \} \,.
  \ee
Note that e.g.\ $(\one,\omega)\cdot(u,e) \eq (u,\omega)$, so that
the two new duality defects form one orbit under 
the left/right $\mathcal{G}_B$-action.


\subsection{Phase changing defects}\label{sec:M56-phasechange}

Topological defects that separate a tetracritical Ising phase \CFTA\ of a 
world sheet from a three-states Potts phase \CFTB\ are described by 
$B$-$A$-bimodules (or equivalently, by $A$-$B$-bimodules).
Because of $A \eq \one$ these are nothing but left $B$-modules. The 
isomorphism classes of simple $B$-$A$-bimodules can thus be labelled as
  \be
  \mathcal{K}_{BA} = \JJ_B = 
  \{ \one,u,e,\omega, \hat\one,\hat u,\hat e,\hat\omega\} \,.
  \ee
Let us compute the $\mathcal{G}_B{\times}\mathcal{G}_A$-action on 
$\mathcal{K}_{BA}$ to see how many orbits, i.e.\ phase-changing defects not 
linked by a symmetry, there are. For the right action we get
$M_\kappa \oti U_w \Cong M_\kappa$ as $B$-$A$-bi\-modules, for all $\kappa 
\iN \mathcal{K}_{BA}$. Since $B$ is commutative, a possible isomorphism is
  \bea \begin{picture}(240,81)(0,51)
  \put(20,0)     {\includeourbeautifulpicture 60 }
  \put(20,0){
     \setlength{\unitlength}{.38pt}\put(0,0){
     \put( 30,-23) {\scriptsize$ \dot M_\kappa $}
     \put( 92,-22) {\scriptsize$ U_w $}
     \put( 31,339) {\scriptsize$ \dot M_\kappa $}
     \put( 96,141) {\scriptsize$ B $}
     }\setlength{\unitlength}{1pt}}
  \put(90,58)    {$\in \Hom_{BA}(M_\kappa \oti U_w,M_\kappa)~.$}
  \epicture32 \labl{pic-ffrs5-60}
Thus $\mathcal{G}_A$ acts trivially.  To find the left $\mathcal{G}_B$-action 
we can use the fusion rules of $B$-$B$-bimodules and forget the right 
$B$-action. One obtains four $\mathcal{G}_B$-orbits:
  \be
  \{ \one,e,\omega \} \,,~~~ \{ \hat\one,\hat e,\hat\omega\} \,,~~~
  \{ u \} \,,~~~ \{ \hat u \} \,.
  \ee
Next we check whether any of the phase changing defects are duality defects. 
It is enough to do this for a representative of every orbit. The tensor
product of two `hatted' bimodules is never a sum of group-like bimodules,
so the only 
candidates for duality bimodules are the orbits of the the $B$-$A$-bimodules 
labelled by $\one$ and $u$, i.e.\ $B$ and $\indb(U_u)$. For $B$ we obtain
  \be
  B^\vee \OtB B \cong \one \oplus U_w \qquad \text{and} \qquad
  B \oti B^\vee \cong B \oplus B_\omega \,,
  \ee
where $B_\omega$ denotes the bimodule ${}_{\idsmall}B_{\omega}$ as given in 
definition \ref{def:schel-ind-bimod}. To see the last equivalence, note that
$B^\vee \Cong B$ as right $B$-module and that
  \be
  \Delta \in \HomBB(B,B \oti B) \qquad \text{and} \qquad
  (\id_B \oti \omega ) \cir \Delta \in \HomBB(B_\omega,B \oti B)
  \ee
are monomorphisms of $B$ and $B_\omega$ into $B \oti B$, respectively
(note that $\Delta$ is a monomorphism, since $B$ is simple and $\Delta$ is not 
zero). Evaluating the quantum dimension on both sides of $B \OtA B^\vee \Cong 
B \,{\oplus}\, B_\omega$ shows that these are all simple submodules.
Thus $B$ is both a $B$-$A$-duality defect and an $A$-$B$-duality defect.

For $\indb(U_u)$ the analogous calculation yields
  \be
  \indb(U_u)^\vee \OtB\, \indb(U_u) \cong 
  B \oti U_u^\vee \oti U_u = \one \oplus 2\, U_{\!f} \oplus U_w
  \ee
and
  \be \bearll
  \indb(U_u) \oti \indb(U_u)^\vee \!\!\! &\cong 
  \alpha^+_B(U_u \oti U_u^\vee) \oplus \alpha^+_B(U_u \oti U_u^\vee)_\omega
  \\{}\\[-.7em]&
  \cong B \oplus M_{e,e} \oplus M_{\omega,e} \oplus
  B_\omega \oplus M_{e,\omega} \oplus M_{\omega,\omega} \,.
  \eear\ee
The first equivalence follows with the help of the embedding morphisms
  \bea \begin{picture}(420,99)(9,49)
  \put(20,0)     {\includeourbeautifulpicture 61 }
  \put(20,0){
     \setlength{\unitlength}{.38pt}\put(0,0){
     \put( -3,-20) {\scriptsize$ B $}
     \put( 91,-20) {\scriptsize$ U_u $}
     \put(153,-20) {\scriptsize$ U_u^\vee $}
     \put( 32,385) {\scriptsize$ B $}
     \put( 91,385) {\scriptsize$ U_u $}
     \put(153,385) {\scriptsize$ U_u^\vee $}
     \put(217,385) {\scriptsize$ B^\vee $}
     \put( 85,195) {\scriptsize$ \phi $}
     }\setlength{\unitlength}{1pt}}
  \put(155,67)   {$\in \HomBB(\alpha^+_B(U_u \oti U_u^\vee)_\phi ,
                   \indb(U_u) \oti \indb(U_u)^\vee)$}
  \epicture30 \labl{pic-ffrs5-61}
for $\phi \iN \{e,\omega\}$.

We see that $\indb(U_u) \oti \indb(U_u)^\vee$ decomposes into a direct sum 
of group-like $B$-$B$-bimodu\-les, whereas $\indb(U_u)^\vee \OtB \indb(U_u)$ 
is not a direct sum of only group-like $A$-$A$-bimodules. Thus by theorem 
\ref{thm:dual}, $\indb(U_u)$ is an example of a topological
defect that is not itself a $B$-$A$-duality defect even though
$\indb(U_u)^\vee$ is an $A$-$B$-duality defect.

\medskip

Given a simple $B$-$A$-bimodule $X_\mu$, $\mu \iN \mathcal{K}_{BA}$, we obtain 
a simple $A$-$B$-bimodule by taking its dual $(X_\mu)^\vee$. Let us choose these
as representatives for the isomorphism classes of simple $A$-$B$-bi\-modules,
so that $\mathcal{K}_{AB} \eq \mathcal{K}_{BA}$. The above calculation
then shows that the simple $A$-$B$- and $B$-$A$-duality defects are
  \be
  \mathcal{D}_{AB} = \{ 1,e,\omega,u \} \qquad{\rm and}\qquad
  \mathcal{D}_{BA} = \{ 1,e,\omega \} \,,
  \ee
respectively. According to proposition \ref{prop:orbifold}, there exists thus 
one way more to obtain the torus partition function of the tetra-critical 
Ising model from the Potts model than vice versa. 

In more detail, each of the subgroups of $S_3$ is conjugate to either $\{\id\}$,
$\{\id,(12)\} \Cong \zet_2$, $A_3 \Cong \zet_3$, or $S^3$. The above results 
imply that when applied to the subgroups $\{\id \}$ or $A_3$, the orbifold-like 
expression \erf{pic-ffrs5-66} gives again the partition function of the Potts 
model, while applying the same prescription to the subgroups $\zet_2$ or $S_3$ 
results in the partition function of the tetra-critical Ising model.
In the reverse direction, we can only use the $\zet_2$ symmetry 
of the tetra-critical Ising model to obtain the Potts model.

An analogous lattice construction, relating e.g.\ the A$_5$ height model and the
D$_4$ height model via a `non-critical orbifold' was given in \cite{fegi}.

\vskip2em

\noindent
{\bf Acknowledgements.}\\
We thank Urs Schreiber for helpful comments and discussions, and
Natalia Potylitsina-Kube for her skillfull help with
the numerous illustrations.
I.R.\ thanks the Max-Planck-Institute f\"ur Gravitationsphysik in Potsdam,
and C.S.\ the Erwin-Schr\"odinger-Institute in Vienna,
for a stimulating stay during which part of this work was completed.
J.F.\ is supported by VR under project no.\ 621--2003--2385, and I.R.\
by the EU RTN grants `Euclid', contract number HPRN-CT-2002-00325, and
`Superstring Theory', contract number MRTN-CT-2004-512194.

 \newpage

\newcommand\wb{\,\linebreak[0]} \def\wB {$\,$\wb}
 \newcommand\Bi       {\bibitem}
 \renewcommand\J[5]     {{\sl #5\/}, {#1} {#2} ({#3}) {#4} }
 \renewcommand\K[6]     {{\sl #6\/}, {#1} {#2} ({#3}) {#4}~\,[#5]}
 \newcommand\PhD[2]   {{\sl #2\/}, Ph.D.\ thesis (#1)}
 \newcommand\Mast[2]  {{\sl #2\/}, Master's thesis (#1)}
 \newcommand\Prep[2]  {{\sl #2\/}, pre\-print {#1}}
 \newcommand\BOOK[4]  {{\sl #1\/} ({#2}, {#3} {#4})}
 \newcommand\inBO[7]  {{\sl #7\/}, in:\ {\sl #1}, {#2}\ ({#3}, {#4} {#5}), p.\ {#6}}
 \newcommand\inBOxxx[3]  {{\sl #3\/}, in:\ {\sl #1} (#2)}
 \newcommand\Erra[3]  {\,[{\em ibid.}\ {#1} ({#2}) {#3}, {\em Erratum}]}
 \def\jf    {J.\ Fuchs}
 \def\dim   {dimension}  
 \def\adma  {Adv.\wb Math.}
 \def\anop  {Ann.\wb Phys.}
 \def\aspm  {Adv.\wb Stu\-dies\wB in\wB Pure\wB Math.}
 \def\coia  {Com\-mun.\wB in\wB Algebra}
 \def\coma  {Con\-temp.\wb Math.}
 \def\comp  {Com\-mun.\wb Math.\wb Phys.}
 \def\cpma  {Com\-pos.\wb Math.}
 \def\fiic  {Fields\wB Institute\wB Commun.}
 \def\foph  {Fortschritte\wB d.\wb Phys.}
 \def\ijmb  {Int.\wb J.\wb Mod.\wb Phys.\ B}
 \def\ijmp  {Int.\wb J.\wb Mod.\wb Phys.\ A}
 \def\jams  {J.\wb Amer.\wb Math.\wb Soc.}
 \def\jgap  {J.\wb Geom.\wB and\wB Phys.}
 \def\jhep  {J.\wb High\wB Energy\wB Phys.}
 \def\joal  {J.\wB Al\-ge\-bra}
 \def\jomp  {J.\wb Math.\wb Phys.}
 \def\jopa  {J.\wb Phys.\ A}
 \def\josp  {J.\wb Stat.\wb Phys.}
 \def\jpaa  {J.\wB Pure\wB Appl.\wb Alg.}
 \def\jste  {J.\wb Stat.\wb Mech.:\wB Theory\wB Exp.}
 \def\lemp  {Lett.\wb Math.\wb Phys.} 
 \def\maan  {Math.\wb Annal.}
 \def\mpla  {Mod.\wb Phys.\wb Lett.\ A}
 \newcommand\nqma[2] {\inBO{Non-perturbative QFT Methods and Their Applications}
            {Z.\ Horv\'ath and L.\ Palla, eds.} \WS\Si{2001} {{#1}}{{#2}} }
 \newcommand\nqft[2] {\inBO{Nonperturbative Quantum Field Theory}
            {G.\ 't Hooft et al., eds.} \PL\NY{1988} {{#1}}{{#2}} }
 \def\nuci  {Nuovo\wB Cim.}
 \def\nupb  {Nucl.\wb Phys.\ B}
 \newcommand\phgt[2] {\inBO{Physics, Geometry, and Topology}
            {H.C.\ Lee, ed.} \PL\NY{1990} {{#1}}{{#2}} }
 \def\phla  {Phys.\wb Lett.\ A}
 \def\phlb  {Phys.\wb Lett.\ B}
 \def\phrd  {Phys.\wb Rev.\ D}
 \def\phre  {Phys.\wb Rev.\ E}
 \def\phrl  {Phys.\wb Rev.\wb Lett.}
 \def\phrp  {Phys.\wb Rep.}
 \def\phrv  {Phys.\wb Rev.}
 \def\prtp  {Progr.\wb Theor.\wb Phys.}
 \newcommand\qfsm[2] {\inBO{\Q Fields and Strings: A Course for Mathematicians}
            {P.\ Deligne et al., eds.} \AMS\PR{1999} {{#1}}{{#2}} }
 \def\remp  {Rev.\wb Mod.\wb Phys.}
 \def\rvmp  {Rev.\wb Math.\wb Phys.}
 \def\sebo  {S\'emi\-naire\wB Bour\-baki}
 \def\sjnp  {Sov.\wb J.\wb Nucl.\wb Phys.}
 \def\taac  {Theo\-ry\wB and\wB Appl.\wB Cat.}
 \def\trgr  {Trans\-form.\wB Groups}
 \def\zfpc  {Z.\wb Phy\-sik C}
 \def\AMS    {{American Mathematical Society}}
 \def\EMS    {{European Mathematical Society}}
 \def\PL     {{Plenum Press}}
 \def\SV     {{Sprin\-ger Ver\-lag}}
 \def\WS     {{World Scientific}}
 \def\Bo     {{Boston}} 
 \def\PR     {{Providence}}
 \def\Si     {{Singapore}}
 \def\NY     {{New York}}


\small
\begin{thebibliography}{XX} \addtolength{\itemsep}{-6pt}
\bibitem[{\,0\,}]{tft0} {\jf, I.\ Runkel, and C.\ Schweigert,
           \K\nupb{624}{2002}{452}
           {hep-th/0110133} {\Con \corfu s, \frob s and triangulations}}
\bibitem[{\,I\,}]{tft1} {\jf, I.\ Runkel, and C.\ Schweigert,
           \K\nupb{646}{2002}{353} {hep-th/0204148}
           {TFT construction of RCFT correlators I: Partition functions}}
\bibitem[{\,II\,}]{tft2} {\jf, I.\ Runkel, and C.\ Schweigert,
           \K\nupb{678}{2004}{511} {hep-th/0306164}
           {TFT construction of RCFT correlators II: Unori\-en\-ted surfaces}}
\bibitem[{\,III\,}]{tft3} {\jf, I.\ Runkel, and C.\ Schweigert,
           \K\nupb{694}{2004}{277} {hep-th/0403158}
           {TFT construction of RCFT correlators III: Simple currents}}
\bibitem[{\,IV\,}]{tft4} {\jf, I.\ Runkel, and C.\ Schweigert,
           \K\nupb{715}{2005}{539} {hep-th/0412290} {TFT construction of RCFT
           correlators IV: Structure constants and correlation functions}}
\bibitem[{\,V\,}]{tft5} {J.\ Fjelstad, \jf, I.\ Runkel, and C.\ Schweigert,
           \K\taac{16}{2006}{342} {hep-th/0503194}
           {TFT construction of RCFT correlators V: Proof of modular
           invariance and factorisation}}
\Bi{fuRs11}{\jf, I.\ Runkel, and C.\ Schweigert,
           \Prep{math.CT/0511590, to appear in \coma}
           {Ribbon \cats\ and (unoriented) CFT: \frob s, automorphisms, reversions}}
\Bi{vazh}  {F.\ Van Oystaeyen and Y.H.\ Zhang,
           \J\joal{202}{1998}{96} {The Brauer group of a braided monoidal \cat}}
\Bi{woaf}  {E.\ Wong and I.\ Affleck, \J\nupb{417}{1994}{403}
           {Tunneling in \q wires: A boundary \cft\ approach}}
\Bi{card}  {J.L.\ Cardy, \J\nupb{240}{1984}{514}
           {Conformal invariance and surface critical behavior}}
\Bi{osaf}  {M.\ Oshikawa and I.\ Affleck,
           \K\phrl{77}{1996}{2604} {hep-th/9606177}
           {Defect lines in the Ising model and boundary states on orbifolds}}
\Bi{osaf2} {M.\ Oshikawa and I.\ Affleck,
           \K\nupb{495}{1997}{533} {cond-mat/9612187} {Boundary \cft\
           approach to the critical \twodim\ Ising model with a defect line}}
\Bi{bddo}  {C.\ Bachas, J.\ de Boer, R.\ Dijkgraaf, and H.\ Ooguri,
           \K\jhep{0206}{2002}{027}
           {hep-th/0111210} {Permeable \con walls and holography}}
\Bi{qusc}  {T.\ Quella and V.\ Schomerus, \K\jhep{0206}{2002}{028}
           {hep-th/0203161} {Symmetry breaking boundary states and defect lines}}
\Bi{watt8} {G.M.T.\ Watts, \K\nupb{596}{2001}{513} {hep-th/0002218}
           {On the boundary Ising model with disorder operators}}
\Bi{pezu5} {V.B.\ Petkova and J.-B.\ Zuber, \K\phlb{504}{2001}{157}
           {hep-th/0011021}  {Generalized twisted \parfu s}}
\Bi{pezu7} {V.B.\ Petkova and J.-B.\ Zuber, \inBO{Non-perturbative QFT Methods
           and Their Applications} {Z.\ Horvath and L.\ Palla, eds.}
           \WS\Si{2001} {1~\,[hep-th/0103007]} {\Con \bc s and what they teach us}}
\Bi{coSc}  {R.\ Coquereaux and G.\ Schieber,
           \K\jgap{42}{2002}{216} {hep-th/0107001} {Twisted
           \parfu s for ADE boundary \cfts\ and Ocneanu \alg s of \q sym\-me\-tries}}
\Bi{grwa}  {K.\ Graham and G.M.T.\ Watts, \K\jhep{0404}{2004}{019}
           {hep-th/0306167} {Defect lines and boundary flows}}
\Bi{ffrs3} {J.\ Fr\"ohlich, \jf, I.\ Runkel, and C.\ Schweigert,
           \K\phrl{93}{2004}{070601}
           {cond-mat/0404051} {Kramers\hy Wannier duality from \con defects}}
\Bi{cmop}  {C.H.O.\ Chui, C.\ Mercat, W.P.\ Orrick, and P.A.\ Pearce,
           \K\phlb{517}{2001}{429} {hep-th/0106182}
           {Integrable lattice realizations of \con twisted \bc s}}
\Bi{otpe2} {C.H.\ Otto Chui and P.A.\ Pearce,
           \K\jste{0506}{2005}{P008} {hep-th/0505085} {Lattice realizations
           of the open descendants of twisted \bc s for $sl(2)$ A-D-E models}}
\Bi{baGa}  {C.\ Bachas and M.\ Gaberdiel, \K\jhep{0411}{2004}{065}
           {hep-th/0411067} {Loop operators and the Kondo problem}}
\Bi{krwa}  {H.A.\ Kramers and G.H.\ Wannier, \J\phrv{60}{1941}{252}
           {Statistics of the \twodim\ ferromagnet, Part I.}}
\Bi{savi}  {R.\ Savit,
           \J\remp{52}{1980}{453} {Duality in \ft\ and statistical systems}}
\Bi{drwa}  {K.\ Dr\"uhl and H.\ Wagner, \J\anop{141}{1982}{225} {Algebraic
           formulation of duality transformations for abelian lattice models}}
\Bi{petk}  {V.B.\ Petkova, \J\ijmp3{1988}{2945} {Two-dimensional
           (half-) integer spin conformal theories with central charge $c{<}1$}}
\Bi{fugp}  {P.\ Furlan, A.Ch.\ Ganchev, and V.B.\ Petkova,
           \J\ijmp5{1990}{2736} {Fusion
           matrices and $c{<}1$ (quasi)local conformal theories}}
\Bi{gaCa}  {A.\ Gamsa and J.\ Cardy, \K\jopa{39}{2006}{12983} {math-ph/0606065} 
           {Correlation
           functions of twist operators applied to single self-avoiding loops}}
\Bi{kawi}  {A.N.\ Kapustin and E.\ Witten, \Prep{hep-th/0604151}
           {Electric-magnetic duality and the geometric Langlands program}}
\Bi{pezu6} {V.B.\ Petkova and J.-B.\ Zuber, \K\nupb{603}{2001}{449}
           {hep-th/0101151} {The many faces of Ocneanu cells}}
\Bi{ruel'5}{Ph.\ Ruelle, \K\phrl{95}{2005}{225701}
           {cond-mat/0504758} {Kramers-Wannier dualities via sym\-me\-tries}}
\Bi{boek}  {J.\ B\"ockenhauer, D.E.\ Evans, and Y.\ Kawahigashi,
           \K\comp{208}{1999}{429} {math.OA/9904109} {On
           $\alpha$-induction, chiral generators and \modinv ts for subfactors}}
\Bi{ostr}  {V.\ Ostrik, \K\trgr8{2003}{177}
           {math.QA/0111139} {Module \cats, weak Hopf \alg s and \modinv ts}}
\Bi{TUra}  {V.G.\ Turaev, \BOOK{\Q Invariants of Knots and $3$-Manifolds}
           {de Gruyter}\NY{1994}}
\Bi{ffrs}  {J.\ Fr\"ohlich, \jf, I.\ Runkel, and C.\ Schweigert, \K\adma{199}
           {2006}{192} {math.CT/0309465} {Correspondences of ribbon \cats}}
\Bi{huan24}{Y.-Z.\ Huang, \K\coma{391}{2005}{135} {math.QA/0502558}
           {Vertex \oa s, \furu s and modular transformations}}
\Bi{joSt5} {A.\ Joyal and R.\ Street,
           \J\adma{88}{1991}{55} {The geometry of tensor calculus, I}}
\Bi{fuSc16}{\jf\ and C.\ Schweigert, \K\fiic{39}{2003}{25}
           {math.CT/0106050} {Category theory for \con \bc s}}
\Bi{stre8} {R.\ Street,
           \J\jomp{45}{2004}{3930} {Frobenius monads and pseudomonoids}}
\Bi{laud}  {A.D.\ Lauda, \K\taac{16}{2006}{84}
           {math.QA/0502550} {Frobenius \alg s and ambidextrous adjunctions}}
\Bi{muge8}{M.\ M\"uger, \K\jpaa{180}{2003}{81} {math.CT/0111204} {From subfactors
           to \cats\ and topology I.\ \frob s in and Morita equivalence of \tcs}}
\Bi{yama12}{S.\ Yamagami,
           \J\fiic{43}{2004}{551} {\frob s in \tcs\ and bimodule extensions}} 
\Bi{scfr2} {C.\ Schweigert, \jf, and I.\ Runkel, \inBO{{\rm Proceedings of the}
           International Congress of Mathematicians 2006} {M.\ Sanz-Sol\'e, J.\
           Soria, J.L.\ Varona, and J.\ Verdera, eds.} \EMS{}{2006}
           {443~\,[math.CT/0602079]} {Categorification and \corfu s in \cft}}
\Bi{stre3} {R.\ Street, \inBOxxx{Proceedings CT95} {Halifax 1995}
           {Low-dimensional topology and higher-order categories}}
\Bi{laud2} {A.D.\ Lauda, \Prep{math.QA/0508349} {Frobenius algebras
           and planar open string topological field theories}}
\Bi{huko}  {Y.-Z.\ Huang and L.\ Kong, \K\comp{250}{2004}{433}
           {math.QA/0308248} {Open-string vertex algebras, \tcs\ and operads}}
\Bi{caet}  {D.\ Calaque and P.I.\ Etingof,
           \Prep{math.QA/0401246} {Lectures on \tcs}}
\Bi{chmp}  {C.H.O.\ Chui, C.\ Mercat, and P.A.\ Pearce,
           \K\jopa{36}{2003}{2623} {hep-th/0210301} {Integrable 
           and \con twisted \bc s for $sl(2)$ $A$-$D$-$E$ lattice models}}
\Bi{ppc}   {P.A.\ Pearce, private communication}
\Bi{dofa2} {Vl.S.\ Dotsenko and V.A.\ Fateev,
           \J\nupb{251}{1985}{691} {Four-point \corfu s
           and \oa\ in 2D \con invariant theories with central charge $c\le1$}}
\Bi{fegi}  {P.\ Fendley and P.\ Ginsparg,
           \J\nupb{324}{1989}{549} {Non-critical orbifolds}}
\Bi{grrw2} {K.\ Graham, I.\ Runkel, and G.M.T.\ Watts,
           \K\nupb{608}{2001}{527}
           {hep-th/0101187} {Minimal model boundary flows and $c=1$ CFT}}
\Bi{pare14}{B.\ Pareigis, \J{Publ.\wb Math.\wB Debrecen}{25}{1978}{177}
           {Non-additive ring and module theory III.\ Morita theorems}}
\Bi{pare16}{B.\ Pareigis, \J\coia9{1981}{1455}
           {Morita equivalence of module \cats\ with tensor products}}
\Bi{scsW}  {U.\ Schreiber, C.\ Schweigert, and K.\ Waldorf, \Prep{hep-th/0512283}
           {Unoriented WZW models and holonomy of bundle gerbes}}
\Bi{ruve}  {Ph.\ Ruelle and O.\ Verhoeven, \K\nupb{535}{1998}{650}
           {hep-th/9803129} {Discrete symmetries of unitary minimal \con theories}}
\Bi{ruel'3}{Ph.\ Ruelle, \K\jopa{32}{1999}{8831}
           {hep-th/9904100} {Symmetric \bc s in boundary critical phenomena}}
\Bi{fffs2} {G.\ Felder, J.\ Fr\"ohlich, \jf, and C.\ Schweigert, \K\phrl{84}
           {2000}{1659} {hep-th/9909140} {\Con \bc s and three-\dim al \tft}}
\Bi{witt27}{E.\ Witten, \J\comp{121}{1989}{351}
           {Quantum field theory and the Jones polynomial}}
\Bi{frki}  {J.\ Fr\"ohlich and C.\ King, \J\ijmp4{1989}{5321}
           {Two-\dim al \cft\ and three-\dim al topology}}
\Bi{retu}  {N.Yu.\ Reshetikhin and V.G.\ Turaev, \J\comp{127}{1990}1
           {Ribbon graphs and their invariants derived from \qg s}}
\Bi{scya}  {A.N.\ Schellekens and S.\ Yankielowicz, \J\nupb{327}{1989}{673}
           {Extended chiral \alg s and \modinv t \parfu s}}
\Bi{scya6} {A.N.\ Schellekens and S.\ Yankielowicz,
           \J\ijmp5{1990}{2903} {Simple currents, \modinv ts, and fixed points}}
\Bi{fusw}  {\jf, C.\ Schweigert, and J.\ Walcher, \K\nupb{588}{2000}{110}
           {hep-th/0003298} {Projections in string theory and 
           boundary states for Gepner models}}
\Bi{scfr}  {C.\ Schweigert, \jf, and I.\ Runkel, \Prep{math.CT/0602077}
           {Twining characters and Picard groups in rational \cft}}
\Bi{naid}  {D.\ Naidu, \Prep{math.\linebreak[0]QA/0605530}
           {Categorical Morita equivalence for group-theoretical \cats}}
\Bi{jf18}  {\jf, \J\ijmb6{1992}{1951} {WZW quantum dimensions}}
\Bi{garw}  {T.\ Gannon, P.\ Ruelle, and M.A.\ Walton, \K\comp{179}{1996}{121}
           {hep-th/9503141} {Automorphism \modinv ts of current \alg s}}
\Bi{DIms}  {P.\ di Francesco, P.\ Mathieu, and D.\ Senechal,
           \BOOK{Conformal Field Theory} \SV\NY{1996}}
\end{thebibliography}
\end{document}